\documentclass[12pt]{article}
\pdfoutput=1
\usepackage{amsmath,amssymb,epsfig,amsfonts}
\usepackage{graphicx,subfigure}
\usepackage[usenames, dvipsnames]{color}
\usepackage{multirow}
\usepackage[backref]{hyperref}
\usepackage[nosort]{cite}
\usepackage{lscape}
\usepackage{rotating}
\usepackage[normalem]{ulem}

\usepackage{extarrows}

\usepackage{verbatim} 

\makeatletter

\newcommand{\be}{\begin{equation}}
\newcommand{\ee}{\end{equation}}
\newcommand{\ba}{\begin{aligned}}
\newcommand{\ea}{\end{aligned}}
\newcommand{\su}{\mathfrak{su}}
\newcommand{\so}{\mathfrak{so}}

\newcommand{\lp}{\left(}
\newcommand{\rp}{\right)}
\newcommand{\nn}{\nonumber}
\newcommand{\bea}{\begin{eqnarray}}
\newcommand{\eea}{\end{eqnarray}}
\newcommand{\p}{\partial}
\newcommand{\eee}{\mathbf{e}}

\newcommand{\wat}{\widehat}

\def\abcd#1#2#3#4{{\left(\begin{array}{cc}#1&#2\\#3&#4\end{array}\right)}}

\newcommand{\Z}{{\mathbb Z}}

\newcommand{\ti}[1]{\tilde{#1}}

\def\Tr{\mathop{\mathrm{Tr}}\nolimits}

\def\p{\partial}

\def\unit{{1\kern-.65ex {\rm l}}}
\def\1{{1\kern-.65ex {\rm l}}}

\def\CA{{\cal A}}
\def\CB{{\cal B}}
\def\CC{{\cal C}}
\def\CD{{\cal D}}
\def\CE{{\cal E}}

\def\CL{{\cal L}}

\def\CN{{\cal N}}

\def\bbE{{\mathbb{E}}}

\def\bbP{{\mathbb{P}}}

\def\bbR{{\mathbb{R}}}

\def\bbZ{{\mathbb{Z}}}

\newcount\hour \newcount\minute
\hour=\time \divide \hour by 60
\minute=\time
\def\now{%
\ifnum \hour<13
  \ifnum \hour=0 \advance \hour by 12 \number\hour:\else \number\hour:\fi%
     \ifnum \minute<10 0\fi%
     \number\minute%
\ A.M.%
\else \advance \hour by -12 \number\hour:%
  \ifnum \minute<10 0\fi%
  \number\minute%
  \ P.M.%
\fi%
}

\makeatother

\begin{document}

\baselineskip=18pt  
\numberwithin{equation}{section}  


%
%


\thispagestyle{empty}

\vspace*{-2cm} 
\begin{flushright}
{\tt  CERN-TH-2016-216 } 
\end{flushright}
%
\vspace*{0.8cm} 
\begin{center}
{\Huge Six-dimensional Origin of \\
$\CN=4$ SYM with Duality Defects \\
 }

 \vspace*{1.5cm}
Benjamin Assel$^{1,3}$ and Sakura Sch\"afer-Nameki$^{2, 3}$\\

 \vspace*{1.0cm}

 {\it $^1$ CERN, Theoretical Physics Department, \\
 1211 Geneva 23, Switzerland }\\
 
{\it $^2$ Mathematical Institute, University of Oxford,\\
Woodstock Road, Oxford, OX2 6GG, UK}\\

{\it $^3$ Department of Mathematics, King's College \\
 The Strand, London WC2R 2LS, UK }\\
\smallskip

 {\tt {gmail:$\,$  benjamin.assel, sakura.schafer.nameki}}\\

\vspace*{0.8cm}
\end{center}
\vspace*{.1cm}
\noindent 
We study the topologically twisted compactification of the 6d $(2,0)$ M5-brane theory on an elliptically fibered K\"ahler three-fold preserving two supercharges. We show that upon reducing on the elliptic fiber, the 4d theory is $\mathcal{N}=4$ Super-Yang Mills, with varying complexified coupling  $\tau$, in the presence of defects. 
For abelian gauge group this agrees with the so-called duality twisted theory, and we determine a non-abelian generalization to $U(N)$. 
When the elliptic fibration is singular, the 4d theory contains 3d walls (along the branch-cuts of $\tau$) and 2d surface defects, around which the 4d theory undergoes  $SL(2,\mathbb{Z})$ duality transformations. Such duality defects carry chiral fields, which from the 6d point of view arise as modes of the two-form $B$ in the tensor multiplet. Each duality defect has a flavor symmetry associated to it, which is encoded in the structure of the singular elliptic fiber above the defect. 
Generically 2d surface defects will intersect in points in 4d, where there is an enhanced flavor symmetry. 
The 6d point of view provides a  complete characterization of this 4d-3d-2d-0d `Matroshka'-defect configuration.

\newpage


\tableofcontents


\section{Introduction}

S-duality is one of the profound characteristics of $\mathcal{N}=4$ Super-Yang Mills (SYM) theory in 4d with gauge group $G$. It dates back to the Montonen-Olive duality \cite{Montonen:1977sn, Goddard:1976qe}, and in the context of $\mathcal{N}=4$ SYM, much evidence exists in its favor, in particular the seminal work \cite{Vafa:1994tf}, and, more recently, its relation to the geometric Langlands program in \cite{Kapustin:2006pk}. 
For the theory with a $U(N)$ gauge group, the conjectured duality states that the theory, with the complexified coupling constant $\tau$ defined in terms of the $\theta$-angle and gauge coupling $g$ as 
\be
\tau = {\theta \over 2 \pi} + i {4 \pi \over g^2} \,,
\ee
is self-dual under  $\tau \rightarrow \gamma \tau$, with $\gamma\in SL(2,\mathbb{Z})$.

The $SL(2, \mathbb{Z})$ duality group of $\mathcal{N}=4$ SYM with gauge group $U(N)$ has a geometric realization in terms of the 6d $(2,0)$ superconformal theory of type $A_{N-1}$, i.e. the theory on a stack of M5-branes in M-theory, by compactification on an elliptic curve with complex structure given by the modular parameter $\tau$. By shrinking the torus fiber, the 6d $(2,0)$ theory reduces at low energies to 4d $\mathcal{N}=4$ SYM. S-duality is then identified with $SL(2,\mathbb{Z})$ modular transformations acting on $\tau$ \cite{Witten:2009at}.

In string theory, a closely related duality is the S-duality of Type IIB string theory, acting on the axio-dilaton $\tau_{IIB} = C_0 + i e^{-\phi}$, which when acted upon on a D3-brane in Type IIB results precisely in the S-duality of the effective $\mathcal{N}=4$ SYM on its world-volume. 
In this case, a formulation, which allows for a manifest realization of $SL(2,\mathbb{Z})$-duality is provided by F-theory  \cite{Vafa:1996xn}, which allows the study of string theory vacua, where the axio-dilaton varies. 

Our goal here is to determine a formulation of $\mathcal{N}=4$ SYM with varying complexified coupling constant in analogy to the F-theoretic approach to IIB string theory. We find that this setup arises naturally from the
 6d $(2,0)$ theory on an elliptically fibered complex three-fold, by dimensional reduction along the elliptic fiber.
 
Supersymmetry can be preserved,  if no topological twist is applied to the 6d theory, if the elliptic three-fold is Calabi-Yau. 
More generally, supersymmetry can be preserved for an arbitrary elliptically fibered K\"ahler three-fold if the theory is topologically twisted. We will study both situations in detail. 
We will find that imposing K\"ahlerity is tied to $\tau$ being a holomorphic function on the base $B_2$.
 This setup is of particular interest, when the fibration becomes singular, i.e. whenever the complex structure degenerates. Singularities in the fiber happen generically over 2d loci, which are complex curves in the base of the fibration. 
 We will find that there are additional chiral degrees of freedom localized at these 2d defects. 
  
\begin{figure}
  \centering
  \includegraphics[width=6cm]{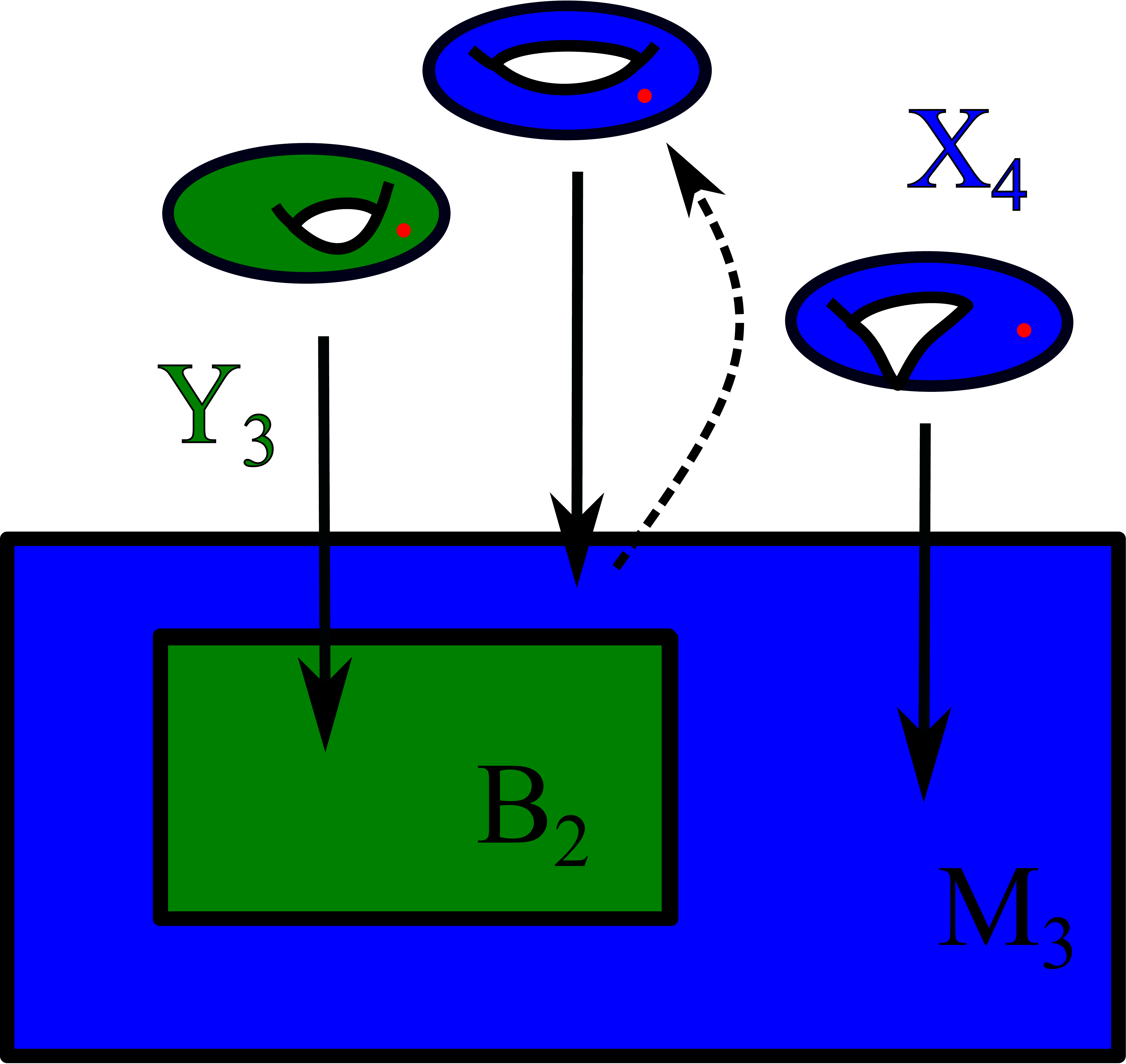} 
  \caption{M-theory setup: $Y_3$ is the world-volume of the M5, and is an elliptic fibration over a base $B$. 
  This elliptic three-fold is embedded into an elliptically fibered Calabi-Yau four-fold $X_4$, with base $M$, which contains $B_2$. The elliptic fibration of $X_4$ restricts to that of $Y_3$. We assume that the fibration has a section, i.e. there is a map from the base to a marked point in the fiber, which is origin of the fiber elliptic curve.
\label{fig:BasicSetup}}
\end{figure}

The complete M-theory setup corresponds to the geometry $X_4 \times \bbR^{1,2}$, where $X_4=CY_4$ is an elliptically fibered Calabi-Yau four-fold, with base three-fold $M_3$. M5-branes wrap an elliptic K\"ahler three-fold $Y_3$, that is a {K\"ahler surface} $B_2\subset M_3$ and the elliptic fiber restricted to this subspace -- see figure \ref{fig:BasicSetup} for a  depiction of the setup. 

An alternative approach to studying this $\mathcal{N}=4$ SYM theory with varying coupling is directly in terms of D3-instantons in F-theory, which was pursued in \cite{Martucci:2014ema}. This 4d point of view is restricted to the abelian $\mathcal{N}=4$ theory, which once twisted by the so-called topological duality twist, defines the theory with varying $\tau$ on a K\"ahler surface. 
This duality twist  is a combination of a standard topological twist, combined with a non-trivial background connection for the gauge field 
\be
\mathcal{A} = - {d\tau_1 \over 4 \tau_2} \,,
\ee
associated to a $U(1)_D$ action on the abelian $\mathcal{N}=4$ SYM theory, which is called ``bonus symmetry" \cite{Intriligator:1998ig, Kapustin:2006pk}. 

One of the main results in this paper is the derivation of this duality twisted theory from 6d, by performing a standard topological twist of the 6d $(2,0)$ theory on an elliptic three-fold and reducing along the elliptic fiber. 
An  advantage of taking the 6d point of view is that it predicts a generalization of the $U(1)$ model in \cite{Martucci:2014ema} to a non-abelian 4d theory with varying coupling, which is the reduction of the type $A_{N-1}$ 6d $(2,0)$ theory on the elliptic fibration. We will present the non-abelian extension of the 4d theory with varying coupling. 

\begin{figure}
  \centering
  \includegraphics[width=8cm]{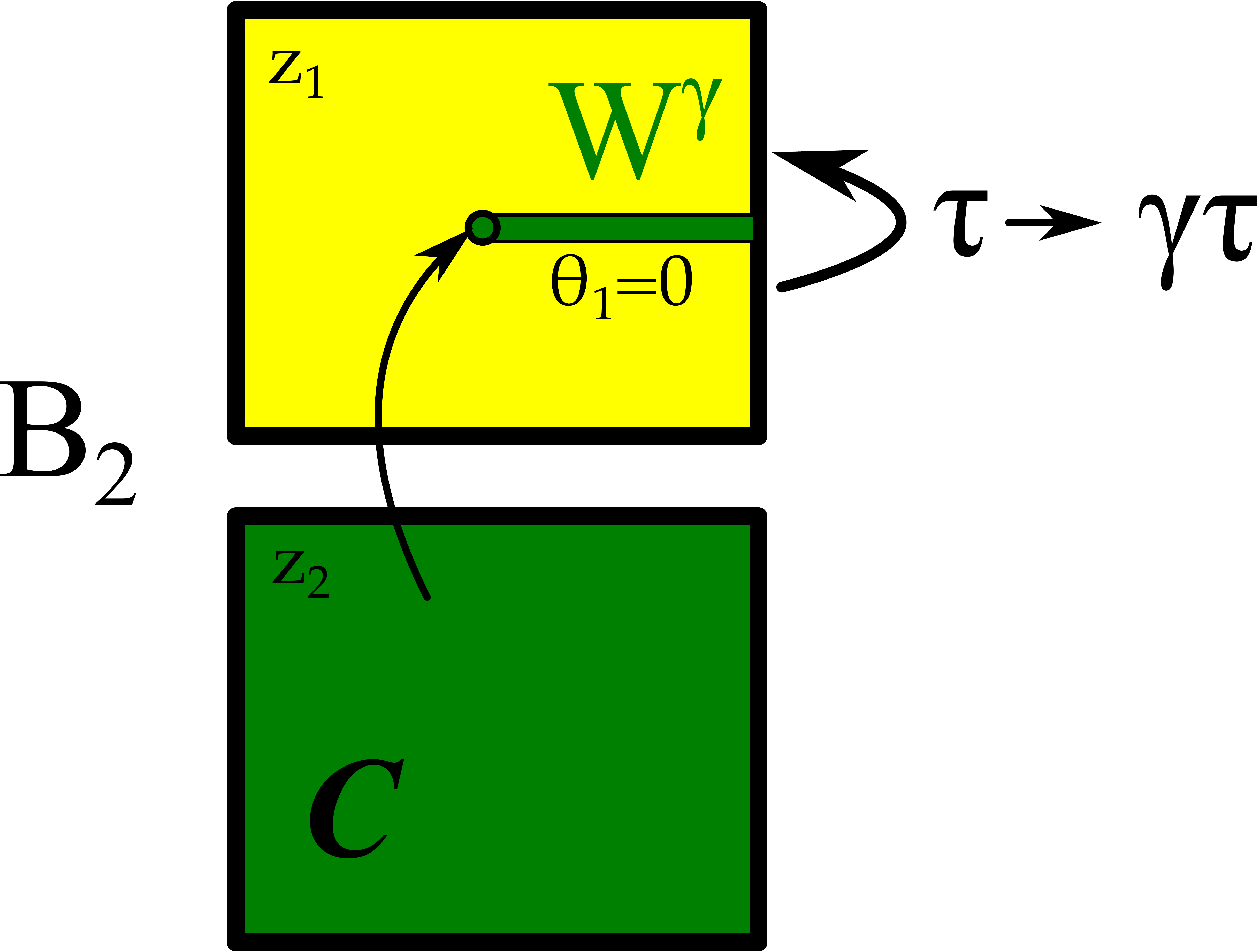} 
  \caption{Setup of 4d theory on $B_2$ with local coordinates $z_1$ and $z_2$. The curve $\mathcal{C}$ is given by $z_1=0$, and the wall $W^{\gamma}$ by $\theta_1 =0$, where $\theta_1$ is the angular coordinate of $z_1$. As one crosses the branch-cut $W^\gamma$, the coupling undergoes an $SL(2,\mathbb{Z})$ monodromy $\gamma$. 
\label{fig:642Setup}}
\end{figure}

One of the salient features of these theories is the presence of 2d defects. Geometrically these correspond to complex one dimensional subspace in $B_2$ above which the fiber degenerates. 
Taking a varying holomorphic coupling $\tau$ over a base manifold $B_2$, can result in singularities in the fiber, above 2 real-dimensional surfaces/complex curves $\mathcal{C}\subset B$, along which $\tau$ degenerates. 
Locally around each such curve $\CC$, the function $\tau$ has a monodromy $\tau \to \tau'$ given by an $SL(2,\bbZ)$ modular transformation $\gamma$ of the elliptic curve
\be\label{sl2z}
\tau' = \gamma.\tau = \frac{a\tau + b}{c\tau + d} \,, \quad ab - cd =1 \,, \quad a,b,c,d \in \bbZ\,.
\ee
There are branch-cuts emanating from the curve  $\CC$, which are real codimension one in the 4d theory on $B_2$ and thus appear as  3d walls $W$, across which the coupling jumps by an $SL(2,\bbZ)$ duality transformation. We will sometimes refer to these as duality walls. This setup is shown in figure \ref{fig:642Setup}.

In IIB/F-theory, the 2d singular loci are the intersections of the D3-branes, on which the 4d SYM theory is realized, with 7-branes, and the chiral modes arise from 3-7 strings. Thus one way to study the defects is in terms of the spectrum of $[p,q]$ strings between the D3- and 7-branes. 
From the 6d $(2,0)$ theory point of view, we will see that these modes arise naturally from the reduction of the two-form field $B_{\mu\nu}$ along the $\bbP_1$s associated to the singular fibers. On the other hand, the 3d walls carry no physical degrees of freedom, and only guarantee continuity of the theory across the entire 4d spacetime. 

The surface defects are labeled by elements $\gamma$ of $SL(2,\mathbb{Z})$. For $\gamma=T^n$, i.e. $\tau \rightarrow \tau + n$, a complete Lagrangian description is possible for the non-abelian theory \cite{Buchbinder:2007ar}. The 3d walls in this case carry a  level $n$ Chern-Simons term, $n \int_W \Tr [A\wedge dA + \frac{2}{3}A\wedge A \wedge A]$. On the defect itself, the theory is a chiral WZW model, which cancels the gauge anomaly of the 3d coupling. Alternatively the defect theory can be described in term of $n$ chiral fermions or $n$ chiral bosons.  

An interesting property that we notice is that the 2d defects generically intersect in points. 
From the point of view of the 2d chiral theories, this corresponds to loci of enhanced flavor symmetries, which can be determined by studying the mechanics of the singular fibers which have an enhanced singularity above these points \cite{EY, MS, Krause:2011xj, Grimm:2011fx, Hayashi:2013lra, Hayashi:2014kca}. 

In summary, the complete setup could be termed a {\it 4d-3d-2d-0d Matroshka-defect} configuration: starting in 6d we reduce on an elliptic fiber to 4d SYM with varying coupling constant. There are 3d walls, which end on a network of 2d surface defects, on which chiral modes are localized, and have a flavor symmetry that is dictated by the singularities in the elliptic fiber. These 2d defects, in turn, intersect at points, which enjoy an enhanced flavor symmetry.

We should mention some related compactifications of the 6d $(2,0)$ theory. The dimensional reduction of the 6d theory on a circle gives rise to maximally supersymmetric 5d Yang-Mills. 
The case of circle-fibration was discussed in \cite{Witten:2009at} and in more detail in 
\cite{Linander:2011jy, Ohlsson:2012yn} including the Taub-NUT geometry, where the circle-fiber can degenerate.
The Taub-NUT reduction will play an important role in our analysis of the 6d reduction. 

In F-theory, a related configuration are D3-branes compactified on a curve in the base of a  Calabi-Yau three-fold including the duality twist \cite{Haghighat:2015ega}. A comprehensive analysis of D3-branes on curves in the base of Calabi-Yau compactifications to dimension 6, 4, and 2, respectively, in F-theory, will appear in \cite{LST}. Unlike the case studied here, which could be understood in terms of D3-instantons with duality surface defects, in those cases, the theories are 2d chiral supersymmetric theories, with point-like defects on the compactification curve, where the D3-branes intersect 7-branes in F-theory.  More recently a discussion of supergravity backgrounds for 4d $\mathcal{N}=4$ SYM appeared in \cite{Maxfield:2016lok}. Also, some duality defects for $\mathcal{N}=2$ theories were proposed in \cite{Gadde:2014wma}. 
Furthermore, there are related configurations in the D3-instanton literature in F-theory: our setup can be viewed from the bulk F-theory or M-theory theory point of view as studying D3- or M5-instantons, for instance in \cite{Witten:1996bn, Ganor:1996pe, Heckman:2008es, Marsano:2008py, Cvetic:2011gp, Martucci:2015oaa}, see \cite{Blumenhagen:2009qh} for a review. 

The plan of this paper is as follows. In section \ref{sec:Setup} we provide the setup for studying the 6d $(2,0)$ theory on an elliptically fibered three-fold. We carry out the reduction to 4d of the abelian 6d theory defined on an elliptically fibered Calabi-Yau three-fold and extend it to the non-abelian theory. We also review the so-called bonus-symmetry in $\mathcal{N}=4$ SYM. In section \ref{sec:DualityTwistFrom6d} we construct the twisted 6d $(2,0)$ abelian theory on an elliptic K\"ahler three-fold and perform the reduction along the elliptic fiber to 4d. We find a twisted 4d $\mathcal{N}=4$ SYM theory with varying coupling, matching the topological duality twisted theory. Furthermore, we determine the non-abelianization of this 4d SYM theory with varying coupling. In section \ref{sec:Walls} we discuss the 3d walls and 2d defects and explain how the defect degrees of freedom arise from 6d. We also discuss the flavor symmetry enhancement at the point defects localized at the intersections of the 2d defects, in relation to the singularity enhancement of the elliptic fiber. 
We summarize various topological twists of the 6d theory, present conventions and spinor manipulations, and provide details on computations in the appendices \ref{app:conventions} to  \ref{app:TwistForms}.


\section{Elliptic fibrations, M5-branes and 4d $\CN=4$ SYM}
\label{sec:Setup}

We begin with some background for studying the 6d $(2,0)$ super-conformal theory on an elliptic fibration, by reviewing some basic facts about such geometries, the relation of the 6d  theory to $\CN=4$ SYM in 4d, as well as the bonus symmetry of the abelian 4d $\CN=4$ theory. All these parts will come together in the following sections.


\subsection{Elliptic Three-folds}
\label{ssec:Elliptic3folds}

We consider the abelian 6d $(2,0)$ theory on an elliptically fibered variety $Y \, (\equiv Y_3)$ of complex dimension 3
\be
\ba
\mathbb{E}_\tau \quad  \rightarrow\quad  & Y_3 \cr
& \downarrow \cr
& B_2
\ea
\ee
where $B \, (\equiv B_2)$ is the four-dimensional base of the fibration. 
Denote by $\tau = \tau_1 + i \tau_2$ the complex structure of the elliptic fiber and by $V$ its volume. Given this setup we consider the Euclidean 6d theory. 
In the case of trivial fibrations, the space is endowed with the direct product metric 
\be
ds^2_{\mathbb{E}_\tau \times B}= {V\over \tau_2} \left( (dx + \tau_1 dy)^2 + \tau_2^2 dy^2  \right)  + ds^2_B \,,
\ee
where $ds^2_B$ is line element on the base. 
Here, $(x,y)$ are periodic coordinates on the elliptic curve with periodicity $1$.
To obtain a generic fibration, we let  $\tau =\tau (b^\mu)$ and $V = V(b^\mu)$ depend on the coordinates $b^\mu$ on the base $B$, and introduce two connections
$\eta = \eta_\mu db^\mu $ and $\rho = \rho_\mu db^\mu$, which are 1-forms on $B$, with metric given by 
\be
ds^2_{Y}= {V\over \tau_2} \left( \left(dx+ \eta_\mu db^\mu   + \tau_1 dy \right)^2 + \tau_2^2  (dy + \rho_\mu db^\mu)^2  \right)  + g^B_{\mu\nu} db^\mu db^\nu \,. 
\ee
Here $g^B$ is the metric on the base. 
When the elliptic fibration has a (holomorphic zero-) section, as we shall assume, the off-diagonal terms in the metric vanish, which is equivalent to
requiring that $\rho = \eta = 0$  \cite{Witten:1996bn, Anderson:2014yva}. Furthermore, we set the volume $V$ to be constant. Later on, taking the limit $V\rightarrow0$ corresponds to the F-theory limit, where the M5-brane reduces to the theory on the D3-brane. 
A convenient frame for this specialization is
\be\label{frameR}
e^a \, , \quad  a =0,1,2,3 \, , \qquad e^4 = \sqrt{\frac{V}{\tau_2}} \lp dx + \tau_1 dy  \rp 
\, , \quad e^5 = \sqrt{V\tau_2} \,  dy   \, ,
\ee
where $e^a$ is the frame on the base, associated to $g^B$.
The inverse frame is given by 
\be\label{InvframeR}
e_a^{\mu}\p_\mu \, , \quad  a =0,1,2,3 \, , \qquad e_4^{\underline{\mu}}\p_{\underline{\mu}} = \sqrt{\frac{\tau_2}{V}} \p_x  
\, , \quad e_5^{\underline{\mu}}\p_{\underline{\mu}} = \frac{1}{\sqrt{V\tau_2}} \,  ( - \tau_1 \p_x + \p_y )   \, ,
\ee
where $\underline{\mu}$ and $\mu$ are curved Lorentz indices on the total space and on the base respectively.
The spin connection $\Omega^{AB}$, $A,B=0,1, \cdots , 5$, is given by 
\be\ba
\Omega^{ab}  &= \omega^{ab} \,, \quad  a,b=0,1,2,3 \,,  \qquad  \Omega^{45}  =  -\frac{\p_a  \tau_1}{2 \tau_2} \, e^a  \,,\cr
\Omega^{a4}  &=   \frac{\p^a  \tau_2}{2 \tau_2} \, e^4 - \frac{\p^a  \tau_1}{2 \tau_2} \, e^5 \, , \qquad
\Omega^{a5}  =   -\frac{\p^a  \tau_1}{2 \tau_2} \, e^4 - \frac{\p^a  \tau_2}{2 \tau_2} \, e^5 \, ,
\ea\ee
where $\omega^{ab}$ is the spin connection on $B$.

In order to preserve some supersymmetries, we will topologically twist the theory, which requires $Y$ to be K\"ahler manifold, with reduced $U(3)_L \subset SO(6)_L$ holonomy.
From the results in \cite{MR1109635} it is known
that all allowed base surfaces, that support (minimal) elliptic fibrations are
Hirzebruch surfaces $\mathbb{F}_m$ for $m = 0,2,3,\cdots,7,8,12$,
projective space $\mathbb{P}^2$, the Enriques surface, or blow ups thereof.
We assume that this fibration has a section, which is precisely given by a copy of the base $B$, and can be represented thereby by a Weierstrass model
\be\label{Weier}
y^2 = x^3 + fx w^4 + g w^6 \,,
\ee
with $f$ and $g$ sections $K_B^{-2/-3}$, respectively. Furthermore, we allow for singularities along a complex codimension one locus $\Sigma$ in the case $B$, characterized by 
\be\label{Discriminant}
\Delta = 4 f^3 + 27 g^2  =0 \,,
\ee
which we will characterize in terms of the vanishing of a local coordinate in $B$. Locally around each singularity, the function $\tau$ has a monodromy $\tau \to \tau'$ given by an $SL(2,\bbZ)$ modular transformation $\gamma$ as in (\ref{sl2z}). 

To obtain a K\"ahler metric we start by introducing complex coordinates $z_i$, $i=1,2$, on the base $B$, which has to be K\"ahler, and $w = x + \tau y$ on the torus,  and pick $\tau = \tau(z_i)$ holomorphic in $z_1, z_2$. The metric now takes the form
 \begin{align}
 ds^2_{Y}= {V\over \tau_2} \left( \left(dx   + \tau_1 dy \right)^2 + \tau_2^2  dy^2  \right)  + g^B_{i \bar j} \, dz_i d\bar z_{\bar j} \,,
 \label{metricK} 
 \end{align}
 with $g^B_{i \bar j}$ a hermitian metric on $B$.
If $\mathcal{K}_B$ is the K\"ahler potential of the base,  the K\"ahler potential for $Y$ is given by
\be 
 \mathcal{K}_Y = \mathcal{K}_B -  V \frac{(w - \bar{w})^2}{2 \tau_2}  
 = \mathcal{K}_B - 2 V \tau_2 y^2 \, .
\ee
We introduce the complex frame, which we refer to as  {\it canonical frame}
\be
\eee^I = e^{2I} + i e^{2I+1} \, , \quad \bar\eee^{\bar I} = e^{2I} - i e^{2I+1} \, , \quad  I =0, 1 ,2 \, ,
\label{frameC}
\ee
with real components (\ref{frameR}).
The holomorphicity of $\tau$, $\bar\p \tau =0$, can be recast into $\p_0 \tau_1 - \p_1 \tau_2 = \p_0 \tau_2 + \p_1 \tau_1 = \p_2 \tau_1 - \p_3 \tau_2 = \p_2 \tau_2 + \p_3 \tau_1 = 0$. Using this relations we find the vanishing components
\be
\Omega^{04} - \Omega^{15} \  = \  \Omega^{24} - \Omega^{35} \  = \ \Omega^{14} + \Omega^{05} \  = \ \Omega^{34} + \Omega^{25} \  = \ 0 \,.
\ee
The K\"ahler condition for the base manifold $B$ implies that the four-dimensional spin connection $\Omega_{ab}$ has also two vanishing components, which in this frame are $\Omega^{02} - \Omega^{13} = \Omega^{03}+\Omega^{12}=0$. This makes six components of the spin connection vanish, signaling the reduction of the holonomy  $SU(4)_L \simeq SO(6)_L \rightarrow U(3)_L $ of the six-manifold, as it should be for a K\"ahler space. More precisely the embedding of $U(3)_L$ inside $SO(6)_L$ can be described as follows (see e.g.~ \cite{green1987superstring}, chap. 15.3): consider a $3\times 3$ unitary matrix $U = (u_{ij})$ and replace each component by a $2\times 2$ real matrix $u_{ij} \rightarrow $Re$(u_{ij}) \mathbf{1}_2 + $Im$(u_{ij}) \epsilon$, where $\mathbf{1}_2$ is the $2\times 2$ identity matrix and $\epsilon$ is the antisymmetric matrix with component $\epsilon_{12} = - \epsilon_{21} = 1$. The resulting matrix is a 6d orthogonal matrix. The same operation maps an anti-hermitian $3\times 3$ matrix to an anti-symmetric $6\times 6$ matrix and describes an embedding of the $\mathfrak{u}(3)$ algebra into the $\so(6)$ algebra. The $\so(6)$ matrices in the image of this map obey precisely the six relations above, showing that $\Omega$ belongs to this $\mathfrak{u}(3)$ subalgebra.


\subsection{Abelian 6d $(2,0)$ Theory}

We consider the abelian 6d $(2,0)$ theory, which describes the theory on a single M5-brane, in Euclidean signature. The symmetry group is the 6d Lorentz group and the R-symmetry, $SU(4)_L \times  Sp(4)_R$.
 The theory consists of a free tensor multiplet with fields $(H_3, \lambda^m, \Phi^{mn})$, where $H_3$ is a self-dual three-form, $\lambda^m$, $m=1,\cdots, 4$ are Weyl spinors of positive chirality transforming in the $({\bf 4},{\bf 4})$ of $SU(4)_L \times Sp(4)_R$ and $\Phi^{mn}$ are scalars transforming in the $({\bf 1},{\bf 5})$. They obey $\Phi^{mn} = - \Phi^{nm}$ and $\Omega_{mn} \Phi^{mn} =0$, where $\Omega_{mn}$ is the anti-symmetric $Sp(4)$ invariant tensor.
 Our conventions on spinors are provided in appendix \ref{app:conventions}.
 
In Lorentzian signature the bosonic fields $H_3$ and $\Phi^{mn}$ are real and the fermions $\lambda^m$ obey a Symplectic-Majorana condition. In Euclidean signature, which is the case of interest here, we cannot enforce self-duality on a real three-form\footnote{In Euclidean signature $\star^2 = -1$ on a three-form.}. Instead we can consider the modified self-duality condition
 \begin{align}
 \star H_3 = i \, H_3 \,,
 \end{align}
 which forces us to consider a complex three-form $H_3$.
 Moreover there is no reality condition one can impose on the fermions $\lambda^m$. For consistency with supersymmetry we can also think of the scalar $\Phi^{mn}$ as complex scalars, so that the number of degrees of freedom is doubled compared to the Lorentzian theory.
After reduction to four dimensions, reality conditions can be imposed on the 4d fields, leading to a Euclidean SYM theory with the usual number of degrees of freedom.

The conformal equations of motion of the abelian theory in curved space are given by  \begin{align}
 dH_3 &= \ 0 \, ,\quad  \star H_3 = i \, H_3 \,, \quad
\Gamma^A \nabla_A \lambda^m \ = \ 0 \, , \quad
\nabla^2 \Phi^{mn} - \frac{R}{5}  \Phi^{mn} \ = \  0 \, ,
\label{6dEOM}
\end{align}
where $R$ is the 6d Ricci scalar. 
The equation on $H_3$ tells us that it can be written locally as the differential of a two-form $H_3 = dB_2$.
These equations are invariant under Weyl transformations with Weyl weights $w(H_3) = 0$, $w(\lambda^m) = \frac 52$ and $w(\Phi^{mn}) = 2$.
The equations of motion of the spinor and scalar fields can be derived from the actions
\begin{align}
S^{\rm (6d)}_{\lambda} &= \int d^6x \sqrt g  \, \bar\lambda_m \Gamma^A \nabla_A \lambda^m \,, \nn\\
S^{\rm (6d)}_{\Phi} &= \frac{1}{16} \int d^6x \sqrt g \,  \lp \p^A \Phi_{mn}  \p_A \Phi^{mn} + \frac{1}{5} R \,  \Phi_{mn}\Phi^{mn}\rp \,.
\label{6dAction}
\end{align}
We will also label the five scalars $\Phi^{\wat i}$, $i =1, \cdots, 5$ corresponding to the vector representation of $SO(5)_R$, with
\begin{align}
\Phi^{mn} &= \Phi^{\wat i} \, \lp \Gamma^{\wat i} \rp{}^{mn} \,, \quad \Phi^{\wat i} = \frac 14 \, \Phi^{mn} \lp \Gamma^{\wat i} \rp{}_{mn} \,,
\end{align}
using the R-symmetry Gamma matrices defined in appendix \ref{app:Rsym}.

The supersymmetry transformations of the Euclidean theory are generated by negative chirality spinors $\mathcal{E}^m$, of Weyl weight $w=-\frac 12$, satisfying the Killing spinor equations
\be
\nabla_A \mathcal{E}^m - \frac 16 \Gamma_A \Gamma^B \nabla_B \mathcal{E}^m = 0 \,,
\label{KSE}
\ee
and are given by
\be\label{6dTwistedSUSY}
\ba
\delta B_{AB} &=  - \bar{\mathcal{E}}^m \Gamma_{AB} \lambda_m \cr 
\delta \lambda^m &= \frac{1}{48} H_{ABC} \Gamma^{ABC} \mathcal{E}^m - \frac 14 \Gamma^A \p_A \Phi^{\wat i} \,  ( \Gamma^{\wat i} )^{m}{}_{n} \mathcal{E}^n  - \frac 16 \Phi^{\wat i}  ( \Gamma^{\wat i} )^{m}{}_{n} \Gamma^A \nabla_A \mathcal{E}^n  \cr 
\delta \Phi^{\wat i} &=  \bar{\mathcal{E}}_m ( \Gamma^{\wat i} )^{m}{}_{n}\lambda^n \,.
\ea
\ee
The theory, defined in this way, preserves supersymmetries only when the six-dimensional geometry admits solutions to \eqref{KSE}. We note that the Killing spinor equations imply the relation
\be
\nabla^2 \mathcal{E}^m + \frac{R}{20} \mathcal{E}^m =0 \,.
\ee
The supersymmetry transformations leave the equation of motions \eqref{6dEOM} invariant and  the super-algebra closes on-shell.


\subsection{M5-branes on an elliptic Calabi-Yau}
\label{ssec:CYRed}

In this section we consider the abelian $(2,0)$ theory on an elliptic fibration admiting Killing spinors and we perform the reduction to four dimensions. 
We will show that, without any additional twisting of the 6d $(2,0)$ theory, this implies that the three-fold has to be a Calabi-Yau variety. In four dimensions the reduced theory will be the abelian $\CN=4$ Super-Yang-Mills theory on a K\"ahler space with varying complex coupling.  Having obtained the abelian theory we will promote it to the  non-abelian theory with $U(N)$ gauge group and comment on issues related to monodromies of the coupling.

We wish to reduce the six-dimensional theory on $\bbE_\tau \rightarrow Y \rightarrow B$ to a four-dimensional theory on the base $B$, by shrinking the elliptic fiber $\bbE_\tau$. This corresponds to taking the constant volume $V$ of $\bbE_\tau$ very small compared to the size of $B$ (or considering the theory at energies small compared to $V^{-1/2}$ if $B$ is non-compact).
We start by studying the constraints imposed by supersymmetry on the geometry.
We assume the existence of a 6d Killing spinor $\CE$ independent on the fiber coordinates $x,y$, with negative chirality. Such a spinor can be decomposed, as explained in appendix \ref{app:FermionRed}, into
\be
\CE = \CE_- \otimes \eta_+ + \CE_+ \otimes \eta_- \,,
\ee
where $\CE_{+}, \CE_-$ are 4d spinors of positive and negative 4d chirality respectively and $\eta_{\pm}$ are constant 2d chiral spinors which serve as a basis for the decomposition. The 6d Killing spinor equations \eqref{KSE} then reduce to the 4d Killing spinor equations,
\be\ba
 \nabla_a \CE_- - \frac{i \p_a\tau_1}{4\tau_2}\CE_- &= 0 \,, \quad \nabla_a \CE_+ + \frac{i \p_a\tau_1}{4\tau_2}\CE_+ = 0 \,, \cr
 \p_a\tau \gamma^a\CE_- &= 0 \,, \quad  \p_a\bar\tau \gamma^a\CE_+ = 0 \,.
\label{KSE4d}
\ea\ee
These equations, written back in 6d language, imply that the 6d Killing spinor $\CE$ is covariantly constant, $\nabla_A \CE =0$, so that the six-dimensional geometry has to be Calabi-Yau. We find therefore that elliptic fibrations  with a section, which admit a Killing spinor constant along the elliptic fiber, correspond exclusively to Calabi-Yau three-folds. The equations on the second line of \eqref{KSE4d} imply that $\tau$ is holomorphic in appropriate complex coordinates on the base $B$. We will come back to this when we discuss the twisted theory. 

On a generic Calabi-Yau manifold the reduced holonomy group is $SU(3)_L \subset  SU(4)_L$ and a Weyl spinor in the ${\bf 4}$ of $SU(4)_L$ decompose into ${\bf 3} \oplus {\bf 1}$ of $SU(3)_L$ (see appendix \ref{app:CYsupercharges}). Therefore a generic Calabi-Yau three-fold has a single covariantly constant spinor $\CE$, which reduces in four dimensions to a single pair of Killing spinors of opposite chirality $(\CE_+, \CE_-)$ satisfying \eqref{KSE4d}. Since the Killing spinor equations for $\CE_+$ and $\CE_-$ are decoupled, it must be that, for a generic Calabi-Yau, either $\CE_+ = 0$ or $\CE_- =0$, otherwise we could consider these spinors as independent and they would generate more supersymmetry in 4d than we had in 6d. 
We can argue about this point as follows.
If we have both $\CE_+$ and $\CE_-$ non-zero, we can construct $\CE'_- = \CB (\CE_-)^\ast$ \footnote{See appendix \ref{app:conventions} for the definition of the matrix $\CB$.} which satisfies the same equations as $\CE_+$, but which has negative chirality\footnote{This is true only in Euclidean signature. It can be worked out from the explicit conventions given in appendix \ref{app:conventions}.}. In particular we have $\p_a\bar\tau \gamma^a\CE_+ =0$ and $\p_a\bar\tau \gamma^a\CE'_-=0$, which together imply $\p_a\bar\tau=0$ and therefore $\p_a\tau=0$. The coupling $\tau$ is then constant, corresponding to the Calabi-Yau being a direct product $\bbE_\tau \times K^3$. This is a non-generic Calabi-Yau, with further reduced holonomy. So, on a generic Calabi-Yau, $\CE_+=0$ or $\CE_-=0$.

The supersymmetries of the six-dimensional theory are parametrized by the four spinors $\CE^m$, each one being a copy of the Killing spinor. The supersymmetries of the four-dimensional theory are similarly parametrized by the four spinors $\CE_+^m$ or $\CE^m_-$.
In the following we will assume $\CE^m_- =0$ and parametrize the supersymmetry transformations with $\CE_+^m$.

In order to  reduce to a four-dimensional theory on the base manifold $B$,  we decompose the three-form $H_3$ into
\be\label{eqn:3formdecomp}
H_3 = E_3 + F_2 \wedge dx + G_2 \wedge dy  + D_1 \wedge dx \wedge dy \,,
\ee
where $E_3, F_2, G_2, D_1$ are forms on $B$, taken independent of the fiber coordinates $x,y$.
 The equations of motions for $H_3$ \eqref{6dEOM} are
\be
dH_3 =0 \,,\qquad  \star H_3 = i \,  H_3 \,.
\ee
The latter equation can be used to solve for $E_3$ and $G_2$ in terms of $F_2$ and $D_1$. 
Some useful properties to work this out are
\be
\ba
& \star \lp \omega \wedge e^4 \rp  =  -(\star_4 \, \omega) \wedge e^5 \, , \quad
\star \lp \omega \wedge e^5 \rp  =  (\star_4 \, \omega) \wedge e^4 \, , \\
& \star \lp \theta \wedge e^4 \wedge e^5 \rp  = (\star_4 \, \theta)  \, ,
\ea
\ee
with $\omega$ a 2-form and $\theta$ a 1-form on B.
We obtain
\be
\ba
 E_3 & = -\frac iV \, \star_4  D_1 \,, \quad 
G_2 = \tau_1 \, F_2 + i \, \tau_2 \, \star_4  F_2  \, .
\ea
\label{EandG}
\ee
The closeness relation $dH_3=0$ becomes
\be\label{eqn:ablclosed}
\ba
& dF_2 = dD_1 =0  \,, \cr
& dE_3 = dG_2=0 \,.
\ea
\ee
The equations on the first line remain in the reduced theory and allow to write locally $F_2 = dA$ and $D_1 = dC$, with $A$ a one-form and $C$ a scalar, which is periodic $C\rightarrow C + 2 \pi n$ with $n \in \mathbb{Z}$, due to the large gauge transformations of the potential $B$. The equations on the second line, after replacing with \eqref{EandG}, lead to the equations of motion for $F_2$ and $D_1$ of the four-dimensional theory,
\begin{align}
  dE_3 &= -\frac{i}{V}d\star_4D_1 = 0 \,, \cr
  dG_2 &= d\tau_1 \wedge F_2 + i \, d\tau_2 \wedge \star_4 F_2 + i \, \tau_2 d \star_4 F_2
  = 0 \,.
\end{align}
It is now desirous to integrate the equations of motion for $F_2$ to find a
Lagrangian description of that sector of the 4d theory. The theory is that
of an abelian connection $A$ with action
\begin{align}
S_F &= {1 \over 4 \pi} \int_B \lp  \tau_2  F \wedge \star_4 F - i \tau_1   F\wedge F \rp  \,,
\end{align}
where $F \equiv F_2$ is the curvature of $A$ and $\tau$ varies over the base $B$, which we assumed to be without boundary.

The equations of motion for the scalar field $C$ defined by $D_1 = dC$ reduces simply to
\begin{align}
\nabla^2_{\rm (4d)} C = 0 \,.
\end{align}
It integrates to the action
\begin{align}
S_{C} &= \int d^4b\,  \sqrt{g_B} \, \p_a C \p^a C \, .
\end{align}

Next we consider the scalar fields and take them independent of the $x$ and $y$ coordinates,
\be
 \Phi^{mn} = \frac{\phi^{mn}}{\sqrt{V}} \,, 
 \ee
 where the rescaling by $\sqrt V$ is convenient to simplify later formulas. It also gives the scalars $\phi^{mn}$ the canonical 4d scalar scaling dimension one.
  The equations of motion \eqref{6dEOM} reduce to
\begin{align}
\nabla^2_{\rm (4d)} \phi^{mn} - \frac 15 \lp R_{\rm (4d)} - \frac{\p_a \tau_1 \p^a \tau_1 + \p_a \tau_2 \p^a \tau_2}{2 \tau_2^2}\rp \phi^{mn} = 0 \, ,
\end{align}
where we used the relation between 6d and 4d Ricci scalars $R_{\rm (6d)} = R_{\rm (4d)} - \frac{\p_a \tau_1 \p^a \tau_1 + \p_a \tau_2 \p^a \tau_2}{2 \tau_2^2}$. This equation simplifies with the assumption that the six-dimensional geometry is Calabi-Yau. We have $R_{\rm (6d)}=0$ and the equations of motions are simply
\be
\nabla^2_{\rm (4d)} \phi^{mn}  = 0 \, .
\ee
 This equation integrates to the action
\begin{align}
S_{\phi} &= \int d^4b\,  \sqrt{g_B} \, \lp \p_a \phi^{\wat i} \p^a \phi^{\wat i}  \rp \,,
\end{align}
where we have changed the scalar index $\phi^{\wat i}= \frac 14 (\Gamma^{\wat i} )_{mn} \phi^{mn}  $, with ${\wat i}=1, \cdots , 5$ the index of the ${\bf 5}$ of $SO(5)_R \simeq Sp(4)_R$ \footnote{Throughout the paper we use the notation  $G_1 \simeq G_2$ for two groups $G_1$ and $G_2$, which have the same algebra, but may differ as groups. For instance $SU(2) \simeq SO(3)$.}. The index $\wat i$ is implicitly summed over in the action.

Finally we consider the fermions $\lambda^m$. Following the analysis of appendix \ref{app:FermionRed} $\lambda^m$ can be decomposed into
\begin{align}
\lambda^m &= \frac{1}{\sqrt{V}} \lp \lambda_+^m \otimes \eta_+  + \lambda_-^m \otimes \eta_- \rp \, ,
\label{SpinorDecomp}
\end{align}
where $\lambda_+^m$ is a positive chirality spinor, $\lambda_-^m$ is a negative chirality spinor in four dimensions, and $\eta_+ = \binom{1}{0}$, $\eta_- = \binom{0}{1}$. The rescaling by $\sqrt V$ gives the spinors the canonical 4d spinor scaling dimension 3/2.
The spinors $\lambda_+^m$ and $\lambda_-^m$ transform in the $({\bf 2}, {\bf 1}, {\bf 4}) \oplus ({\bf 1}, {\bf 2}, {\bf 4}) $ of $SU(2)_1 \times SU(2)_2 \times Sp(4)_r$, where $SU(2)_1 \times SU(2)_2 \simeq SO(4)$ is the four-dimensional structure group.
The equations of motion \eqref{6dEOM} reduce to
\begin{align}
\gamma^a \nabla_{\rm (4d)}{}_a \lambda_+^m - i \frac{\p_a \tau_1}{4 \tau_2} \gamma^a \lambda_+^m = 0\, , \quad
\gamma^a \nabla_{\rm (4d)}{}_a \lambda_-^m  + i \frac{\p_a \tau_1}{4 \tau_2} \gamma^a \lambda_-^m  = 0 \, ,
\end{align}
and can be derived from the action
\begin{align}
S_{\lambda} &= \int d^4b\,  \sqrt g_B \, \lp \bar\lambda_{+\, m} \gamma^a \nabla_a \lambda_-^m +   i \frac{\p_a \tau_1}{4 \tau_2} \,  \bar\lambda_{+\, m} \gamma^a \lambda_{-}^m  \rp \, .
\end{align}

The supersymmetry transformations, obtained from six-dimensions, are generated by four Killing spinors $\CE_+^{m}$ or $\CE_-^m$, each one of them being equal to the solutions to the Killing spinor equation \eqref{KSE4d} 
\be\ba
\delta A_a &= \frac{i}{\sqrt{\tau_2}} \bar\CE^m_+\gamma_a \lambda_{- \, m}   \,, \cr
\delta\lambda^m_+ &= \frac 18  \sqrt{\tau_2} F_{ab}\gamma^{ab}\CE^m_+   \,, \cr
\delta\lambda^m_- &=  - \frac{i}{4} \p_a \phi^6 \gamma^a \CE^m_+ + \frac{1}{4} \p_a \phi^{\wat i} ( \Gamma^{\wat i})^{mn} \gamma^a \CE_{+ \, n}  \,, \cr
\delta \phi^{\wat i} &= - i \bar\CE_{+ \, m} (\Gamma^{\wat i})^{mn} \lambda_{+ \, n} \,, \cr
\delta \phi^{6} &= \bar\CE_{+}^m  \lambda_{+ \, m}  \,,
\ea\ee
where we replace the scalar $C$ by  $\phi^6 \equiv \frac{C}{\sqrt{V}}$.

The R-symmetry of the $(2,0)$ theory, $Sp(4)_R$, is
enhanced to become the $SU(4)_r$ R-symmetry of the $4d$ $\CN=4$ theory. The
embedding of $Sp(4)$ inside $SU(4)$ is the special maximal embedding, where the
representations of interest decompose as
\be\ba\label{Nis4R}
  SU(4) &\quad \rightarrow\quad  Sp(4) \cr
\bar{\bf 4} \hbox{ and }  {\bf 4} &\quad \rightarrow\quad {\bf 4}   \cr
  {\bf 6} &\quad \rightarrow\quad {\bf 1 \oplus 5} \,.
\ea\ee
The scalar field $\phi^6$ is in
the trivial representation of the $Sp(4)_R$ R-symmetry, inherited from the
three-form.  
The five scalars $\phi^{\wat i}$, which transform in the
${\bf 5}$ of $Sp(4)_R$ combine with $\phi^6$ and transform in the ${\bf 6}$ of $SU(4)_r$.
Note that in the limit $V\rightarrow 0$ (i.e. the F-theory limit), the compact scalar $\phi^6 = C/\sqrt{V}$ decompactifies and is therefore on the same footing as the non-compact scalars $\phi^{\wat i}$. 
The fermions $\lambda_-^m, \lambda_+^m$ transform in the ${\bf 4}$ of
$Sp(4)_R$. Under the enhanced R-symmetry, they transform as the ${\bf 4}$ and $\bar{\bf 4}$ of $SU(4)_r$. More precisely $\lambda_+^{\dot m}$ and $\lambda_-^m$ transform in the $({\bf 2}, {\bf 1}, \bar{\bf 4}) \oplus ({\bf 1}, {\bf 2}, {\bf 4}) $ of the 4d flat space  symmetry group $SU(2)_1 \times SU(2)_2 \times SU(4)_r$. 

The complete action takes the $SU(4)_r$ invariant form
\be
\ba\label{AbelianAction}
S_{\rm 4d} &= S_F + {1 \over 4 \pi}  S_C + {1 \over 4 \pi}  S_\phi + {1 \over 2 \pi}  S_\lambda \cr
&= {1 \over 4 \pi} \int \lp  \tau_2  F \wedge \star_4 F - i \tau_1   F\wedge F \rp  \cr
 & \ + \frac{1}{4\pi} \int d^4b\,    \sqrt{g_B} \Big[  \p_a \phi^{\wat i} \p^a \phi^{\wat i} + 8i \Big( 2 \bar\lambda_{+ \, m} \gamma^a \nabla_a \lambda_{-}^m +   i \frac{\p_a \tau_1}{2 \tau_2} \,  \bar\lambda_{+\, m} \gamma^a \lambda_-^m \Big)  \Big] \,,
\ea\ee
with the $SU(4)_r$ covariant  supersymmetry transformations
\be\ba\label{4dAbelianSusyTransfo}
\delta A_a &= - \frac{i}{\sqrt{\tau_2}}  \bar\CE_{+\, m}\gamma_a \lambda^m_{-}  \,, \cr
\delta\lambda_{+}^{\dot m} &= \frac 18  \sqrt{\tau_2} F_{ab}\gamma^{ab}\CE_{+}^{\dot m}  \,, \cr
\delta\lambda^m_- &=  - \frac{1}{4} \p_a \phi^{\wat i} ( \Gamma^{\wat i})^{m }{}_{\dot n} \gamma^a \CE_{+}^{\dot n }  \,, \cr
\delta \phi^{\wat i} &=  i \bar\CE_{+ \, m} (\Gamma^{\wat i})^{m}{}_{\dot n} \lambda_{+}^{\dot n} \,.
\ea\ee
The index $\wat i$ now goes from 1 to 6, labeling the representation ${\bf 6}$ of $SU(4)_r$, and we have defined $\bar \Gamma^{\wat i} \equiv  - \Gamma^{\wat i}$, for $\wat i = 1 , \cdots , 5$, and $(\Gamma^{\hat 6})^{m}{}_{\dot n} = (\bar \Gamma^{\hat 6})^{\dot m}{}_{n} \equiv i \delta^{m}_{n}$. Indices are now raised and lowered with $\Omega^{m \dot n}$ and $\Omega_{\dot m n}$.

\subsection{Non-abelian Generalization}

The four-dimensional action can be promoted to a non-abelian theory by letting the fields take value in the $\mathfrak{u}(N)$, or $\su(N)$ algebra, with $A_\mu$ the non-abelian gauge potential and $F = dA + A\wedge A$, and taking the action to be
\be
\ba\label{NonAbelianAction}
S_{\rm 4d} 
&= {1 \over 4 \pi} \int  \Tr \Big[  \tau_2  F \wedge \star_4 F - i \tau_1   F\wedge F \Big]  \cr
 & \ + \frac{1}{4\pi} \int d^4b\,    \sqrt{g_B} \Tr \Big[  D_a \phi^{\wat i} D^a \phi^{\wat i} + 8i \Big( \bar\lambda_{+ \, m} \gamma^a D_a \lambda_{-}^m +   i \frac{\p_a \tau_1}{4 \tau_2} \,  \bar\lambda_{+\, m} \gamma^a \lambda_-^m \Big) \Big] \cr
 & \ + \frac{1}{4\pi} \int d^4b\,    \sqrt{g_B} \Tr \Big[ -\frac{4i}{\sqrt{\tau_2}} \bar\lambda_{+\, m} (\Gamma^{\wat i})^{m}{}_{\dot n} [\phi^{\wat i}, \lambda_{+}^{\dot n}] + \frac{4i}{\sqrt{\tau_2}} \bar\lambda_{-\, \dot m} (\bar\Gamma^{\wat i})^{\dot m}{}_{n} [\phi^{\wat i}, \lambda_{-}^{n}] 
  - \frac{1}{\tau_2} [\phi^{\wat i}, \phi^{\wat j}] [\phi^{\wat i}, \phi^{\wat j}] \Big] \,.
\ea\ee
The covariant derivative is $D_a = \nabla_a + [A_a , . \,]$.
The supersymmetry transformations of the bosons are as in \eqref{4dAbelianSusyTransfo}. The transformations of the fermions become 
\be\ba
\delta\lambda_{+}^{\dot m} &= \frac 18  \sqrt{\tau_2} F_{ab}\gamma^{ab}\CE_{+}^{\dot m}  - \frac{1}{8\sqrt{\tau_2}} [\phi^{\wat i}, \phi^{\wat j}] (\bar\Gamma^{\wat i \wat j})^{\dot m}{}_{ \dot n} \CE_+^{\dot n}  \,, \cr
\delta\lambda^m_- &=  - \frac{1}{4} D_a \phi^{\wat i} ( \Gamma^{\wat i})^{m }{}_{\dot n} \gamma^a \CE_{+}^{\dot n }  \,, \cr
\ea\ee
with $\Gamma^{\wat i \wat j} \equiv \Gamma^{[\wat i} \bar\Gamma^{\wat j]}$ and  $\bar\Gamma^{\wat i \wat j} \equiv \bar \Gamma^{[\wat i}\Gamma^{\wat j]}$.
The non-abelian theory with gauge algebra $\su(N)$  is then interpreted as the dimensional reduction of the 6d $(2,0)$ theory of type $A_{N-1}$ on an elliptic Calabi-Yau three-fold.

We have derived a Euclidean action with complex fields in four-dimensions. We could impose reality conditions, demanding that the bosonic fields are hermitian and the fermions obey a Symplectic-Majorana condition (see appendix \ref{app:IrredSpinors}).

In the flat space and constant $\tau$ limit, namely the 6d flat space limit where the elliptic fibration is a direct product $\bbE_\tau \times \bbR^4$, we recover the standard abelian $\CN=4$ Super-Yang-Mills action as expected \cite{Witten:1995zh}.

Finally we must comment on the fact that such Calabi-Yau elliptic fibrations have singular loci where the modular parameter $\tau$ degenerates, as mentioned in Section \ref{ssec:Elliptic3folds}. In the four-dimensional theory the profile of the holomorphic coupling $\tau$ has surface defects, where $\tau$ degenerates, from which emanate branch cuts. These branch cuts appear as domain-walls across which the coupling jumps by an $SL(2,\bbZ)$ duality transformation, which we may call {\it duality walls}. We will discuss these duality walls and the surface defects in section \ref{sec:Walls}. 
For now, we only notice that, due to the presence of these walls, the supersymmetry variation of the action is not vanishing, but instead picks boundary terms for each duality wall $W$
\be
\ba
\delta S_W &=  \left( \int_{W^+} - \int_{W^-}\right)  \ d^3 x \sqrt{g_3} n^a \Tr\Big[ - \frac{\bar\tau}{\sqrt{\tau_2}} \epsilon_a{}^{bcd} \bar\CE_{+ \, m} \gamma_b \lambda_-^m F_{cd} \cr
& \phantom{\int_{W^+} - \int_{W^+}  \ d^3 x \sqrt{g_3} n^a \Tr } + 2i (\Gamma^{\wat i})^m{}_{\dot n} \bar\CE_{+ \, m} \lambda_+^{\dot n} D_a\phi^{\wat i} 
+ \frac{i}{\sqrt{\tau_2}} \bar\CE_{+ \, m} \gamma_a \lambda_-^n (\Gamma^{\wat i \wat j})^m{}_n [\phi^{\wat i}, \phi^{\wat j}] \Big] \,,
\ea
\ee
where $\int_{W^+} - \int_{W^-}$ computes the difference of the integrals with the fields and coupling $\tau$ evaluated on one side of the wall (at $W^+$) and the fields and coupling $\tau'$ evaluated across the wall (at $W^-$). Here, $n^\mu$ denotes a unit vector normal to the 3d wall.


\subsection{$\CN=4$ SYM and Bonus Symmetry}
\label{sec:Bonus}

The abelian $\CN=4$ SYM theory has an additional transformation, which we refer to as  $U(1)_{\CD}$, also known as ``bonus symmetry" in \cite{Intriligator:1998ig}. 
It arises when we consider $SL(2, \bbZ)$ electric-magnetic duality transformations in $\CN=4$ SYM. 
Such a transformation is parametrized by an $SL(2, \bbZ)$ element
\begin{align}
\gamma &= \left(
\begin{array}{cc}
a & b \\
c & d 
\end{array}
\right) \,, \quad  a,b,c,d \in \bbZ \,, \quad \quad ad - bc =1 \,.
\end{align} 
As found in \cite{Intriligator:1998ig} and \cite{Kapustin:2006pk} (using different approaches), the supercharges transform under the transformation $\gamma$. The bosonic symmetries of the theory are the Lorentz group $SO(4)_l \simeq SU(2)_1 \times SU(2)_2$ and the R-symmetry group $SU(4)_r$. The supercharges of $\CN=4$ SYM are $Q^{\dot m}_{\alpha}, \ti Q^{m}_{\dot\alpha}$ transforming in the representation $({\bf 2}, {\bf 1}, \bar{\bf 4}) \oplus ({\bf 1}, {\bf 2}, {\bf 4}) $ of $SU(2)_1 \times SU(2)_2 \times SU(4)_r$. Under an $SL(2, \bbZ)$ duality the supercharges transform by a phase rotation, which is identified with a $U(1)_\CD$ transformation,
\be
\ba
 Q^{\dot m} \  &\rightarrow \  e^{-\frac i2 \alpha(\gamma)}  Q^{\dot m} \cr 
\ti Q^{m} \ &\rightarrow \  e^{\frac i2 \alpha(\gamma)} \ti Q^{m} \, ,
\label{Qrotation}
\ea
\ee
with 
\be
e^{i\alpha(\gamma)} = \frac{c\tau + d}{|c\tau +d|}\,.
\ee
The fields of the SYM theory transform under this $U(1)_{\mathcal{D}}$ as
\be
\phi^{\wat i} \ \rightarrow \ \phi^{\wat i}
\,,\qquad 
\lambda_{+}^{\dot m} \ \rightarrow \  e^{-\frac i2 \alpha(\gamma)} \lambda_{+}^{\dot m}\,,\qquad
\lambda_{-}^{m} \ \rightarrow \  e^{\frac i2 \alpha(\gamma)} \lambda_{-}^{m} \,.
\ee
The gauge field $A_\mu$ does not have a definite transformation under  $U(1)_{\CD}$. In the abelian theory the field strength components do have definite $U(1)_{\CD}$ transformations,
\be
F^{(\pm)}_{\mu\nu}  \ \rightarrow \  e^{\mp  i \alpha(\gamma)} F^{(\pm)}_{\mu\nu} \,,
\ee
where we defined $F^{(\pm)} \equiv \sqrt{\tau_2} \lp \frac{F \pm \star F}{2} \rp$.
In \cite{Intriligator:1998ig} this $U(1)_{\CD}$ transformation was described as a symmetry of the equations of motion and of the supersymmetry variations of the abelian  theory and the charges of the fields under $U(1)_{\CD}$ were given. However it is not a symmetry of the SYM action. In the non-abelian case the symmetry does not survive, even at the level of the equations of motion, however it still acts on the sector of protected operators. 

In \cite{Martucci:2014ema} this $U(1)_\CD$ action was used to perform the so-called ``topological duality twist" in the abelian SYM theory, allowing $U(1)_\CD$ to mix with the Lorentz symmetry, leading to a theory with varying coupling constant and duality defects. The approach that we will take, starting in 6d, will allow for a definition of the twisted theory even in the absence of a clear definition of a 4d $U(1)_{\CD}$ for non-abelian gauge groups.


\section{Duality Twist of 4d $\CN=4$ SYM from 6d}
\label{sec:DualityTwistFrom6d}

We consider now the dimensional reduction on an elliptic K\"ahler three-fold $Y_3$ (which is notnecessarily a Calabi-Yau). In order to preserve some supersymmetries, we need to topologically twist the six-dimensional theory, which means turning on an appropriate R-symmetry gauge field background.\footnote{Notice that we cannot preserve supersymmetry by twisting on a manifold with generic holonomy $SU(4)_L \simeq SO(6)_L$, since the R-symmetry group is only $Sp(4)_R \simeq SO(5)_R$.} This is achieved by twisting the $U(1)_L =$diag$( U(3)_L)$ residual holonomy with a $U(1)_R$ subgroup of the R-symmetry group. 
In this section we describe the six-dimensional abelian twisted theory on a K\"ahler three-fold. We then carry out the reduction on the elliptic fiber and obtain a twisted version of $\CN=4$ SYM theory with varying coupling constant.  In particular, we identify in this way the topologically twist of the M5-brane theory with the duality twist of $\CN=4$ SYM described in \cite{Martucci:2014ema}. 

Interestingly we show that the bonus symmetry, or rather ``non-symmetry", of $\CN=4$ SYM in 4d, has its origin in terms of the geometric $SO(2)_{45}$ rotation symmetry in the $(x,y)$ plane in the flat space M5-brane theory.

We also provide the non-abelian version of the topologically twisted 4d theory and point out difficulties related to the description of the walls and surface defects associated to the inherent coupling constant monodromies.

\subsection{Twist in 6d, Duality Twist and Bonus Symmetry}
\label{ssec:6dTo4dTwist}

We first explain the twist in the six-dimensional theory and work out its relation to the duality twist in four dimensions, which was studied in \cite{Martucci:2014ema}. In the process we identify the six-dimensional origin of the bonus symmetry.

A generic K\"ahler six-manifold has reduced holonomy $U(3)_L \simeq SU(3)_L \times U(1)_L \subset SU(4)_L$ and the twisting consists in mixing the $U(1)_L$ transformation with a $U(1)_R$ R-symmetry transformation, to obtain supercharges transforming trivially under the new Lorentz symmetry. In appendix \ref{sec:6dTwists} we detail the various inequivalent twistings, which correspond to different choices of embedding of $U(1)_R$ inside $Sp(4)_R$.

In the following we consider the topological twisting, which preserves the maximal amount of supersymmetries, namely the twisting described in appendix \ref{sec:6dTwists} as ``Twist 1", which preserves two supercharges on an arbitrary K\"ahler manifold. It can be described as follows. Consider the $SU(2)_R \times U(1)_R$ R-symmetry subgroup associated to the following decomposition of the representation ${\bf 4}$ of $Sp(4)_R$
\be
\ba  Sp(4)_R &\quad \rightarrow \quad SU(2)_R \times U(1)_R \cr
  {\bf 4} &\quad \rightarrow \quad  {\bf 2}_1 \oplus {\bf 2}_{-1} \,.
\ea
\ee
The supercharges decompose under the Lorentz and R-symmetry subgroups as
\begin{equation}
\ba
SU(4)_L\times  Sp(4)_R &\quad \rightarrow \quad    SU(3)_L \times SU(2)_R \times U(1)_L \times U(1)_R \cr 
({\bf 4}, {\bf 4}) &\quad \rightarrow \quad  ({\bf 3}, {\bf 2})_{1,1} \oplus ({\bf 3}, {\bf 2})_{1,-1} \oplus
  ({\bf 1}, {\bf 2})_{-3,1} \oplus ({\bf 1}, {\bf 2})_{-3,-1}\,.
 \ea
\end{equation}
We now choose an alternate Lorentz group
 by redefining the $U(1)_L$ generator as
\begin{equation}
T'_L = T_L - 3T_R \,,
\label{6dTwist}
\end{equation}
where $T_L$ and $T_R$ are the generators of the Lorentz and R-symmetry
$U(1)$s. The decomposition of the supercharges under the twisted symmetry group is then
\be
\ba
SU(4)_L\times  Sp(4)_R &\quad \rightarrow \quad  
  SU(3)_L \times SU(2)_R \times U(1)^\prime_L \times U(1)_R \cr
 ({\bf 4}, {\bf 4}) &\quad \rightarrow \quad    ({\bf 3}, {\bf 2})_{-2,1} \oplus ({\bf 3}, {\bf 2})_{4,-1} \oplus 
  ({\bf 1}, {\bf 2})_{-6,1} \oplus ({\bf 1}, {\bf 2})_{0,-1} \,.
\ea
\ee
The last term corresponds to supercharges that are trivial under the twisted
Lorentz group and transform in the ${\bf 2}_{-1}$ under the residual $SU(2)_R \times U(1)_R$
R-symmetry group. These correspond to two scalar supercharges in the twisted theory.

Anticipating the result of the reduction, we would like to make immediately the connection with the duality twist in 4d. For this purpose it is useful to understand more precisely which component of the spin connection will be affected by the twisting. 
At the end of section \ref{ssec:Elliptic3folds}, and following \cite{green1987superstring}, we explained how the reduced holonomy algebra $\mathfrak{u}(3)_L$ is embedded into $\so(6)_L$. 
Under this embedding, the generator of the $U(1)_L$ subgroup is given, up to normalization, by the $\so(6)$ matrix $T_L \equiv J = - \textrm{diag}(\epsilon, \epsilon, \epsilon)$.\footnote{The analysis in \cite{green1987superstring} shows that the $U(1)_L$ generator $J = (J^A{}_B)$ can be identified with a chosen complex structure of the K\"ahler manifold. This is made explicit in appendix \ref{app:TwistForms}.} The spin connection component which gauges $U(1)_L$ is  given by 
\be
\Omega^{U(1)_L} = \ \frac{1}{12} \, \Omega^B{}_A \, J^A{}_B \ = \ - \frac{1}{6} \lp \Omega^{01} + \Omega^{23} + \Omega^{45} \rp \,,
\ee
with a normalization chosen such that supercharges have have the $U(1)_L$ charges given above. 
The twist of the 6d theory \eqref{6dTwist} is then implemented by turning on a $U(1)_R$ connection 
\be\label{ARDef}
A^{U(1)_R}  = \  - 3  \, \Omega^{U(1)_L} \ = \  \frac 12 \lp \Omega^{01} + \Omega^{23} + \Omega^{45} \rp  \, .
\ee
We can now see what is effect of this twist from the four-dimensional point of view.
 The base $B$ is K\"ahler, so its holonomy group is also reduced
\be\ba
  \hbox{Holonomy of $B_2$} &  \ \subset  U(1)_\ell \times SU(2)_\ell  \ \subset U(1)_L \times SU(3)_L \,,
\ea\ee
with $U(1)_\ell$ a combination of $U(1)_L$ and a certain $U(1) \subset SU(3)_L$. More precisely the generators of the $U(1)_L$, $U(1)_\ell$ and $SO(2)_{45}$ rotation are related by $T_L = T_\ell + 2 T_{45}$. The $U(1)_\ell$ component of the 4d spin connection is given in our frame by
\begin{align}
\omega^{U(1)_\ell} &= - \frac14 \lp  \omega^{01} + \omega^{23} \rp  \,.
\end{align}
The $SO(2)_{45}$ rotation symmetry in the $(xy)$ plane of the 6d theory is broken when we place the theory on the torus. In standard dimensional reduction, such a symmetry is restored in the infrared theory where it appears as an R-symmetry. In our case the 4d theory does have an enlarged R-symmetry, but it does not come from the $SO(2)_{45}$ rotation, which does not survive as a genuine symmetry of the 4d theory. We will come back to this point shortly.  We propose to identify it with the $U(1)_\CD$ bonus symmetry reviewed in section \ref{sec:Bonus}, which, similarly to R-symmetries, acts on the supercharges of $\CN=4$ SYM, i.e. 
\be
SO(2)_{45} \simeq  U(1)_\CD \,.
\ee
The $SU(2)_R \times U(1)_R$ R-symmetry subgroup can be identified with a subgroup of the 4d $SU(4)_r$ enhanced R-symmetry as follows: we consider the subgroup 
\be
\ba
\hbox{4d R-symmetry}:\quad &  SU(4)_r \supset SU(2)_A \times SU(2)_B \times U(1)_r   \,,
\ea
\ee
and we identify $U(1)_r \equiv U(1)_R$ and $SU(2)_R =$diag$(SU(2)_A \times SU(2)_B)$. 

Having found the origin of all the relevant symmetries in 4d, we find that the six-dimensional twisting \eqref{6dTwist}, which corresponds to turning on the $U(1)_R$ connection \eqref{ARDef}, is expressed in 4d by turning on the corresponding $U(1)_r$ connection
\begin{align}
A^{U(1)_r} &=   - 2  \omega^{U(1)_\ell}  +  \omega^{U(1)_\CD} \,,
\end{align}
with $\omega^{U(1)_\CD} \equiv  \frac 12 \Omega^{45}$.
This corresponds to the four-dimensional twisting described by the generator redefinitions
\be\ba 
T'_\ell &=  T_\ell - 2 \, T_r \,, \quad T'_\CD = T_\CD + T_r \,,
\ea\ee
where $T_\ell, T_r$ and $T_\CD = -2 T_{45}$ are the generators or $U(1)_\ell, U(1)_r$ and $U(1)_\CD$, respectively.
We find
$\omega^{U(1)_\CD}  = -\frac{\p_a  \tau_1}{4 \tau_2} \, e^a $,
in agreement  with the $U(1)_\CD$ connection described in \cite{Martucci:2014ema}\footnote{There is a difference by a factor of $-2$ compared to the conventions in the reference, due to a different normalization of the $U(1)$ charges.}.
For later use we also call $\CA$ the $U(1)_{\CD}$ connection 
\begin{align}
\CA \equiv  \omega^{U(1)_\CD}  =  -\frac{\p_a  \tau_1}{4 \tau_2} \, e^a    \,.
\end{align}

Let us now discuss in more details the identification of the $SO(2)_{45}$ rotations with the $U(1)_\CD$ bonus symmetry. Here we only consider the $(2,0)$ theory on the direct product flat elliptic fibration $Y = E_{\tau} \times \bbR^4$.
First we need to point out a peculiarity of the reduction of the $(2,0)$ theory to $\CN=4$ SYM. In generic dimensional reduction from a $d$-dimensional theory to a $(d-2)$-dimensional theory, the $SO(2)$ Lorentz rotation in the two dimensional plane on which we reduce becomes a $U(1)$ R-symmetry of the reduced theory. This follows from the fact that the $SO(2)$ Lorentz rotation becomes a global symmetry in the lower-dimensional theory, which acts on the supercharges. It then becomes an R-symmetry. When we define the theory on the torus, we break this rotation symmetry, but in the low-energy limit, or zero size limit of the torus, it is recovered, so we still naively expect it to become a $U(1)$ R-symmetry in the lower-dimensional theory.   
However this cannot be the case in our setup since, if it were true, the 4d theory would have a least a $Sp(4)_R \times U(1)$ R-symmetry, but $\CN=4$ SYM has $SU(4)_{r}$ R-symmetry which does not contain $Sp(4)_R \times U(1)$ as a subgroup. This means that the $SO(2)_{45}$ Lorentz symmetry, broken by the torus geometry, is not recovered in the low-energy limit as an R-symmetry in 4d. 
How is this possible? This is possible if the 6d  theory does not have a local Lorentz invariant action. In this way we avoid the contradiction since the 4d theory does not inherit the symmetry by dimensional reduction of a 6d action. The non-existence of this  $U(1)$ R-symmetry in $\CN=4$ SYM can actually be seen as an argument for ruling out the existence of a 6d Lorentz invariant action for the $(2,0)$ theory.
The R-symmetry enhances instead from $Sp(4)_{R}$ in 6d to $SU(4)_r$ in 4d as explained around (\ref{Nis4R}).

We can identify the $SO(2)_{45}$ rotations with the $U(1)_\CD$ transformations by using the interpretation of the $SL(2,\bbZ)$ dualities in 4d as diffeomorphisms of the elliptic fiber in 6d, which act the complex structure parameter $\tau$.
The diffeomorphisms are labeled by an $SL(2,\bbZ)$ element $\gamma$ and are explicitly given by
\begin{align}
\left\lbrace
\ba
x' &= a x - b y \\
y' &= -c x + d y 
\ea
\right. \, , \quad 
\left( 
\begin{array}{cc}
a & b \\
c & d
\end{array}\right) \equiv \gamma \in SL(2,\bbZ)
 \,,
\end{align}
preserving the coordinate periodicities $(x',y') \sim (x' + 1, y' +1)$. The metric \eqref{metricK} transforms to
 \begin{align}
 ds^2_{Y}& = {V\over \tau'_2} \left( \left(dx'   + \tau'_1 dy' \right)^2 + \tau'_2{}^2  dy'{}^2  \right)  + g^B_{i \bar j} \, dz_i d\bar z_{\bar j} \,, \nn\\
 \textrm{with} & \quad \tau' \ = \ \frac{a\tau + b}{c\tau + d} \ \equiv \ \gamma.\tau \, ,
 \end{align}
which is the same as the initial metric with $\tau$ replaced by $\gamma.\tau$.
The new complex coordinate $w' = x' + \tau' y'$ is related to the initial complex coordinate $w = x + \tau y$ by a rescaling
\begin{align}
w' &= \frac{w}{c\tau + d} \,.
\end{align}
The {\it canonical} complex frame in the new coordinates $(\eee'{}^0,\eee'{}^1, \eee'{}^2)$ is related to the initial complex frame \eqref{frameC} by
\begin{align}
\eee^0 \ = \  \eee'{}^0 \, , \quad \eee^1 \ = \ \eee'{}^1 \, , \quad 
\eee^2 \ = \ \frac{c\tau + d}{| c\tau + d |} \, \eee'{}^2 \ = \ e^{i \alpha(\gamma)} \, \eee'{}^2 \, ,
\label{frameRotation}
\end{align}
namely the new frame is obtained by an $SO(2)_{45}$ Lorentz rotation with rotation parameter $\alpha(\gamma)$.
This means that the $SL(2, \bbZ)$ diffeomorphism contains an $SO(2)_{45}$ rotation with parameter $\alpha(\gamma)$ \footnote{An $SL(2,\bbR)$ transformation in the plane has an associated Iwasawa decomposition into a product of three linear transformations, one of which is a rotation of the plane.}. This corresponds to a $U(1)_{\CD}$ rotation with parameter $-\frac{\alpha(\gamma)}{2}$. 

The six-dimensional supercharges transform under the representation $({\bf 4}, {\bf 4})$ of $SU(4)_L \times Sp(4)_R$. Under dimensional reduction, they are mapped to supercharges  $Q^{\dot m}_{\alpha}, \ti Q^{m}_{\dot\alpha}$ of $\CN=4$ SYM, transforming as $({\bf 2}, {\bf 1}, \bar{\bf 4})_{-1} \oplus ({\bf 1}, {\bf 2}, {\bf 4})_{1} $ of $SU(2)_1 \times SU(2)_2 \times SU(4)_r \times U(1)_{\CD} $, where we have indicated the charges under $U(1)_{\CD}$ inherited from 6d. 
The $U(1)_{\CD}$ rotation accompanying the $SL(2,\bbZ)$ elliptic fiber diffeomorphisms in 6d acts on the 4d supercharges precisely as in \eqref{Qrotation}.

We conclude that the uncanny 4d duality twist, is nicely explained from the 6d perspective by identifying the ``bonus symmetry" $U(1)_{\CD}$ in 4d with the $SO(2)_{45}$ Lorentz rotation in 6d. Moreover the so-called duality twist is simply expressed as a standard topological twist of the $(2,0)$ theory.

Finally we would like to provide the $U(1)_{\CD}$ charges of the 4d fields. These are inherited from the $SO(2)_{45}$ charges of the 6d fields. $H_3$ and $\Phi^{\wat i}$ are invariant under $SO(2)_{45}$. The 4d scalars $\phi^{\wat i}$ are thus invariant under $U(1)_{\CD}$. We can rewrite the three-form decomposition \eqref{eqn:3formdecomp} as
\be
\ba
& H_3 = \frac{1}{\sqrt V} \, F^{(+)}_2 \wedge \eee^2 + \frac{1}{\sqrt V} \, F^{(-)}_2 \wedge \bar\eee^2 - \frac{i}{V} \star_4 D_1 + \frac{i}{2V} D_1 \wedge \eee^2 \wedge \bar\eee^2 \,, \cr
& \text{with}  \quad F^{(\pm)}_2 = \ \sqrt \tau_2 \, \lp \frac{F_2 \pm \star_4 F_2}{2} \rp    \,.
\label{DecompBis}
\ea
\ee
The complex frame components $\eee^2, \bar\eee^2$ transform with charge $2$ and $-2$ under $U(1)_{\CD}$. The 4d fields $F^{(+)}_2$ and $F^{(-)}_2$ then transform with opposite charges $-2$ and $2$ respectively. Similarly we deduce that the one-form $D_1=dC$ is uncharged under $U(1)_{\CD}$, implying that $C \simeq \phi^6$ is also uncharged, as are the other scalars $\phi^{\wat i}$.
Finally the 6d spinors decomposes into components of charge $\pm \frac 12$ under $SO(2)_{45}$ or $\pm 1$ under $U(1)_{\CD}$. More precisely in the decomposition \eqref{SpinorDecomp}, the component $\lambda_+^{\dot m} \otimes \eta_+ $ has charge $-1$ and the component $\lambda_-^m \otimes \eta_- $ has charge $1$. We obtain that $\lambda_+{}_m$ has charge $-1$and $\lambda_-^m$ has charge $1$. The charges are summarized in table \ref{tab:U1Dcharges}. They match the charges under the bonus symmetry given in \cite{Intriligator:1998ig}.

\begin{table}
\begin{center}
\begin{tabular}{c|ccccc}
 &  $F^{(+)}_2$ & $F^{(-)}_2$ & $\phi^{\wat i}$ & $\lambda_+{}_m$ & $\lambda_-^m$  \\ [5pt]
 \hline 
$U(1)_{\CD}$ & $-$2 & 2 & 0 & $-$1 & 1  
\end{tabular}
\end{center}
\caption{$U(1)_{\CD}$ charges of fields of 4d $\CN=4$ SYM.}
\label{tab:U1Dcharges}
\end{table}

\subsection{Twisted 6d theory}
\label{sec:Twisted6d}

We now describe the six-dimensional theory on a K\"ahler three-fold $Y_3$ with the topological twist described above. The twisting can be done on any K\"ahler three-fold, so the theory that we describe in this section does not require the K\"ahler manifold to be elliptically fibered. We will restrict to this case in the next section when we study the reduction to four-dimensions.

In the twisted theory the fields transform in representations of the twisted symmetry group $SU(3)_L \times U(1)'_L \times SU(2)_R$\footnote{They also have a $U(1)_R$ symmetry charge that is not relevant for this section.} corresponding to (anti)holomorphic forms. To work out the decomposition of the various fields into form components, it is useful to introduce several geometric objects. The twisted theory possesses a covariantly constant spinor of positive chirality, that we denote $\eta$,
\be
(\nabla_{M} - i A^{U(1)_R}_M )\eta = 0  \,,
\ee
where $M = 0,1, \cdots, 5$ is the spacetime index and the background $U(1)_R$ connection is fixed by the twisting \eqref{ARDef}. From bilinears of this covariantly constant spinor, we can construct a K\"ahler form $J_{(1,1)}$, a holomorphic three-form $\Omega_{(3,0)}$ and an anti-holomorphic three-form $\bar\Omega_{(0,3)}$ on $Y_3$. The details are presented in appendix \ref{app:TwistForms}.

The imaginary self-dual 3-form $H_3$ is in the representation ${\bf 10}$ of $SU(4)_L$ and is not charged under the R-symmetry group. Under the twisted symmetry group  it decomposes to the sum of irreducible representations
\be
\ba
SU(4)_L \times Sp(4)_R \quad \rightarrow \quad &SU(3)_L \times  SU(2)_R\times U(1)'_L \cr 
({\bf 10}, {\bf 1}) \quad \rightarrow \quad & ({\bf 6}, {\bf 1})_{2} \oplus ({\bf 3}, {\bf 1})_{-2} \oplus    ({\bf 1}, {\bf 1})_{-6}  \,,
 \label{H3RepDecomp}
\ea
\ee
corresponding to an imaginary self-dual $(1,2)$-form $h_{(1,2)}$, a holomorphic one-form $h_{(1,0)}$ and a holomorphic three-form $h_{(3,0)}$. 
The twist charge $q_{L'}/2$ indicates the difference in the holomorphic and anti-holomorphic degree: $q_{L'}/2 = q-p$ for a form $\Lambda^{(p,q)}$. For instance $(1,1)_{-6}$ is a $(3,0)$ form. The transformation under the $SU(3)_L$ furthermore fixes the actual form degrees: in the case of ${\bf 6}$ with charge $q_{L'}/2 =1$, the  form is of type $\Lambda^{(p,p+1)}$ with $p=1$. Likewise ${\bf 3}_{-2}$ is a $\Lambda^{p+1,p}$   form with $p=0$. 
Indeed, we can decompose the three-form $H_3$ as
\be
H_3 =   h_{(1,2)} + h_{(1,0)} \wedge J  + h_{(3,0)} \,, \quad {\rm with} \quad J  \wedge h_{(1,2)} = 0 \,,
\ee
where $J$ is the K\"ahler form of $Y_3$, defined in appendix \ref{app:TwistForms}.
Each component is  imaginary self-dual (see appendix \ref{app:DiffGeo}) and corresponds to one of the irreducible pieces in the decomposition \eqref{H3RepDecomp}. The equations of motion $dH_3=0$ become
\begin{align}
\p h_{(1,2)} + \bar\p h_{(1,0)} \wedge J = 0 \,, \quad \bar\p h_{(1,2)} = 0 \,, \quad \p h_{(1,0)} \wedge J + \bar\p h_{(3,0)} = 0 \,.
\end{align}
These equations can be solved locally by expressing the three-forms in terms of the two-form potentials $b_{(1,1)}, b_{(2,0)}, b_{(0,2)}$ as
\begin{align}
h_{(1,2)} &= \bar\p b_{(1,1)} + \p b_{(0,2)} \,, \quad   h_{(1,0)} \wedge J  = \p b_{(1,1)} + \bar\p b_{(2,0)}  \,, \quad h_{(3,0)} = \p b_{(2,0)} \,,
\end{align}
and demanding that $\bar\p b_{(0,2)}=0$, $J \wedge (\bar\p b_{(1,1)} + \p b_{(0,2)}) = 0$ and $\p b_{(1,1)} + \bar\p b_{(2,0)}$ is imaginary self-dual. $h_{(1,0)}$ can be expressed directly in terms of the form-potentials, using the relation $\star ( h_{(1,0)} \wedge J \wedge J ) = 2 i \, h_{(1,0)}$. This leads to $ h_{(1,0)} = -\frac i2 \star( \p b_{(1,1)} \wedge J + \bar\p b_{(2,0)} \wedge J )$.
\smallskip

Next we consider the fermions. The positive chirality spinors $\lambda^m$ transform as the supercharges, in the $({\bf 4}, {\bf 4})$, and  decompose  as follows 
\be  \label{SpinorRepDecomp}
\ba
SU(4)_L  \times Sp(4)_R &\quad \rightarrow \quad SU(3)_L  \times SU(2)_R\times U(1)'_L \cr 
({\bf 4}, {\bf 4}) &\quad \rightarrow \quad   ({\bf 3}, {\bf 2})_{-2} \oplus ({\bf 1}, {\bf 2})_{-6} \oplus ({\bf 3}, {\bf 2})_{4} \oplus    ({\bf 1}, {\bf 2})_0 \,.
\ea
\ee
We can identify these irreducible representations with the holomorphic one- and three-forms $\Lambda^\alpha_{(1,0)}$, $\Lambda^\alpha_{(3,0)}$,  the anti-holomorphic two-form $\Lambda^\alpha_{(0,2)}$ and the scalar $\Lambda^\alpha_{(0,0)}$, respectively. Here $\alpha \in \{1,2\}$ denotes the index for the ${\bf 2}$ of the $SU(2)_R$ R-symmetry. This follows directly from the twist charges: $q_{L'}/2$ gives the degree of the form, and the sign$(q_{L'}) = \pm $ determines whether the form is holomorphic or anti-holomorphic. 
The equations of motion for the twisted fermions are determined in appendix \ref{app:TwistForms}, using the explicit decomposition of the spinors into form components,  to be
\be
\ba
 \bar\p \star \Lambda^\alpha_{(1,0)} \ = \ 0 \,, \quad \bar\p \Lambda^\alpha_{(0,2)} \ &= \ 0\cr 
  \p \Lambda^\alpha_{(1,0)} -  i \star \bar\p \Lambda^\alpha_{(3,0)} \ &= \ 0 \cr 
 \bar\p \Lambda^\alpha_{(0,0)} -  \star \p \star \Lambda^\alpha_{(0,2)} \ &= \ 0    \,.
\ea
\ee
These equations of motion can be integrated to the action 
\be\ba\label{Twisted6dFermAction}
S^{\rm 6d \, twist}_{\Lambda} &= \int 4 \, \epsilon_{\alpha\beta} \lp \Lambda^\alpha_{(0,0)} \bar\p \star \Lambda^\beta_{(1,0)}
 -  \Lambda^\alpha_{(0,2)} \wedge \star \p \Lambda^\beta_{(1,0)} 
  +  i \, \Lambda^\alpha_{(0,2)} \wedge \bar\p \Lambda^\beta_{(3,0)}    \rp \,,
\ea\ee
up to total derivatives.
Note that  the forms $\Lambda^\alpha_{(n,m)}$ are fermionic (Grassmann-odd).

We now turn to the scalar fields $\Phi^{\wat i}$ which transform in the ${\bf 5}$ of $Sp(4)_R$. Under the subgroup $SU(2)_R \times U(1)_R$ they decompose into
\begin{align}
{\bf 5} \ \rightarrow \ {\bf 3}_{0} + {\bf 1}_{-2} + {\bf 1}_{2} \,.
\end{align}
In the twisted theory these fields transform in the representation
\be
\ba
SO(6)_L \times Sp(4)_R &\quad \rightarrow \quad SU(3)_L \times  SU(2)_R\times U(1)'_L\cr 
({\bf 1}, {{\bf 5}}) &\quad \rightarrow \quad {(\bf 1,\bf 3)}_{0} + {(\bf 1,\bf 1)}_{-6} + {(\bf 1,\bf 1)}_{6} \,,
\ea
\ee
where the twist charges again identify these in terms of the forms as follows:
an $SU(2)_R$-triplet of scalars $\varphi^{\alpha \beta}_{(0,0)}$ (with $\varphi^{\alpha \beta} = \varphi^{\beta \alpha}$), a holomorphic three-form $\sigma_{(3,0)}$ for $q_{L'}/2= -3$ and an anti-holomorphic three-form $\bar\sigma_{(0,3)}$ for  $q_{L'}/2= 3$.
The scalars $\varphi^{\alpha \beta}_{(0,0)}$ can be identified with $\Phi^{\wat k}$, $\wat k=1, 2,3$, via
\begin{align}
\varphi^{\alpha \beta} &= \sum_{\wat k=1}^3 \Phi^{\wat k}\, \big( \sigma^{\wat k} \big)_{\alpha \beta} \,, 
\end{align}
with $\sigma^{\wat k} $ the standard Pauli matrices, generating the $SU(2)_R$ symmetry.
With the remaining scalars $\Phi^4, \Phi^5$, we can build  $\Phi = (\Phi^4 - i \Phi^5)/ 2$, $\bar\Phi = (\Phi^4 + i \Phi^5)/ 2$ which have charges $-2$ and $2$ respectively under $U(1)_R$. In the twisted theory they carry a charge $-6$ and $6$ under $U(1)'_L$ and are identified with the (single) components of holomorphic and anti-holomorphic three-forms $\sigma_{(3,0)}$, $\bar\sigma_{(0,3)}$: 
\begin{align}
\sigma_{(3,0)} &=  \ \, \Phi \, \Omega_{(3,0)} \,, \quad  \bar\sigma_{(0,3)}   \ = \  \,\bar\Phi \, \bar \Omega_{(0,3)} \,,
\end{align}
where $\Omega_{(3,0)}$ and $\bar \Omega_{(0,3)}$ are holomorphic and anti-holomorphic three-forms respectively, which can be constructed in the twisted theory, as explained in appendix \ref{app:TwistForms}.
The inverted relations are
\begin{align}
\Phi &= \   \frac{1}{48} \,  \sigma_{ABC} \bar \Omega^{ABC} \, , \quad  \bar\Phi \ = \  \frac{1}{48} \, \bar \sigma_{ABC} \Omega^{ABC} \,.
\end{align}
In the 6d Euclidean theory we are considering a doubled number of fields and $\Phi^{\wat i}$ are complex scalars. Then $\varphi^{\alpha \beta}$ are again complex scalars, while $\sigma$, $\bar\sigma$ are independent (anti-)holomorphic three-forms.

The equations of motion  for the scalars in flat space are $\p^M \p_M \Phi^{\wat i} = 0$. In the twisted theory on curved space we need to covariantize the derivatives $\p_M \to D_M \equiv \nabla_M - i q_R A^{U(1)_R}_M$ and possibly to add curvature terms, including terms involving the field strength of the background connection $A^{U(1)_R}$.
We will simply guess the form of these curvature terms, requiring some cancellations, and verify that the supersymmetries are preserved.\footnote{Another approach would be to consider the twisted background in the context of rigid supergravity. We could use the maximal superconformal gravity of \cite{Bergshoeff:1999db}, find the appropriate background (auxiliary) fields going along with the twisting and extract the equations of motion for the scalars.}
So the equations of motion  for the scalars are generically
\be
D^A D_A \Phi^{\wat i} + \textrm{curv}  =  0 \,,
\ee
where ``curv" denotes extra curvature terms.
The equations of motion for the scalars $\varphi^{\alpha\beta}$ are found to be  $\nabla^2 \varphi^{\alpha\beta}   =  0$, without extra curvature terms. It can be re-written
\be 
(\p \star \bar \p + \bar \p \star \p )\varphi^{\alpha\beta} =0 \,.
\ee
The equations of motion for $\sigma_{(3,0)}$ are
\be\ba
0 &= D^A D_A \sigma_{BCD} \, \bar\Omega^{BCD} +  \textrm{curv}  \cr
&= \lp 4 D^A D_{[A}\sigma_{BCD]} + 3 D_A D_{[B} \sigma^A{}_{CD]} \rp \bar\Omega^{BCD} +  \textrm{curv} \cr
&= \lp 4 D^A D_{[A}\sigma_{BCD]} + 3  D_{B} D_A\sigma^A{}_{CD}  + 3 R_{AB}{}^{AE} \sigma_{ECD} + 6 R_{ABC}{}^{E} \sigma^A{}_{ED}\rp \bar\Omega^{BCD} +  \textrm{curv} \,. \nonumber
\ea\ee
With the curvature terms canceling in this equation, this leaves us with
\be\ba
  & \lp 4 D^A D_{[A}\sigma_{BCD]} + 3  D_{B} D_A\sigma^A{}_{CD} \rp \bar\Omega^{BCD} =0 \cr
 \Rightarrow  & \quad  (\star d \star d  + d \star d \star) \, \sigma_{(3,0)} \wedge \bar\Omega_{(0,3)}  =   0  \cr
 \Rightarrow &\quad \p \star \bar\p \, \sigma_{(3,0)}  =  0 \,,
\ea\ee
where we used  the imaginary self-duality properties of holomorphic three-forms. Similarly we find  $\bar \p \star \p \, \bar\sigma_{(0,3)} =  0$.
The equations of motion that we end up with, assuming nice cancellations of curvature terms, are then\footnote{The equations of motion for $\varphi^{\alpha\beta}$ can also be simplified using  $\bar \p \star \p \varphi^{\alpha\beta} = \p \star \bar \p \varphi^{\alpha\beta}$, if needed.}
\begin{align}
(\p \star \bar \p + \bar \p \star \p )\varphi^{\alpha\beta} =0 \,, \quad  \p \star \bar\p \, \sigma_{(3,0)}  =  0 \,, \quad \bar \p \star \p \, \bar\sigma_{(0,3)}  =  0 \,.
\end{align}
These equations of motion can be integrated to the six-dimensional action
\begin{align}\label{Twisted6dScalarAction}
S^{\rm 6d \, twist}_{\varphi, \sigma} &= \frac 14 \int   \, \bar\p \varphi^{\alpha\beta} \wedge \star \p \varphi_{\alpha\beta}  +  \bar\p\sigma_{(3,0)} \wedge \star \p \bar\sigma_{(0,3)} \,,
\end{align}
which can also be  derived from \eqref{6dAction} after adding extra appropriate curvature terms. 

The full set of equations of motion of the twisted theory is then
\be\ba
& \p h_{(1,2)} + \bar\p h_{(1,0)} \wedge J = 0 \,, \quad  \bar\p h_{(1,2)} = 0 \,, \quad \p h_{(1,0)} \wedge J + \bar\p h_{(3,0)} = 0 \,, \cr
  \bar\p \star \Lambda^\alpha_{(1,0)}  &=  0 \,, \quad \bar\p \Lambda^\alpha_{(0,2)}  =  0    \,, \quad
  \p \Lambda^\alpha_{(1,0)} -  i \star \bar\p \Lambda^\alpha_{(3,0)}  =  0   \,, \quad
 \bar\p \Lambda^\alpha_{(0,0)} -  \star \p \star \Lambda^\alpha_{(0,2)}  =  0   \,, \cr
& (\p \star \bar \p + \bar \p \star \p )\varphi^{\alpha\beta} =0 \,, \quad  \p \star \bar\p \, \sigma_{(3,0)}  =  0 \,, \quad \bar \p \star \p \, \bar\sigma_{(0,3)}  =  0 \,.
\ea\ee




In the twisted theory two supersymmetries are preserved. They are generated by two covariantly constant spinors of negative chirality and thus are proportional to $\eta^c$, which is the charge conjugate of $\eta$ (see appendix \ref{app:TwistForms}). These correspond to two specific spinors out of the four spinors $\mathcal{E}^m$ parametrizing the flat space supersymmetry, 
\begin{align}
\mathcal{E}^m \ \rightarrow  \mathcal{E}^\alpha_{-1} \equiv   \Upsilon^\alpha \, \eta^c  \, , \quad \mathcal{E}^\alpha_{1} = 0 \, ,
\end{align}
where $\Upsilon^\alpha$ are constant (Grassmann-odd) spacetime scalars and transforming as a doublet of $SU(2)_R$ and parametrizing the preserved supersymmetries. The index $\pm 1$ again matches the $U(1)_R$ charge of the spinors preserved by the twisiting.  The reduction from the $Sp(4)_R$ indices $m$ to the $SU(2)_R \times U(1)_R$ indices $\alpha, \pm 1$  is explained in appendix \ref{app:Rsym}.
The supersymmetry transformations of the twisted theory on the K\"ahler space are obtained form the flat space transformations by covariantizing the derivatives with respect to the curved metric and R-symmetry connection. They are given by
\be\ba\label{SusyTwisted6d}
& \delta b_{(1,1)} =   i \, \Upsilon_\alpha \Lambda^\alpha_{(0,0)} \,  J  \,, \quad 
\delta b_{(0,2)} = 2 \,,  \Upsilon_\alpha \Lambda^\alpha_{(0,2)} \, \quad  \delta b_{(2,0)} = 0 \,, \cr
& \delta \Lambda^\alpha_{(0,0)} = 0 \,, \quad \delta \Lambda^\alpha_{(0,2)} = - \frac{1}{4} \Upsilon^\alpha \, \star \p \bar\sigma_{(0,3)}    \,,   \cr
& \delta \Lambda^\alpha_{(1,0)} = \frac i4 \, h_{(1,0)} \Upsilon^\alpha  - \frac 14 \, \p \varphi^{\alpha \beta} \Upsilon_\beta \,, \quad  
\delta \Lambda^\alpha_{(3,0)} = \frac{1}{4} \, h_{(3,0)} \Upsilon^\alpha \,, \cr
& \delta \varphi^{\alpha\beta} = \Upsilon^\alpha \Lambda^\beta_{(0,0)} + \Upsilon^\beta \Lambda^\alpha_{(0,0)} \,, \quad 
 \delta \sigma_{(3,0)} = 4 i  \, \Upsilon_\alpha \Lambda^\alpha_{(3,0)} \,, \quad \delta \bar\sigma_{(0,3)} = 0 \,.
\ea\ee
The transformation of the three-form field strength are
\begin{align}
\delta h_{(1,2)} = \Upsilon_\alpha \lp  i \bar\p \Lambda^\alpha_{(0,0)} \wedge J + 2 \p \Lambda^\alpha_{(0,2)} \rp  \,, \quad
\delta h_{(1,0)} =  i  \Upsilon_\alpha \p \Lambda^\alpha_{(0,0)}  \,, \quad   \delta h_{(3,0)} = 0 \,.
\end{align}
The transformation $\delta h_{(1,2)}$ does not respect $\delta h_{(1,2)} \wedge J =0$, so that the supersymmetry transformation breaks the self-duality condition of $h_{(1,2)}$, however we do have $\delta h_{(1,2)} \wedge J =0$ after imposing the eom. Similarly we have a non-vanishing supersymmetry variation for the $(0,3)$ component of $H_3 =dB_2$, which obeys the wrong self-duality condition:
$\delta h_{(0,3)} = 8 \Upsilon_\alpha \bar\p \Lambda_{(0,2)}^{\alpha}$.
Again we have $\delta h_{(0,3)}=0$ upon imposing the eom.
These supersymmetry transformations for the three-form $H_3$ are thus understood modulo equations of motion. 

%
%

\subsection{Reduction to 4d}

We now specialize to the case when the K\"ahler three-fold is elliptically fibered. This allows us to dimensionally reduce the 
twisted six-dimensional theory on the elliptic fiber to a topologically twisted four-dimensional theory with varying complexified coupling $\tau$. 
In section \ref{sec:DualityTwistFrom6d} we already showed that the six-dimensional R-symmetry twist descends to the Topological Duality Twist of \cite{Martucci:2014ema}. 
We show now that the theory in four dimensions obtained by reducing on the elliptic fiber coincides with topologically duality twisted $\CN=4$ SYM theory described in \cite{Martucci:2014ema}. 


To reduce the theory on the torus we first decompose the six-dimensional K\"ahler form $J$ (defined by \eqref{CanonicalForms}) into the sum of the K\"ahler forms $j$ on the base and $- e^4 \wedge e^5$ on the torus
\begin{align}
J &= j - e^4 \wedge e^5 \,, \quad j = - e^0\wedge e^1 - e^2 \wedge e^3 \,.
\end{align}
The three-forms field of the twisted theory decompose into a collection of forms on the base
\be\ba
 h_{(1,2)} &=  f_{(0,2)} \wedge \frac{\eee^2}{\sqrt V}  + f_{(1,1)} \wedge \frac{\bar\eee^2}{\sqrt V} + \frac{i}{V} \,  d_{(0,1)} \wedge (j +  e^4 \wedge e^5)  \cr
  h_{(1,0)} &= f_{(0,0)}  \frac{\eee^2}{\sqrt V}  - \frac{i}{V} \,  d_{(1,0)}  \,, \quad   h_{(3,0)} = f_{(2,0)} \wedge \frac{\eee^2}{\sqrt V}   \,,
  \label{hDecomp}
\ea
\ee
with the constraint $f_{(1,1)} \wedge j =0$, following from $h_{(1,2)} \wedge J =0$. We used here the complex frame components $\eee^2= e^4 + i e^5$, $\bar\eee^2 = e^4-ie^5$ on the torus.
The subscripts $(m,n)$ refer to the (anti-)holomorphic indices in four dimensions. In particular $j \equiv j_{(1,1)}$.
The equations of motion reduce to
\begin{align}
& \p_{\CA} f_{(0,2)} + \bar\p_{\CA} f_{(0,0)} \wedge j  -i \frac{\bar\p \bar\tau}{2\tau_2} \wedge f_{(1,1)} =0  \,, \quad  
\p_{\CA} f_{(1,1)} + i  f_{(0,0)} \frac{\p \tau}{2\tau_2} \wedge j = 0 \,, \,, \nn\\
&  \bar\p_{\CA} f_{(1,1)}  +  i  f_{(0,2)} \wedge \frac{\p \tau}{2\tau_2}   =0 \,, \quad
    \p_{\CA} f_{(0,0)} \wedge j + \bar\p_{\CA} f_{(2,0)}   = 0 \nn\\
&  \p d_{(1,0)}  = \bar\p d_{(0,1)} = 0 \,, \quad  \bar\p d_{(1,0)} + \p d_{(0,1)} = 0 \,, \quad  \bar\p \star d_{(1,0)} + \p \star d_{(0,1)} = 0  \,,
\end{align}
where we denoted $\p_{\CA}, \bar\p_{\CA}$ the four-dimensional (anti)holomorphic differentials  covariantized with respect to the $U(1)_\CD\simeq SO(2)_{45}$ connection $\CA$. Explicitly we have
\be\ba
d_\CA &= d - i q_\CD \CA \,, \cr
\p_{\CA}  &=  \p + i q_{\CD} \frac{\p\tau_1}{4\tau_2} \,, \quad
\bar\p_{\CA}  = \bar\p  - i q_{\CD} \frac{\bar\p\tau_1}{4\tau_2} \,,
\ea\ee
with $q_{\CD}$ the $U(1)_\CD$ charge.
The one-forms $d_{(1,0)}$ and $d_{(0,1)}$ are not charged under $U(1)_\CD$, so the differentials acting on them are simply $\p, \bar\p$.
The forms $h_{(1,2)}, h_{(1,1)}$ and $h_{(3,0)}$ are not charged under $U(1)_{\CD}$, however the complex frame one-forms $e^4 \pm i e^5$ have charges $\pm 2$, so that the forms $f_{(2,0)}, f_{(0,2)}$ and $f_{(0,0)}$ have charge $-2$ under $U(1)_{\CD}$ and $f_{(1,1)}$ has charge $+2$, and the gauge field $\CA$ appears in their covariant derivatives.
When deriving the above equations we have used the identities
\be\ba
\p_{\CA} \eee^2 = 0 \,, \quad \bar\p_\CA \eee^2 = i \frac{\p \tau}{2\tau_2} \wedge \bar\eee^2 \,, \quad
\p_\CA \bar\eee^2 = -i \frac{\bar\p \bar\tau}{2\tau_2} \wedge \eee^2  \,, \quad \bar\p_{\CA} \bar\eee^2= 0 \,,
\ea\ee
which follow from the the frame definition \eqref{frameR}.\footnote{We also used the relations $j \wedge \p d_{(0,1)} = -i \p \star d_{(0,1)}$, $j \wedge \bar\p d_{(1,0)} = i \bar\p \star d_{(1,0)}$.}
The equations of motion can be expressed more compactly using the (anti)self-dual two form $F_2^{(\pm)}$  and the one-form $D_1$,
\be
F_2^{(+)} = f_{(2,0)} +  f_{(0,2)} + f_{(0,0)} j \,, \quad F_2^{(-)} = f_{(1,1)} \,, \quad D_1 = d_{(1,0)} + d_{(0,1)} \,.
\ee
The equations of motion are then
\begin{align}
&  d_{\CA} F_2^{(+)} =i \frac{\bar\p \bar\tau}{2\tau_2} \wedge F_2^{(-)} \,, \quad  d_{\CA} F_2^{(-)} = -i   \frac{\p \tau}{2\tau_2} \wedge F_2^{(+)}  \,, \quad d D_1 = 0 \,, \quad d \star D_1 = 0  \,.
\end{align}
Defining $F_2$ in terms of $F_2^{(\pm)}$ as in \eqref{DecompBis}, we obtain 
\begin{align}
dF_2 = 0 \,, \quad d\star F_2 + \frac{d\tau_2}{\tau_2} \wedge \star F_2 - i \frac{d\tau_1}{\tau_2} \wedge  F_2  =0  
\,, \quad d D_1 = 0  \,, \quad d \star D_1 = 0   \,.
\end{align}
The closedness relations $dF_2=0$ and $dD_1=0$ are solved by introducing local potentials 
\be\label{DefC}
F_2= dA \,, \qquad D_1 = dC\,,
\ee
and the remaining equations of motion can be integrated to the actions
\be\ba
S_{F} &= {1 \over 4 \pi} \int_B   \tau_2  F_2 \wedge \star F_2 - i \tau_1   F_2\wedge F_2    \,, \cr
S_{C} &= {1 \over 4 \pi} \int_B   \bar\p C \wedge \star \p C    \,.
\ea\ee
We recover the standard Yang-Mills action, up to the fact that $\tau_1$ and $\tau_2$ vary along the base $B$ in the present case. This part of the reduction is actually identical to the reduction on a Calabi-Yau three-fold presented in section \ref{ssec:CYRed}.

The fermions of the twisted theory decompose as
\be\ba
& \Lambda_{(1,0)}^{\alpha} = \frac{1}{\sqrt V} \big( \psi_{(1,0)}^{\alpha} + \ti\chi_{(0,0)}^{\alpha} \eee^2 \big) \,, \quad 
\Lambda_{(3,0)}^{\alpha} = \frac{1}{\sqrt V} \big( \ti\rho_{(2,0)}^{\alpha} \wedge \eee^2 \big)  \cr
& \Lambda_{(0,0)}^{\alpha} = \frac{2}{\sqrt V} \chi_{(0,0)}^{\alpha} \,, \quad \Lambda_{(0,2)}^{\alpha} = \frac{1}{\sqrt V} \big( 2 \rho_{(0,2)}^{\alpha} + \ti\psi_{(0,1)}^{\alpha} \wedge \bar\eee^2 \big) \,.
\label{spinorDecomp}
\ea\ee
The dimensional reduction of the 6d fermionic action \eqref{Twisted6dFermAction} leads to
\begin{align}
S_{\psi,\chi,\rho} &= - 8 \int_B  \epsilon_{\alpha\beta} \Big[   \bar\p \star_4 \psi_{(1,0)}^{\alpha} \wedge  \chi_{(0,0)}^{\beta}    - \p \psi_{(1,0)}^{\alpha} \wedge  \rho_{(0,2)}^{\beta}  -  \p_{\CA} \star_4 \ti\psi_{(0,1)}^{\alpha} \wedge   \ti\chi_{(0,0)}^{\beta} +  \bar\p_{\CA}  \ti\psi_{(0,1)}^{\alpha} \wedge \ti\rho_{(2,0)}^{\beta} \Big] \,.
\end{align}
In the reduction we have used $\star_6 \omega_{(p)}= \star_4 \omega_{(p)} \wedge e^4 \wedge e^5$ for $\omega_{(p)}$ a $p$-form on the base $\CB$, $\star_6 (\omega_{(1)} \wedge \eee^2 ) = - i \star_4 \omega_{(1)} \wedge \eee^2$ and $\star_4 \omega_{(0,2)} = \omega_{(0,2)}$.

Finally the scalars of the twisted theory decompose as
\begin{align}
& \varphi^{\alpha\beta}|_{\rm 6d} = \frac{1}{\sqrt V} \varphi^{\alpha\beta}|_{\rm 4d} \,, \quad \sigma_{(3,0)} = \frac{i}{\sqrt V}\sigma_{(2,0)} \wedge \eee^2 \,, \quad 
\bar\sigma_{(0,3)} = \frac{i}{\sqrt V} \ti\sigma_{(0,2)} \wedge \bar\eee^2 \,.
\label{scalarDecomp}
\end{align}
The dimensional reduction of the 6d scalar action \eqref{Twisted6dScalarAction} yields
\begin{align}
S_{\varphi, \sigma} &=  \frac 14 \int   \, \bar\p \varphi^{\alpha\beta} \wedge \star \p \varphi_{\alpha\beta}  +   2 \bar\p_{\CA} \sigma_{(2,0)} \wedge \star \p_{\CA} \ti\sigma_{(0,2)}  \,.
\end{align}
We used $\star_6 (\omega_{(3)} \wedge \bar\eee^2 ) = i \star_4 \omega_{(3)} \wedge \bar\eee^2$.

The four-dimensional theory has an enlarged R-symmetry $SU(2)_{A} \times SU(2)_{B}$ and we denote doublets of $SU(2)_A$, resp. $SU(2)_B$, with the indices $\alpha, \beta$, resp. $\dot\alpha, \dot\beta$. The scalar field $C$ can be combined with the three scalars $\varphi^{\alpha \beta}$ into the four scalars $\varphi^{ \alpha \dot\alpha}$ with the explicit relations 
\be\label{varphiC}
\varphi^{1 \dot 1} = \varphi^{11} \,,\qquad 
\varphi^{2 \dot 2} = \varphi^{22} \,,\qquad
\varphi^{1 \dot 2} = \varphi^{12} + \frac{C}{\sqrt V} \,,\qquad 
\varphi^{2 \dot 1} = \varphi^{21} - \frac{C}{\sqrt V}\,.
\ee
The total action is
\be\ba
S_{\rm total} &=  S_F + \frac{2}{V} S_C  + \frac{1}{\pi} S_{\psi,\chi,\rho} -  \frac{1}{\pi} S_{\varphi, \sigma}   \cr
&= {1 \over 4 \pi} \int_B   \tau_2  F_2 \wedge \star F_2 - i \tau_1   F_2\wedge F_2 \cr
& \  + \frac{8}{\pi}  \int_B  \,  \bar\p \star \psi_{(1,0)}^{\alpha} \,   \chi_{(0,0)}{}_{\alpha}     -    \p \psi_{(1,0)}^{\alpha}  \wedge  \rho_{(0,2)}{}_{\alpha}  -  \p_{\CA} \star \ti\psi_{(0,1)}^{\dot\alpha} \,  \ti\chi_{(0,0)}{}_{\dot\alpha} +    \bar\p_{\CA} \ti\psi_{(0,1)}^{\dot\alpha} \wedge \ti\rho_{(2,0)}{}_{\dot\alpha} \cr
& \ - \frac{1}{4\pi} \  \int_B   \,   \bar\p \varphi^{\alpha \dot\alpha} \wedge \star \p \varphi_{\alpha\dot\alpha}  +  2 \bar\p_{\CA} \sigma_{(2,0)} \wedge \star \p_{\CA} \ti\sigma_{(0,2)} \,.
\label{4dActionTot}
\ea\ee
The relative coefficients between the terms were fixed by requiring invariance under the supersymmetry transformations inherited from six dimensions, which are given by
\begin{align}
& \delta a_{(1,0)} = 0 \,, \quad  \delta a_{(0,1)} =  \frac{2}{\sqrt \tau_2} \Upsilon_{\dot\alpha} \ti\psi_{(0,1)}^{\dot\alpha} \nn\\
& \delta \chi_{(0,0)}^{\alpha} = 0 \,, \quad \delta \ti\chi_{(0,0)}^{\dot\alpha} = \frac i4 \Upsilon^{\dot\alpha} f_{(0,0)} \nn\\
& \delta \psi_{(1,0)} ^{\alpha} = - \frac 14 \Upsilon_{\dot\alpha} \p \varphi^{\alpha \dot\alpha} \,, \quad 
\delta \ti\psi_{(0,1)}^{\dot\alpha} = \frac 14 \Upsilon^{\dot\alpha} \star \p_{\CA} \ti\sigma_{(0,2)} \nn\\
& \delta \rho_{(0,2)}^{\alpha} = 0 \,, \quad \delta \ti\rho_{(2,0)}^{\dot\alpha} = \frac 14 \Upsilon^{\dot\alpha} f_{(2,0)} \nn  \\
& \delta \varphi^{\alpha \dot\alpha} = 4 \Upsilon^{\dot\alpha} \chi_{(0,0)}^{\alpha} \,, \quad  
\delta \sigma_{(2,0)} = 4 \Upsilon_{\dot\alpha} \ti\rho_{(2,0)}^{\dot\alpha} \,, \quad  \delta \ti\sigma_{(0,2)} = 0  \,,
\end{align}
where we have defined the holomorphic and anti-holomorphic component of the gauge field $A = a_{(1,0)} + a_{(0,1)}$. The relation $dA = F_2 =  (F_2^{(+)} + F_2^{(-)})/ \sqrt \tau_2$ yields $f_{(2,0)} = \sqrt\tau_2 \p a_{(1,0)}$, $f_{(0,2)} = \sqrt\tau_2 \bar\p a_{(0,1)}$ and $f_{(1,1)} + f_{(0,0)} \wedge j = \sqrt\tau_2(\bar\p a_{(1,0)} + \p a_{(0,1)})$.
To obtain these transformations we have used the equations of motion of the theory.\footnote{For instance we obtain from the dimensional reduction $\delta f_{(0,2)} = - 2 i \Upsilon_{\dot\alpha} \frac{\bar\p\tau_1}{\tau_2} \wedge \ti\psi_{(0,1)}^{\dot\alpha} = 2 \Upsilon_{\dot\alpha} \bar\p_{\CA} \ti\psi_{(0,1)}^{\dot\alpha} $+ eom, that we simply replace by $\delta f_{(0,2)} =  2\Upsilon_{\dot\alpha} \bar\p_{\CA} \ti\psi_{(0,1)}^{\dot\alpha}$. } This allows us to derive the supersymmetry transformations of the gauge fields $a_{(1,0)}, a_{(0,1)}$ and the fourth scalar $C$.

The transformations on the field strength forms are
\begin{align}
\delta f_{(2,0)} = 0 \,, \quad \delta f_{(0,2)} =  2\Upsilon_{\dot\alpha} \bar\p_{\CA} \ti\psi_{(0,1)}^{\dot\alpha} \,, \quad
\delta f_{(0,0)} \, j +  \delta f_{(1,1)} =  2\Upsilon_{\dot\alpha} \p_{\CA} \ti\psi_{(0,1)}^{\dot\alpha} \,.
\end{align}
We should now comment on the $U(1)_{\CD}$ action in the 4d twisted theory. The $U(1)_{\CD}$ charges of the 4d  fields can be read from the decomposition of the 6d fields \eqref{hDecomp}, \eqref{spinorDecomp}, \eqref{scalarDecomp}, and the identification $U(1)_{\CD} \simeq SO(2)_{45}$. They are given in table \ref{tab:U1Dcharges2}.
As in the non-twisted theory studied in section \ref{ssec:CYRed} , $U(1)_{\CD}$ is a symmetry of the equations of motion of the abelian twisted theory, but not a symmetry of the action \eqref{4dActionTot}, due to the non-invariance of the $S_F$ term.

\begin{table}[h]
\begin{center}
\begin{tabular}{c|ccccccccccc}
 &  $F^{(+)}_2$ & $F^{(-)}_2$ & $\varphi^{\alpha \dot\alpha}$ & $\sigma_{(2,0)}$ & $\ti\sigma_{(0,2)}$ & $\chi_{(0,0)}^\alpha$ & $\ti\chi_{(0,0)}^{\dot\alpha}$ & $\psi_{(1,0)}^\alpha$ & $\ti\psi_{(0,1)}^{\dot\alpha}$ & $\rho_{(0,2)}^\alpha$ & $\ti\rho_{(2,0)}^{\dot\alpha}$  \\ [5pt]
 \hline 
$U(1)_{\CD}$ & $-2$ & 2 & 0 & $-2$ & 2 & 0 & $-2$ & 0 & 2 & 0 & $-2$ 
\end{tabular}
\end{center}
\caption{$U(1)_{\CD}$ charges of 4d twisted fields.}
\label{tab:U1Dcharges2}
\end{table}

In four-dimensions $U(1)_{\CD}$ transformations accompany $SL(2, \Z)$ dualities. This implies that under and an $SL(2,\Z)$ transformations $\gamma = \abcd{a}{b}{c}{d}$ the fields transform as
\begin{align}
\Phi' &= e^{- \frac i2 q_{\CD}\alpha(\gamma) } \Phi  \ = \ \lp \frac{c\tau + d}{| c\tau + d |} \rp^{- \frac{q_{\CD}}{2}} \Phi \,,
\label{U1DTransfo}
\end{align}
where  $q_{\CD}$ denotes the $U(1)_{\CD}$ charge of $\Phi$.

As observed for the theory on the elliptic Calabi-Yau in section \ref{ssec:CYRed}, the action \eqref{4dActionTot} is locally invariant under the supersymmetry transformations given above, but not globally. Rather the variation of the Lagrangian $\delta\CL^{\rm (4d)}_{\rm total}$ is a total derivative
\begin{align}
\delta\CL^{\rm (4d)}_{\rm total} &=  \frac{1}{\pi} \,  d \lp - i \frac{\bar\tau}{\sqrt\tau_2} \Upsilon_{\dot\alpha} \ti\psi_{(0,1)}^{\dot\alpha} \wedge F_2 + \Upsilon_{\dot\alpha} \chi_{(0,0)}{}_{\alpha} \star d \varphi^{\alpha \dot\alpha} 
+ 2 \Upsilon_{\dot\alpha} \ti\rho_{(2,0)}^{\dot\alpha} \wedge \star \p_{\CA} \ti\sigma_{(0,2)} \rp  \,.
\label{LagrSusyTransfo}
\end{align}
This is important since, as we already explained, the holomorphic function $\tau$ has monodromies around (real) codimension two surfaces, implying that $\tau$ jumps across three-dimensional walls. 
The supersymmetry variation of the action is not vanishing, but instead picks boundary terms for each duality wall $W$
\be
\ba
\delta S_W &= \frac{1}{\pi} \left( \int_{W^+} - \int_{W^-}\right) \lp - i \frac{\bar\tau}{\sqrt\tau_2} \Upsilon_{\dot\alpha} \ti\psi_{(0,1)}^{\dot\alpha} \wedge F_2 + \Upsilon_{\dot\alpha} \chi_{(0,0)}{}_{\alpha} \star d \varphi^{\alpha \dot\alpha} 
+ 2 \Upsilon_{\dot\alpha} \ti\rho_{(2,0)}^{\dot\alpha} \wedge \star \p_{\CA} \ti\sigma_{(0,2)} \rp  \,,
\label{deltaSWallTwisted0}
\ea
\ee
where $\int_{W^+} - \int_{W^-}$ computes the difference of the integrals with the fields and coupling $\tau$ evaluated on the two sides of the duality wall. These fields, on the two sides of $W$, are related by the $U(1)_\CD$ transformation \eqref{U1DTransfo}, with $\gamma$ the $SL(2,\Z)$ element associated to $W$. The two last terms in \eqref{deltaSWallTwisted0} are $U(1)_\CD$ invariant and therefore cancel, leaving
\be
\ba
\delta S_W &= \frac{1}{\pi} \left( \int_{W^+} - \int_{W^-}\right) \lp - i \frac{\bar\tau}{\sqrt\tau_2} \Upsilon_{\dot\alpha} \ti\psi_{(0,1)}^{\dot\alpha} \wedge F_2 \rp  \,.
\label{deltaSWallTwisted}
\ea
\ee
We will discuss how to cure the remaining non-invariance in section \ref{sec:Walls}.

\subsection{Non-abelian theory}
\label{sec:NonAb}

Despite the lack of a Lagrangian formulation, it is believed that non-abelian 6d $(2,0)$ theories exist, with gauge algebras of ADE type. These non-abelian theories can in principle be defined on a K\"ahler elliptic fibration with the topological twist that we have described, and reduced to four-dimensions on the elliptic fiber, as we did for the abelian theory. We expect therefore that the
 four-dimensional twisted theory that we found can be promoted to a non-abelian (interacting) theory, with all fields valued in a Lie algebra $\mathfrak{g}$ of ADE type. Since the profile of the coupling $\tau$ has monodromies/duality walls, the theory only makes sense
if for each defect, that is labeled by an $SL(2,\bbZ)$ element $\gamma$, that corresponds to the total monodromy, the gauge group in the local patch is self-dual under $\gamma$. This is satisfied, for instance, for the gauge groups $U(N)$ and $SO(2N)$, which are self-dual under the full $SL(2,\bbZ)$.
In this section we take the fields to be valued in the $\mathfrak{u}(N)$ gauge algebra, corresponding to the reduction of the $A_{N-1}$ type $(2,0)$ theory, with center $\mathfrak{u}(1)$ undecoupled, or equivalently the infrared worldvolume theory of a stack of $N$ M5-branes on the elliptic fibration $Y$ (see discussion in section \ref{ssec:DDfromM5}).

We can promote the gauge field $A$ to a $\mathfrak{u}(N)$ valued  connection, with field strength $F = dA + A \wedge A$.
The action of the non-abelian theory is obtained by covariantizing the derivatives with respect to the gauge connection $d \rightarrow D = d + [A, .]$, taking the trace over the Lie algebra in the action and adding the following terms
\be\ba
S^{NA} &=  \int  \frac{8}{\pi \sqrt{\tau_2}} \Tr \Big[ - \frac{i}{16} f_{(0,0)} [\sigma_{(2,0)} \wedge \ti\sigma_{(0,2)}] - [\ti\psi_{(0,1)}^{\dot\alpha} \wedge \star \psi_{(1,0)}^{\alpha} ] \varphi_{\alpha \dot\alpha}  \cr 
&\phantom{=  \int \frac{8}{\pi \sqrt{\tau_2}} \Tr }  
+ \frac{1}{4} [ \ti\psi_{(0,1)}^{\dot\alpha} \wedge \ti\psi_{(0,1)}{}_{\dot\alpha} ] \wedge \sigma_{(2,0)} 
-  \frac{1}{4} [ \psi_{(1,0)}^{\alpha} \wedge \psi_{(1,0)}{}_{\alpha}] \wedge \ti\sigma_{(0,2)}  \cr
&\phantom{=  \int \frac{8}{\pi \sqrt{\tau_2}} \Tr }  
+  [\ti\chi_{(0,0)}^{\dot\alpha} \wedge \star \chi_{(0,0)}^{\alpha} ] \varphi_{\alpha \dot\alpha} 
 +  [\ti\rho_{(2,0)}^{\dot\alpha} \wedge  \rho_{(0,2)}^{\alpha} ] \varphi_{\alpha \dot\alpha} \cr
&\phantom{=  \int \frac{8}{\pi \sqrt{\tau_2}} \Tr }  
  - [\ti\chi_{(0,0)}^{\dot\alpha} , \ti\rho_{(2,0)}{}_{\dot\alpha}]  \wedge \ti\sigma_{(0,2)} 
+  [\chi_{(0,0)}^{\alpha} , \rho_{(0,2)}{}_{\alpha} ] \wedge \sigma_{(2,0)}  \Big]  \cr
&\phantom{=  \int} + \frac{1}{16\pi \tau_2} \Tr \Big[
   2 [\varphi^{\alpha \dot\alpha}, \sigma_{(2,0)}] \wedge [\varphi_{\alpha \dot\alpha}, \ti\sigma_{(0,2)}] 
+  [\varphi^{\alpha \dot\alpha}, \varphi_{\beta \dot\alpha}] [\varphi^{\beta \dot\beta}, \star \varphi_{\alpha \dot\beta}]  \cr
&\phantom{=  \int \frac{1}{16\pi \tau_2} \Tr }  
+  [\varphi^{\alpha \dot\alpha}, \varphi^{\beta \dot\beta}] [\varphi_{\beta \dot\alpha}, \star \varphi_{\alpha \dot\beta}]   
 +  [\sigma_{(2,0)} \wedge \ti\sigma_{(0,2)} ] \star ([\sigma_{(2,0)} \wedge \ti\sigma_{(0,2)} ])   \Big] \,,
\ea\ee
where, for Lie algebra valued $p$-form and $q$-form $\omega_{(p)}, \ti\omega_{(q)}$, we define
\begin{align}
[ \omega_{(p)} \wedge \ti\omega_{(q)}] \equiv \omega_{(p)} \wedge \ti\omega_{(q)} - (-1)^{pq + \kappa} \,  \ti\omega_{(q)} \wedge \omega_{(p)}  \,,
\end{align}
with $\kappa=1$ if both forms are Grassmann-odd and $\kappa=0$ otherwise.

The supersymmetry transformations receive non-abelian contributions as follows
\be\ba
\delta \ti\chi^{\dot\alpha}_{(0,0)} &=  \frac i4 \Upsilon^{\dot\alpha} f_{(0,0)} -  \frac{1}{16 \sqrt \tau_2} \Upsilon^{\dot\beta} [\varphi_{\alpha \dot\beta} , \varphi^{\alpha \dot\alpha} ] + \frac{1}{16 \sqrt \tau_2} \Upsilon^{\dot\alpha} \star ([\sigma_{(2,0)} \wedge \ti\sigma_{(0,2)} ]) \,, \cr
\delta \rho_{(0,2)}^{\alpha} &=   \frac{1}{8 \sqrt \tau_2} \Upsilon_{\dot\alpha} [\varphi^{\alpha \dot\alpha} , \ti\sigma_{(0,2)} ] \,.
\ea\ee

The topological twist
relies solely on the symmetries of the theory and
can be defined for the non-abelian theories as well. However, since there is no field theoretic description of the non-abelian theory,  we do not know how the fields transform under $SL(2,\Z)$. In particular it is not known how the gauge field, or the field strength, transforms under $SL(2,\Z)$, therefore we are not able to describe the transformations of the fields across the duality walls, with the exception of the $T^k$ duality wall, for which the fields are continuous across the wall, as we shall see in section \ref{sec:Walls}.

The supersymmetry transformation of the action produces the non-abelian total derivative term
\be\ba
\delta\CL^{\rm (4d)}_{\rm total} &=  \frac{1}{\pi} \, \Tr \,   d \Big[ - i \frac{\bar\tau}{\sqrt\tau_2} \Upsilon_{\dot\alpha} \ti\psi_{(0,1)}^{\dot\alpha} \wedge F_2 + \Upsilon^{\dot\alpha} \chi_{(0,0)}^{\alpha} \star D \varphi_{\alpha \dot\alpha} 
+ 2 \Upsilon_{\dot\alpha} \ti\rho_{(2,0)}^{\dot\alpha} \wedge \star D_{\CA} \ti\sigma_{(0,2)}   \cr
 &  \phantom{=  \frac{1}{\pi} \, \Tr \,   d \Big[}  + \frac{1}{\sqrt{\tau_2}} \Upsilon_{\dot\beta} \star\ti\psi^{\dot\alpha}_{(0,1)} [\varphi^{\alpha\dot\beta}, \varphi_{\alpha\dot\alpha}] + \frac{2}{\sqrt{\tau_2}} \Upsilon^{\dot\alpha} \psi^{\alpha}_{(1,0)} \wedge [\varphi_{\alpha \dot\alpha} , \ti\sigma_{(0,2)} ]    \Big] \,.
\label{LagrSusyTransfoNA}
\ea\ee
This term leads to a non-invariance of the action in the presence of 2d surface defects and associated $SL(2,\Z)$ duality walls, as in the abelian theory.
We will see in the next section how to restore full supersymmetry invariance in the abelian theory. However the discussion will not apply for the non-abelian theory in the presence of $SL(2,\Z)$ surface defects, except for the $T^k$ duality defect. It is not obvious how to tackle this issue for the non-abelian theory in general and we leave this for future work.


\section{Duality and Point-Defects}
\label{sec:Walls}

So far we have shown how starting with the twisted 6d $(2,0)$ theory on a K\"ahler elliptic three-fold, we obtain by dimensional reduction on the elliptic fiber, the 4d $\mathcal{N}=4$ SYM theory with varying coupling $\tau$ on the base four-manifold $B$. 
The key ingredient in this construction is the topological twist, which as we have seen, is from the 6d point of view a standard topological twist, and descends in the 4d theory to the duality twist \cite{Martucci:2014ema}, which uses the so-called bonus symmetry of \cite{Intriligator:1998ig}.  

The elliptic fibers can become singular along complex codimension one loci in the base $B$. Such singular loci $\mathcal{C}$ are characterized in terms of the Weierstrass equation of the fibration, as the discriminant locus $\Delta =0$ (see (\ref{Discriminant})) intersecting $B$.  The axio-dilaton, or $\mathcal{N}=4$ SYM coupling, diverges at these loci. The branch-cuts associated to these singularities, which are 3d duality walls, have no direct physical meaning, however, they ensure continuity equations on the fields of the 4d theory. More precisely, as one crosses these cuts, the fields of the 4d theory undergo an $SL(2,\mathbb{Z})$ duality transformation. 

Moreover, the 2d duality defects generically intersect at points in the base, corresponding to singularity enhancement in the fiber of the elliptic three-fold.
We will now discuss how this network of 4d-3d-2d bulk, wall and defect systems arises, first in a gauge theoretic description in 4d, and then from the M5-brane point of view. This setup is depicted in figure \ref{fig:642Setup}. We finally discuss the point defects and describe the local flavor symmetry enhancements associated to them.


\subsection{Gauge-theoretic Description of Walls and Defects}

First we summarize what is known about the  gauge theoretic description of 3d walls and 2d defects in the $\mathcal{N}=4$ SYM with varying coupling in the abelian theory in \cite{Ganor:1996pe, Kapustin:2009av, Martucci:2014ema}.

\subsubsection{3d Duality Walls}

The 3d walls in the elliptic fibration correspond to the branch-cuts of $\tau$, and should not carry any physical degrees of freedom. 
These branch-cuts end at the points where $\tau$ has a log singularity.
The base can be written as $B= \cup B_i$, with $W_{ij}$ a 3d duality wall separating the regions $B_i$ and $B_j$.\footnote{To obtain a global partition $B= \cup B_i$, some of the walls need to be trivial, in the sense that the complex coupling is continuous across them.} 

The gluing conditions of the fields across a duality wall are described by their $U(1)_\CD$ transformation \eqref{U1DTransfo}. For the gauge field, the gluing condition can be described in terms of $F$ and $F_D$ (instead of $F^{(\pm)}$), where
\be
F_D = \tau_1 F + i \tau_2 \star F \,.
\ee
Note that we can write the gauge-part of the 4d action as 
\be\label{FFDAction}
S_{F} = -{i\over 4\pi} \int_{B} F\wedge F_D \,.
\ee
Let $\gamma= \footnotesize{\left(
\begin{array}{cc}
a & b \\
c & d 
\end{array}
\right)} \in SL(2, \mathbb{Z})$ be the transformation associated to the wall $W_{ij}$.
The gluing conditions along the wall are
\begin{align}
\binom{F_D^{(j)}}{F^{(j)}}\Big|_{W_{ij}} &= 
\left(
\begin{array}{cc}
a & b \\
c & d 
\end{array}
\right) 
\binom{F_D^{(i)}}{F^{(i)}} \Big|_{W_{ij}} \,.
\label{GluingGeneral}
\end{align}
This agrees with the transformations of $F^{(\pm)}$ under the $U(1)_{\CD}$ action \eqref{U1DTransfo}, as expected from our analysis.

The simplest walls are the $T^k$ walls, which correspond to the case $a=d=1$, $c=0$ and $b=k \in \Z$ and are associated with shifts of $\tau_1$ by $k$ (i.e.~ shifts of the theta angle by $2\pi k$). The identification of field strengths is
\be
F^{(j)} = F^{(i)} \,,\qquad F_D^{(j)} = F_D^{(i)} + k F^{(i)} \,,
\ee
resulting in the identification of the two gauge fields along the wall $A^{(i)}|_{W_{ij}} = A^{(j)}|_{W_{ij}} \equiv A|_{W_{ij}}$. Requiring the bulk equations of motion to hold at the location of the wall requires the presence of an additional Chern-Simons term along the wall
\begin{align}
S^{T^k}_{W_{ij}} = \frac{i \, k}{4\pi} \int_{W_{ij}} A \wedge F  \,.
\label{TkWall}
\end{align}

More generally, the 3d term canceling the difference between the variations of the two actions on either sides of the wall is easily obtained from (\ref{FFDAction}) \footnote{To see it, one can think of $F=dA$ and $F_D=d A_D$ as independent field strengths.},
\be
S^\gamma_{W_{ij}} = -{i\over 4\pi} \int_{W_{ij}} \left(A^{(i)} \wedge F^{(i)}_D - A^{(j)} \wedge F^{(j)}_D \right) \,.
\label{DualityWallTerm}
\ee
The relative minus sign is related to the fact that the wall $W_{ij}$ is the boundary of both $B_i$ and $B_j$ regions, but with opposite orientations.
Inserting the expression for $F_D$, for $c \neq 0$, this is precisely the expression for the wall obtained in \cite{Ganor:1996pe}
\be
S^\gamma_{W_{ij}} = {i\over 4\pi} \int_{W_{ij}}  \left( {d\over c} A^{(i)}\wedge F^{(i)}  + {a\over c}  A^{(j)} \wedge F^{(j)}  -{1\over c} A^{(j)} \wedge F^{(i)} - {1\over c} A^{(i)} \wedge F^{(j)}    \right) \,.
\label{GeneralWall}
\ee
Moreover, a straighforward calculation reveals that the presence of the 3d duality wall term \eqref{DualityWallTerm} is necessary for supersymmetry. Indeed the supersymmetry variation of \eqref{DualityWallTerm} precisely cancels the 3d term on $W_{ij}$ induced by the supersymmetry variation of the bulk action \eqref{LagrSusyTransfo}, with the identification $(A^{(i)},\tau^{(i)}) = (A, \tau)$, $(A^{(j)},\tau^{(j)}) = (A', \tau')$, and $\tau^{(j)} = \frac{a \tau^{(i)} + b}{c \tau^{(i)} + d}$.
In the presence of a $T^k$ wall ($c=0$), the supersymmetry invariance is also restored by the addition of the 3d term \eqref{TkWall}.

We can generalize to the non-abelian theory in the presence of a $T^k$ wall. To enforce the bulk equations of motion at the location of the wall and restore supersymmetry we must add the non-abelian Chern-Simons term
\be
S^{T^k}_{W_{ij}} = \frac{i \, k}{4\pi} \int_{W_{ij}} \Tr \Big[ A \wedge dA + \frac 23 A \wedge A \wedge A \Big]  \,.
\label{NATkWall}
\ee
For other duality walls in the non-abelian theory, it is not clear how to restore supersymmetry by adding 3d terms to the action. Indeed, in the supersymmetry variation of the action \eqref{LagrSusyTransfoNA}, the commutator terms involve the coupling $\tau_2$, which jumps across the wall, and the gauge field $A$, whose transformation across the wall is not understood. Only for the $T^k$-wall, when $\tau_2$ and $A$ are continuous across the wall, do these commutator terms cancel. It is not obvious that the other duality walls in the non-abelian theory can have a Lagrangian description, since they always involve configurations of large gauge coupling ($\tau_2  < 1$).


\subsubsection{Surface Defects}
\label{sssec:SurfaceDefects}

The 3d walls end of 2d surface defects, which are curves in $B$ (see figure \ref{fig:642Setup}). In the elliptic fibration, these will be the loci above which the fiber becomes singular, i.e. these curves $\CC$ are components of $\Delta \cap B$. As we move around these defects, the theory undergoes an $SL(2,\Z)$ monodromy. This means that the fields charged under $U(1)_\CD$ have twisted periodicity conditions around the defect $\CC_\gamma$ described by the transformation \eqref{U1DTransfo}.

The 3d duality wall term \eqref{DualityWallTerm} is neither gauge invariant nor supersymmetric, when the wall has a boundary. This implies that additional degrees of freedom have to be added to the theory along the surface defect $\CC = \p W$. 

The 2d surface theory on the boundary of a $T^k$ duality wall was studied  in \cite{Buchbinder:2007ar, Martucci:2014ema}.
In this case 
the gauge field $A$ is continuous across the wall and well-defined at the position of the surface defect. 
The defect theory is described by a WZW model at level $k$, which can be obtained by integrating chiral fermions living on the surface defect \cite{Buchbinder:2007ar}.
For $k>0$, this is equivalent to coupling the theory to $k$ chiral bosons living on the surface defect, namely bosonic fields $\beta_i$, $i = 1, \cdots, k$, with $\star_2 d\beta_i = i d\beta_i$, with action \cite{Witten:1996hc}\footnote{The action is given in terms of non-chiral bosons $\beta_i$, with only the chiral component $\p\beta_i$ coupled to the gauge field. For more details, we refer the reader to \cite{Witten:1996hc}. Also, for $k<0$ the overall sign of the surface action must be reversed.}
\begin{align}
S_{\CC} =  \sum_{i=1}^{k} -\frac{1}{8\pi} \int_{\CC} \star_2 (d\beta_i - A) \wedge (d\beta_i - A) - \frac{i}{4\pi} \int_{\CC} \beta_i \, F \,,
\end{align}
which can equivalently be written in terms of 
\be
S_{\CC} =  \sum_{i=1}^{k} -\frac{i}{4\pi} \int_{\CC} \p\beta_i \wedge \bar\p \beta_i - 2 \bar{\p}\beta_i \wedge A_{(1,0)} + A_{(1,0)} \wedge A_{(0,1)} \,.
\label{S2d}
\ee
The surface term $S_{\mathcal{C}}$ is not gauge invariant. Under a gauge transformation $A \to A + d\Lambda$, the chiral bosons transform as $\beta_i \to \beta_i+ \Lambda$ and $S_{\mathcal{C}}$ transforms as
\begin{align}
S_{\mathcal{C}} \to S_{\mathcal{C}} - \frac{i k}{4\pi} \int_{\CC} \Lambda F\,.
\end{align}
This precisely cancels the gauge transformation of the Chern-Simons term \eqref{TkWall} defined on the $T^k$ duality wall $W$ whose boundary is $\CC$. So the total action of the theory is gauge invariant after the addition of the surface action $S_{\mathcal{C}}$.

The 2d theory on a single defect long $\CC$ has two chiral, topological supercharges, corresponding to a type of half-twist of a (0,8) theory \footnote{The flat-space D3-D7 intersection preserves eight chiral supersymmetries in 2d. The duality defect is a twisted version of this, which preserves two chiral, topological supersymmetries.}. The supersymmetry variations of the chiral modes is trivial, $\delta \beta =0$, and they form Fermi multiplets.


\subsection{Duality Defects from M5-branes}
\label{ssec:DDfromM5}

We will now derive the 2d defect theory  from the M5-brane -- or 6d $(2,0)$ theory -- point of view, which will require us to be a bit more specific about the geometry of the fibration. We begin by discussing the geometry of the singular fibers in the M-theory setup 
and its relation to the F-theory setup, through M/F duality. We then identify the chiral modes living on the 2d duality defects as components of the tensor multiplet $B$ field reduced on the curves of the resolved fiber.

\subsubsection{M/F-duality and Singular Fibers}

Consider M-theory on $\bbR^{1,2}\times X_4$, where $X_4$ is an elliptic Calabi-Yau four-fold $\mathbb{E}_\tau \rightarrow X_4 \rightarrow M_3$, with base $M_3$, given in terms of a Weierstrass model (\ref{Weier}). The abelian twisted $(2,0)$ theory, that we defined and studied in this paper, can be described as the low-energy world-volume theory of one M5-brane wrapping a vertical divisor  inside $X_4$, i.e. 
the elliptic three-fold $Y$, which is the world-volume of the M5-brane, is embedded into $X_4$ by restriction of the elliptic fibration to a surface $B_2 \subset M_3$. 
The dual F-theoretic setup is that of a D3-brane (instanton) wrapped on $B$. 

\begin{figure}
  \centering
  \includegraphics[width=10cm]{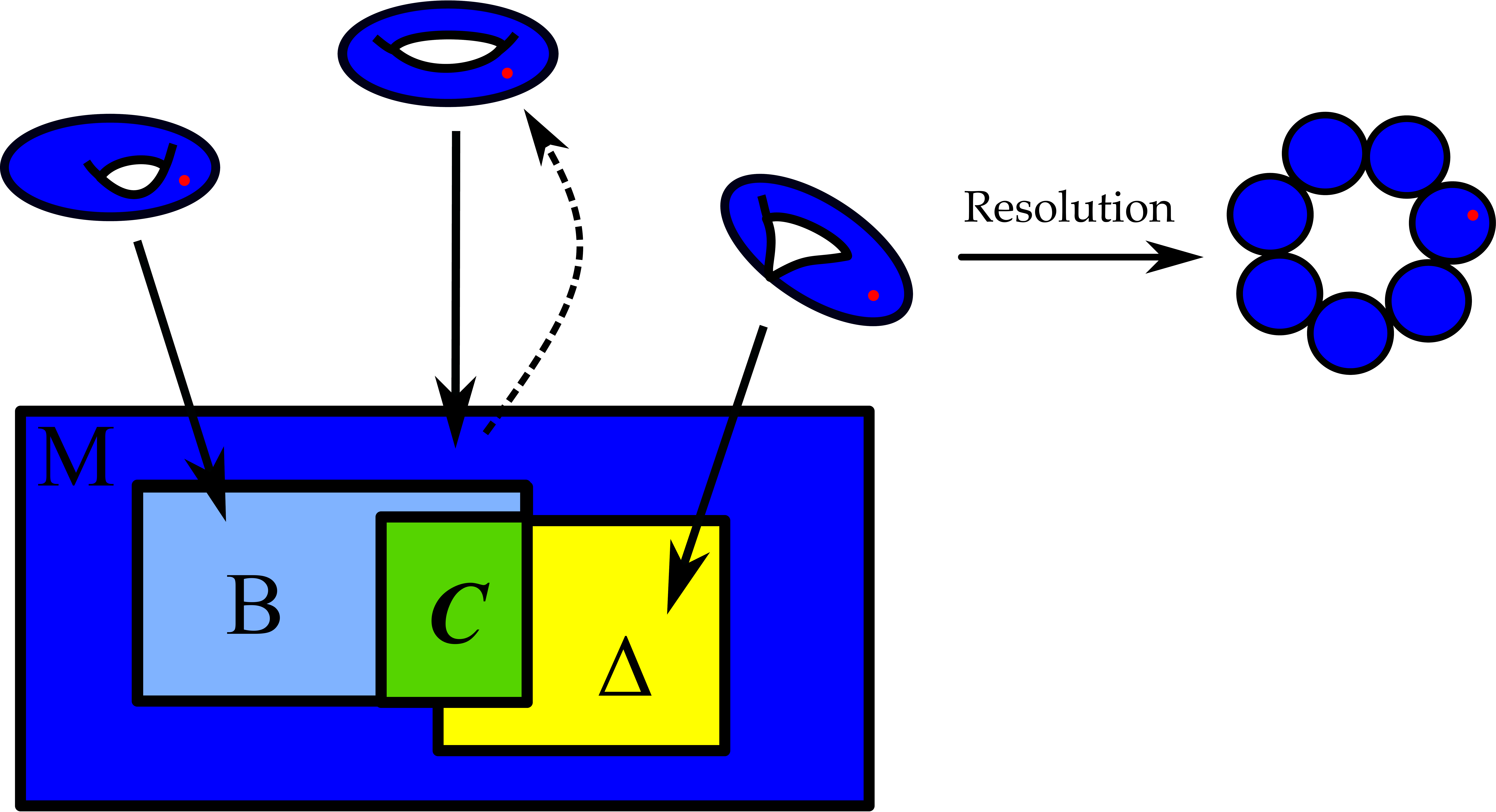} 
  \caption{Schematics of the setup: The total space is the elliptically fibered Calabi-Yau four-fold $X_4$, with base three-fold $M$. The restriction of the fibration to the subspace $B \subset M$ results in the elliptic three-fold $Y_3$, which will be wrapped by the M5-brane. 
 The fiber becomes singular above the locus $\Delta\subset M$, which intersects $B$ in (a collection of) curve(s) $\mathcal{C}$. 
\label{fig:BasicString}}
\end{figure}

The singularities in the fiber are above a complex codimension one locus in $M_3$, given by $\Delta=0$, and, intersected with $B$, give a collection of complex curves in the base. In the F-theory picture this is the intersection of the D3-instanton with the 7-branes that wrap the discriminant locus and the transverse $\mathbb{R}^{1,3}$, which is a collection of curves
\be
\cup_n\mathcal{C}_n =[\Delta] \cap B \subset [\Delta] \subset M_3 \,.
\ee
Along the curves ${\mathcal{C}_n}$ there are localized defect degrees of freedom, which in an F-theory realization correspond to open string modes between the D3-branes and 7-branes. The setup is shown in figure \ref{fig:BasicString}.

From the M5-brane point of view the 3-7 modes arise from studying the reduction of the self-dual three-form $H_3^+$ along harmonic $(1,1)$ forms. In an elliptic Calabi-Yau four-fold, the set of $(1,1)$ forms can either arise from the base $M_3$, the elliptic fiber, and rational sections (corresponding to rational points on the elliptic curve). For our considerations, of dimensional reduction of the M5-brane world-volume on the elliptic fiber to a 4d theory with defects, the fibral contributions are of particular interest. For a smooth elliptic fiber, there is one $(1,1)$ form, which was already discusssed in the bulk reduction to 4d. Above $\mathcal{C}_n \subset [\Delta]$, however, there are additional $(1,1)$ forms, which we will now describe. In the next section we use these to define the chiral defect theory localized on the curves $\mathcal{C}_n$.

Above the singular loci, the elliptic fibers degenerate, and the so-called singular fibers have been classified by Kodaira \cite{Kodaira}. The singularities are best understood by resolution (which retain the Calabi-Yau property of $X_4$, so-called crepant resolutions). Above $\Delta=0$ the resolved fibers are collections of rational curves $\mathbb{P}^1$ intersecting for instance in affine ADE type Dynkin diagrams, possibly with multiplicities. Each rational curve is associated to a simple root of an affine Lie algebra. 
Fibering these $\mathbb{P}^1$s over the dicriminant components, gives divisors in the Calabi-Yau four-fold, which are dual to $(1,1)$ forms. In this way each fibral curve arising in singular fibers has associated a (1,1) form, with a harmonic representative.

 The simplest example is the $I_k$ fiber, whose intersection graph is that of affine $SU(k)$. Let $F_i$, $i= 0, 1, \cdots, k-1$ be the rational curves in the fiber, which intersect as in figure \ref{fig:BasicString}. These curves obey the relation 
\be\label{FibClass}
\sum_{i=0}^{k-1} F_i = \mathcal{F} \,,
\ee
where $\mathcal{F}$ is the fiber class. Fibering each of these $F_i$ over the discriminant locus $\Delta$ gives a complex three-fold, the 
so-called Cartan divisors $D_i$. These intersect with the curves $F_i$ in the negative Cartan matrix $-C_{ij} =  \delta_{i-1,j}-2\delta_{i,j} + \delta_{i+1,j}$, and are Poincar\'e dual to $(1,1)$ forms $\omega_i^{(1,1)}$, $i=0, \cdots, k-1$, which therefore satisfy 
\be\label{Fomega}
\int_{F_i}  \omega_j = - C_{ij} \,.
\ee  
We can identify the $F_i$ with roots and $\omega_i$ with the co-roots of the gauge algebra.  Furthermore we introduce  
\be
\omega_{Z} = \frac{1}{V} e^4 \wedge e^5 \,, \qquad \int_{\mathcal{F}}  \omega_Z =  \int_{F_0}  \omega_Z = 1 \,,
\ee
which is dual to the zero-section  $Z$ (the copy of the base in the fiber, which passes through each of the origins of the elliptic curves). 
These forms are not all independent: the form $\omega_0$ can be expressed as a certain linear combination of $\omega_i$, $i=1, \cdots, k-1$, and $\omega_Z$.

There is an important difference between the reduction of the 11d supergravity modes, including $C_3$, and the reduction of the M5-brane world-volume theory. The expansion $C_3 = \sum_{i=0}^{k-1} A_i \wedge \omega^{(1,1)}_i$ has also a component of $C_3$ that is obtained by reducing along $\omega_0^{(1,1)}$ (which is dual to the zero-section) which in 3d (M-theory on the Calabi-Yau four-fold $X_4$) gives rise to a gauge field, that lifts in F-theory to a component of the metric in 4d -- for an in depth analysis of the F/M duality effective action see \cite{Grimm:2010ks}. We only obtain rank$(G)$ many Cartan gauge fields on the 7-brane, for instance $k-1$ for $I_k$. However, for the M5-brane theory what is important is the distinct number of cycles in the fiber, and independent $(1,1)$ forms, which is $k$ for $I_k$ fibers.

Finally, it will be important to discuss the self-duality of the $(1,1)$ forms. For this, consider the local description of the elliptic fibration, which above a component of the discriminant is given in terms of an ALE-fibration \cite{Hayashi:2008ba, Donagi:2009ra, Marsano:2009gv}, that can be directly constructed from a given elliptic fibration in Weierstrass form \cite{MS}: 
Locally, the Calabi-Yau above a component of the discriminant, $S$,  is given by a Higgs bundle spectral cover, which arises as the BPS equations of 8d SYM on $S$. This defines an ALE-fibration. The rational curves $F_i$ in this case correspond to a basis of $H_2(ALE)$ curves, and are  dual to harmonic representatives of $H^2 (ALE)$. For $A$-type singularities (that describe the local model for an $I_k$ fiber), the local ALE is a $k$-centered Taub-NUT geometry, which has $k$ harmonic $(1,1)$-forms, which are all anti-self-dual \footnote{Our conventions are opposite to the one in the cited references, namely we have anti-self-dual (1,1) forms and self-dual K\"ahler form.} \cite{Gibbons:1979xm, Ruback:1986ag, Sen:1997js, Witten:2009at, NeilHong}. Note, as emphasized in \cite{Witten:2009at, Witten:2009xu}, the $k$ (1,1)-forms are square-integrable, and the holonomy of the associated line bundles is trivial at the boundary of the Taub-NUT space.


\subsubsection{3-7 Modes from M5-branes}
\label{sssec:37Modes}

Our main interest is now the sector of 3-7 strings and resulting 2d defect modes. 
 In the case of a D3-D7-brane intersection, the  3-7 strings give rise at low energies to chiral fermions, which, upon dualization, are mapped to the chiral bosons discussed in section \ref{sssec:SurfaceDefects}. The supergravity background corresponding to this type IIB brane setup, with the defects supporting chiral fermions, has been studied in \cite{Buchbinder:2007ar}.
We will now provide a description of the chiral modes from the reduction of the abelian M5-brane  theory. 

In the last section we have seen that an $I_k$ singularity,  corresponding to $k$ 7-branes, gives rise to $k$  harmonic $(1,1)$ forms which are anti-self-dual in the local 4d ALE fiber defined via the spectral data above $\mathcal{C}$
\be
\star_4\Omega^i_{(1,1)} = -\Omega^i_{(1,1)} \,.
\ee
The local 6d geometry near a generic point in $\mathcal{C}$ is the direct product of a local patch in $\mathcal{C}$ times the ALE, which is   the $k$-centered Taub-NUT space. We identify locally the $k-1$ (1,1) forms $\omega_i$, $i=1, \cdots, k-1$, of the elliptic fibration with $k-1$ Taub-NUT (1,1) forms $\Omega_i$, therefore we have 
\be
\star_4\omega^i_{(1,1)} = -\omega^i_{(1,1)} \,, \qquad i=1, \cdots, k-1 \,.
\ee
The remaining independent (1,1) form $\omega_0$, or $\omega_Z$, being related to the fiber class, has no self-dual property.
We can then expand the (1,1) component of the two-form potential $B_{(1,1)}$ into, 
\be
B_{(1,1)}= B_{(1,1)}'+ b_Z \omega_Z + \mathcal{B}\,,\qquad \mathcal{B}=  \sum_{i=1}^{k-1} b_i \omega_{(1,1)}^i   \,,
\label{BDecomp}
\ee
where $B_{(1,1)}'$ denotes components purely along the base (and not along the fiber).
The new $(1,1)$ forms from the singular fiber give rise to components of $H$ that are $(1,2)$ or $(2,1)$ forms
\be
d\mathcal{B}= \sum_{i=1}^{k-1} \left(
\partial_z b_i dz \wedge \omega_{(1,1)}^i  +  \partial_{\bar{z}} b_i d\bar{z} \wedge \omega_{(1,1)}^i \right)\,.
\ee
From the self-duality of the $\omega^i$  we obtain the relations
\be
\star_6 (dz \wedge \omega^i) = - i dz \wedge \omega^i \,,\qquad 
\star_6 (d\bar z \wedge \omega^i) =  i d\bar z \wedge \omega^i \,.
\ee
The self-duality condition $\star_6 H= i  H$ then implies that 
\be
 \partial_z b_i =0 \,,
\ee
i.e. the $b_i$ are $k-1$ chiral bosons, each associated to a component of $B$ dual to a $\bbP^1$ in the singular fiber. 
The modes arising from $B$ by integrating over the various fiber components are 
\be
\ba
\int_{F_1} B = b_2 - 2 b_1 &\equiv \beta_1 \cr 
\int_{F_i \not= 0, 1, k-1} B = b_{i-1} - 2 b_i + b_{i+1} &\equiv \beta_i \cr 
\int_{F_{k-1}} B = b_{k-2} - 2 b_{k-1} &\equiv \beta_{k-1} \cr 
\int_{F_0} B= b_{1} + b_{k-1} + b_Z & \,.
\ea
\ee
The modes $\beta_i$ defined in this way, $\beta_i = \int_{F_i} B$, $i=1, \cdots, k-1$, are naturally chiral. This is however not true for the integral over $F_0$, as $b_Z$ is not chiral. To get a consistent picture with the 4d approach reviewed in section \ref{sssec:SurfaceDefects} and with the fact that the local Taub-NUT space has $k$ anti-self-dual (1,1) forms $\Omega^i_{(1,1)}$, we need to identify an additional chiral mode from the $b_Z\omega_Z$ component.  We can rewrite the decomposition \eqref{BDecomp}
\be
B_{(1,1)}= B_{(1,1)}''+ b_Z^+  J^+ - b_Z^- J^-  + \mathcal{B} \,,
\ee
with $ J_\pm = j \pm e^4\wedge e^5$. Note that $J_- = J$ is the K\"ahler form of the elliptic fibration, and $B_{(1,1)}''$ denotes again components along the base. We have the relation
\be 
b_Z^+ + b_Z^- = b_Z \,.
\ee
Requiring that the component $d b_Z^+  \wedge J^+ \subset dB$ is imaginary self-dual, independently of the other components of $B$, leads to $\p_z b_Z^+=0$\footnote{This uses the fact that $\p b_Z^+ \wedge J^+$ is anti imaginary self-dual, so we must set $\p b_Z^+=0$, whereas $\bar\p b_Z^+ \wedge J^+$ is imaginary self-dual.}, so that $b_Z^+$ is a chiral boson.
We then define the scalar $\beta_0$ by the relation
\be
\int_{F_0} B = b_{1} + b_{k-1} + b_Z^+ + b_Z^- \equiv \beta_0 + b_Z^- \,,
\ee
where $\beta_0$ is now chiral, satisfying $\partial_z \beta_0=0$. 
We thus obtain $k$ chiral modes $\beta_0, \cdots, \beta_{k-1}$ from $B$.  
The relation on the fiber classes (\ref{FibClass}) translates into 
\be
\int_{\mathcal{F}} B = \int_{\sum_{i=0}^{k-1} F_i} B  = \left(\sum_{i=0}^{k-1} \beta_i \right)+ b_Z^- \,.
\ee

The supersymmetry variations of all the $\beta_i$ is trivial, $\delta\beta_i =0$, since the supersymmetry transformation of $B_{(1,1)}$ is along $J^-$ \eqref{SusyTwisted6d}.

In the description in terms of chiral fermions coming from open strings stretched between the D3- and 7-branes, the chiral modes transform in bi-fundamentals of the D3 and 7-brane gauge groups, respectively. As we decouple the 7-brane gauge degrees of freedom, this symmetry group becomes an $SU(k)$ flavor group. To see this flavor symmetry, note that the 2d action for the chiral bosons $\beta_i$ along $\mathcal{C}$ \eqref{S2d} contains the coupling\footnote{The rest of the chiral bosons action is trivially invariant under the symmetries we describe.}
\be\label{betaF}
S_{\mathcal{C}} \supset \int_{\mathcal{C}} \left(\sum_{i=0}^{k-1} \beta_i \right) F\,,
\ee
where $F$ is the 4d bulk gauge field evaluated on the defect. All chiral modes couple in the same way to $F$. Their transformation under the abelian bulk gauge symmetry was described in section \ref{sssec:SurfaceDefects}: $\beta_i \to \beta_i + \Lambda$ for all $i$, with $\Lambda$ the gauge parameter. In addition this 2d action admits a $U(1)^{k-1}$ abelian flavor symmetry acting by
\be
\beta_i \to \beta_i + \Lambda_i \,, \quad \sum_{i=0}^{k-1} \Lambda_i =0\,,
\ee
where the $\Lambda_i$ are constants.
These must be understood as the Cartan part of the $SU(k)$ global symmetry. The action on the $\beta_i$ of the full $SU(k)$ global symmetry is non-linear and complicated to describe. In order to see it, one must use the description in terms of chiral fermions, which transform in the fundamental representation of $SU(k)$ \footnote{The fermionization maps the bosons and fermions as $\chi_i= :e^{i\beta_i}:$. The term $\sum_i \beta_i \sim \log(\prod_i\chi_i)$ is then an $SU(k)$ invariant. The anti-commutation of fermions is crucial to show that $\prod_i\chi_i$ is a symmetry invariant.}.

An alternative point of view, which allows understanding all these aspects is to describe the singularity in terms of a $k$-centered Taub-NUT geometry. In \cite{Witten:2009at} it was analyzed that this indeed gives $k$ self-dual harmonic $(1,1)$ forms, which were shown to agree with chiral modes. 

Finally, let us consider the non-abelian theory, living on a stack of $N$ M5-branes. In the tensor branch, when the M5s are separated, each one of the tensor multiplets gives rise to $k$ chiral bosons, localized on $\mathcal{C}$. As we move to the origin of the tensor branch, we expect this theory to become a non-abelian 2d chiral SCFT.

\subsection{Point-defects}

Two 2d surface defects intersect generically in a point in the base of the elliptic fibration. Such point-defects have had relatively little attention in the literature. Here we will see, that the geometry dictates precisely how the chiral modes on each of the 2d defects are related, such that the intersection is consistent with an enhanced global symmetry which is induced by the 7-branes.

\subsubsection{Fiber Splittings}

A complex codimension one singularity in the base is a component of $\Delta=0$ given locally by $z_1=0$. In addition there can be codimension two enhancements of the singularities. This is the locus where $z_1=z_2=0$, which in a three-fold base $M_3$ is a complex curve. This codimension two locus intersects $B_2\subset M_3$ in a point. Above this curve, the discriminant of the elliptic fiber vanishes to a higher order than along a generic locus in $z_1=0$
\be
\Delta = c \,  z_1^{n} z_2^{m} + O(z_1^{n+1}) \,.
\ee
 Again, resolving the fiber, gives insight into what type of singularities can happen. This problem has been studied extensively over the last few years \cite{EY, MS, Krause:2011xj, Grimm:2011fx, Hayashi:2013lra}, and the possible codimension two fibers were systematically characterized in \cite{Hayashi:2014kca}.

For clarity, we consider the case of a singularity associated to an $I_k$ fiber (corresponding to an $SU(k)$ gauge group on the 7-branes), which enhances to an $I_{k+m}$ fiber. Let us again denote by $F_i$, $i=0, \cdots, k-1$ the rational curves in the fiber that are one-to-one with the simple roots of $SU(k)$, and $F_0$, the affine node. This is the generic fiber above $z_1=0$. As we pass to $z_1= z_2=0$, the fiber becomes singular. A useful way to think about this is in terms of the intersection of the loci $z_1=0$ and $z_2=0$, above each of which there is a collection of fibral curves $F_i$ and $\tilde{F}_i$. At $z_1 =z_2=0$ these `merge' with relations between these two sets of rational curves.  

As the singularity $z_1= z_2=0$ is codimension two, it is resolved by so-called small resolutions. There is not a unique way to resolve the singularities, but the (relative) K\"ahler cone of the fiber develops a cone structure, with walls along which rational curves can be flopped -- thus passing from one resolution to another topologically inequivalent one. This is precisely the structure of the Coulomb branch of the 3d $N=2$ gauge theory with matter that one obtains by compactifying M-theory on this Calabi-Yau four-fold, and can be systematically described by box graphs \cite{Hayashi:2014kca}. These provide a succinct way of determining the basis of the relative K\"ahler cone of the fiber.  

We consider the simplest case of $m=1$ first and discuss an example for $m>1$ in section \ref{sec:Examp}. Locally, there is an enhancement of $SU(k)$ and $U(1)$ to $SU(k+1)$, which from the point of view of the $SU(k)$ symmetry, corresponds to adding the fundamental ${\bf k}$ representation. 
Above $z_1=0$ there is a collection of rational fibral curves $F_i^{(1)}$, $i=0, \cdots, k-1$ and above $z_2=0$ there is a single curve component ${F}^{(2)}_0$ (the $I_1$ fiber is in fact a nodal curve).
Along $z_1=z_2=0$ one of the curves $F_i$ becomes reducible and splits into two curves $C^{\pm}$
\be\label{SUnF}
F_i^{(1)} \rightarrow C^+ + C^- \,,\qquad F_j^{(1)} \rightarrow F_j^{(1)} \,, \ j\not=i \,,
\ee
where $C^\pm$ are associated to weights of the fundamental representation\footnote{If $L_i$ are the fundamental weights, then the simple roots are $\alpha_i = L_i - L_{i+1}$, and in this sense $F_i^{(1)}$ splits into two rational curves: $C^+$ associated to the weight $L_i$ and $C^-$ associated to the negative weight $- L_{i+1}$.}. 
The intersection graph of the set of rational fibral curves $F_j^{(1)}$ and $C^\pm$ turns out to be simply the affine Dynkin diagram of $SU(k+1)$.

\subsubsection{Point-interactions of 2d Defects}

We can now ask how this splitting of the fibral curves affects the dimensional reduction of the M5-brane on the three-fold $Y_3$. 
For concreteness let $z=0$ be the local equation of $B$, which is wrapped by the M5-brane, inside the base three-fold $M$. 
Reducing the self-dual three-form along $\omega_i$ we have seen that this gives rise to a chiral boson $\beta^{(\alpha)}_i$ along the curve $\mathcal{C}_\alpha = \{z_\alpha= z=0\}$ in $B$. Two such curves intersect in a point inside $B$
\be
P_{\alpha\beta}= \{z_\alpha =z_\beta =z=0 \} = \mathcal{C}_\alpha \cap \mathcal{C}_{\beta}  = B \cap \Delta_\alpha \cap \Delta_\beta\,,
\ee
where the surface $\Delta_{\alpha}$ is the component of the discriminant given by $z_\alpha=0$. 

In the above example of an $I_k$ singularity colliding with an $I_1$, where say $F_i^{(1)} \rightarrow C^+ + C^-$, the components of $B$ above $P$ decompose as follows
\be
\left( \int_{C^+} + \int_{C^-}\right)  B= \int_{F_i^{(1)}} B \qquad \rightarrow \qquad \beta_+ + \beta_- = \beta_i \,,
\ee
where the chiral fields $\beta_\pm$ are obtained from the $(1,1)$ forms associated\footnote{We can locally think of the transverse space above the codimension two singularity as an ALE space associated to $SU(k+1)$, a $k+1$ centered Taub-NUT.} to the curves $C^\pm$.

This results in a constraint at the points where the two 2d defects along $\mathcal{C}_\alpha$ and $\mathcal{C}_\beta$ intersect, which relates the chiral fields in the two theories
\be
\sum_{i} f_i\beta_i^{(\alpha)} =\sum_{j} e_j \beta_j^{(\beta)}\,,
\ee
with the $f_i, e_j$ dependent on the specific choice of K\"ahler cone, or resolution. 
In the above case we obtain $k+1$ chiral bosons, and the basis of effective curves above this point is 
\be
\{F_j^{(1)} \,,\  j\not=i \} \cup \{C_{i}^+ ,\  C_{i+1}^{-}\} \,,
\ee
which from the splitting (\ref{SUnF}) implies the constraints 
\be\ba
\mathcal{C}_{1}:\qquad &
F^{(1)}_i =  C_{i}^+ + C_{i+1}^{-} \cr 
\mathcal{C}_2:\qquad & 
F^{(2)}_0 = \sum_{j \not=i} F_j +  C_{i}^+ + C_{i+1}^{-} \,.
\ea\ee
The chiral boson $\beta^{(1)}_i$ and the chiral boson $\beta^{(2)}_0$ are thus not independent. In fact 
\be
\beta^{(1)}_i|_{P_{12}} = \beta_+ + \beta_-  \,,\qquad  \beta^{(2)}_0|_{P_{12}} =  \sum_{j \not=i} \beta_j^{(1)} +  \beta_+ + \beta_- \,.
\ee

Which of these curves $F_i^{(\alpha)}$ splits depends on the specific (small) resolution, and different resolutions are related by flop-transitions.  However, what remains invariant 
in the case of $I_k$ (and also the $D$-type flavor symmetries that are realized by $I_n^*$ fibers), 
is the fiber type, and thus the complete set of rational curves, above the intersection point. This determines the flavor symmetry, which gets enhanced at the intersection point.
 E.g. in the above case of $I_k$ colliding with $I_1$, the coupling (\ref{betaF}) along $\mathcal{C}_1$ where the $I_k$ fiber is located, restricts at the intersection point $P_{12}$ to
\be
\left. \left(F_{\mathcal{C}_1} \sum_{j=0}^{k-1} \beta^{(1)}_j \right) \right|_{P_{12}} = \left.  F_{\mathcal{C}_1} \bigg( \sum_{j\neq i} \beta^{(1)}_j  + \beta^+ + \beta^- \bigg) \right|_{P_{12}} \,,
\ee
where $F_{\mathcal{C}_1}$ denote the components of the bulk field strength along the defect $\mathcal{C}_1$.
Similarly, along the second defect $\mathcal{C}_2$ where the $I_1$ fiber is located, the coupling (\ref{betaF}) restricts at the intersection point $P_{12}$ to
\be
\left. \left(  F_{\mathcal{C}_2}\beta^{(2)}_0 \right) \right|_{P_{12}} = \left.  F_{\mathcal{C}_2} \bigg( \sum_{j\neq i} \beta^{(1)}_j  + \beta^+ + \beta^- \bigg) \right|_{P_{12}} \,.
\ee
We observe that at the point $P_{12}$, the combination $ \sum_{j\neq i} \beta^{(1)}_j  + \beta^+ + \beta^- $ appears in the action, signaling an $SU(k+1)$ flavor symmetry acting on the $\beta_{j\neq i}$ and $\beta_\pm$ at the point $P_{12}$, as explained in section \ref{sssec:37Modes}.

The point-like defects arising at the intersection of 2d duality defects are thus characterized by the presence of extra localized modes, with continuity relations to the defect fields dictated by the fiber type above these points, and associated to local flavor symmetry enhancements.


\subsubsection{An Example}
\label{sec:Examp}

\begin{figure}
  \centering
  \includegraphics[width=8cm]{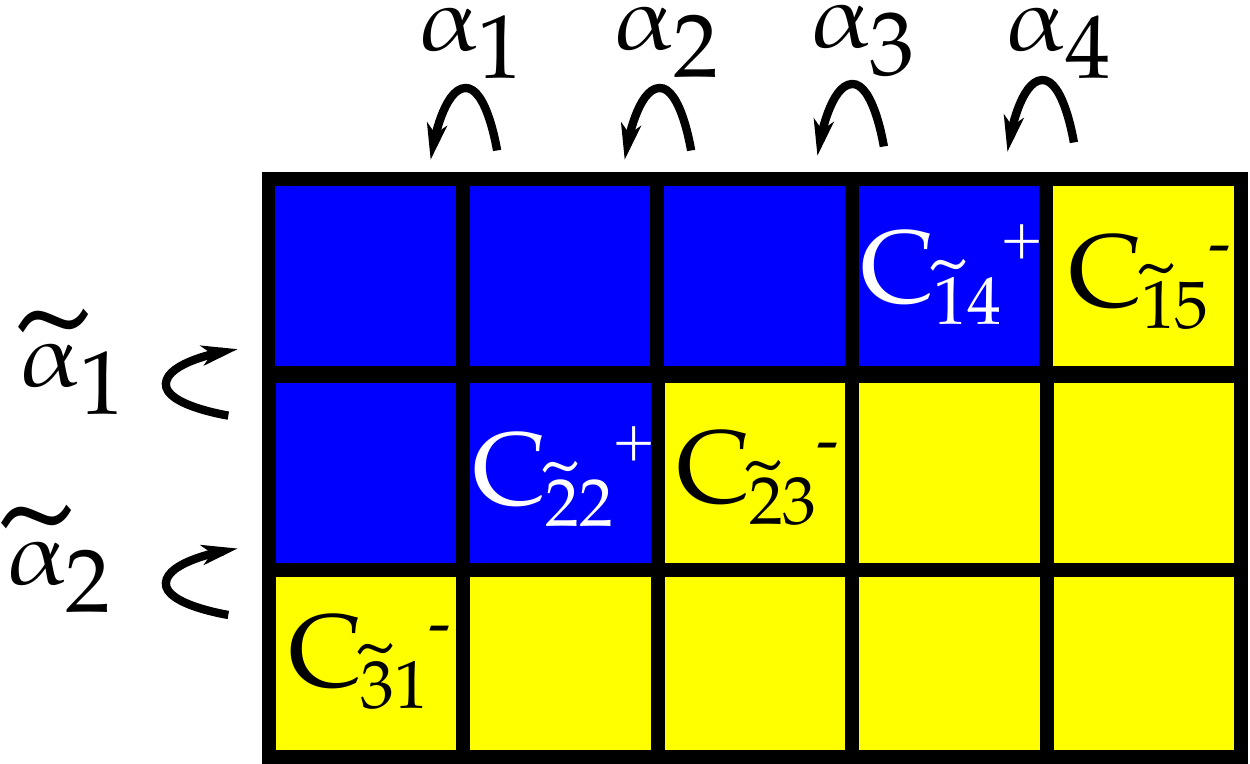} 
  \caption{Example box graph for the codimension two fiber of the collision $I_5$ with $I_3$. The simple roots are $\alpha$ and $\tilde{\alpha}$, and the extremal curves, i.e. generators of the relative Mori cone, are written out explicitly, with $C_{\tilde{j} i}^\pm = \pm(L_{i} + \tilde{L}_j)$. 
\label{fig:I5I3}}
\end{figure}

As another example consider the setup of D3s with the intersection of two 7-brane stacks with $SU(n)$ and $SU(m)$ gauge groups, respectively.  The local enhancement of the 7-brane gauge symmetry at the intersection locus is $SU(n+m)$. Let 
\be
F_i, \quad i=0, 1, \cdots, n-1 \,,\qquad 
\tilde{F}_j,\quad j=0, 1, \cdots, m-1 \,.
\ee 
be the fibral  $\mathbb{P}^1$s over $\mathcal{C}$ and $\tilde{\mathcal{C}}$, where each carries a chiral theory with $n$ and $m$ chiral modes, $\beta_i$ and $\tilde{\beta}_j$. Depending on the resolution of the singularity above $\mathcal{C} \cap \tilde{\mathcal{C}}$, which is characterized in terms of a Coulomb phase or equivalently box graph (see \cite{LawrieTBA} for the discussion of collisions of two non-abelian singularities). An example is shown in figure \ref{fig:I5I3} for $n=5$ and $m=3$, where the graph implies the following splitting
\be
\mathcal{C}\ :\quad \left\{ 
\ba
F_0 \ \rightarrow\  & F_0' + C_{\tilde{3}1}^- \cr 
F_1\ \rightarrow\  & F_1 \cr 
F_2 \ \rightarrow\ & C_{\tilde{2} 2}^+ + C_{\tilde{2} 3}^- \cr 
F_3 \ \rightarrow\  & F_3 \cr 
F_4 \ \rightarrow\  & C_{\tilde{1} 5}^- + C_{\tilde{1}4}^+ 
\ea\right.
\qquad 
\tilde{\mathcal{C}} \ :\quad \left\{ 
\ba
\tilde{F}_0 \ \rightarrow\ & \tilde{F}_0' + C_{\tilde{1}5}^- \cr 
\tilde{F}_1 \ \rightarrow\ & C_{\tilde{1}4}^+ + F_3 + C_{\tilde{2}3}^- \cr 
\tilde{F}_2 \ \rightarrow\ & C_{\tilde{2} 2}^+ + F_1 +C_{\tilde{3} 1}^- 
\ea\right.
\ee
Here, $C_{\tilde{j} i}^\pm$ corresponds to a curve, which carries the fundamental weight $\pm L_{i}$ and  $\pm \tilde L_{j}$ of $SU(5)$ and $SU(3)$ respectively. Also, the non-effective affine nodes split into $F_0 '= C_{\tilde{3}5}^+= \tilde{F}_0' $. 
From the point of view of the chiral 2d theories we have modes 
\be
\mathcal{C}: \qquad \beta_i \,,\qquad i=0, 1,2,3,4  \,,\qquad 
\tilde{\mathcal{C}}:\qquad \tilde\beta_{i}\,,\qquad i=0,1,2 \,.
\ee
Each of these couple to the bulk D3-gauge field as in (\ref{betaF}), and along each curve there is an enhanced $SU(5)$ and $SU(3)$ flavor symmetry, respectively. 
At the intersection point $P= \mathcal{C} \cap \tilde{\mathcal{C}}$, the splitting of the curves dictates that the coupling \eqref{betaF} on $\mathcal{C}$ evaluates to
\be
\left. \left(F_{\mathcal{C}} \sum_{i=0}^4 \beta_i \right) \right|_{P} = \left. F_{\mathcal{C}} \bigg( \beta^+_{\tilde 35} + \beta^-_{\tilde 31} + \beta_1 + \beta^+_{\tilde 22} + \beta^-_{\tilde 23} + \beta_3 + \beta^-_{\tilde 15} + \beta^+_{\tilde 14} \bigg)  \right|_{P} \,,
\ee
where $F_\mathcal{C}$ is the bulk 4d gauge field along $\mathcal{C}$ and the coupling \eqref{betaF} on $\tilde{\mathcal{C}}$ evaluates to
\be
\left. \left(F_{\tilde{\mathcal{C}}} \sum_{i=0}^2 \tilde\beta_i \right) \right|_{P} = \left. F_{\tilde{\mathcal{C}}} \bigg( \beta^+_{\tilde 35} + \beta^-_{\tilde 15} + \beta^+_{\tilde 14} + \beta_3 + \beta^-_{\tilde 23}  + \beta^+_{\tilde 22} + \beta_1 + \beta^-_{\tilde 31}   \bigg)  \right|_{P} \,,
\ee
where the $\beta^{\pm}_{\tilde{j}i}$ are the scalars obtained by reducing the $B$ field along the (1,1) forms dual to the cycles $C_{\tilde{j} i}^\pm$ at the ponit intersection.
We obtain that the flavor symmetry is enhanced at the point $P$ to $SU(8)$ acting on the eight scalars $\beta^+_{\tilde 35}, \beta^-_{\tilde 31}, \beta_1, \beta^+_{\tilde 22}, \beta^-_{\tilde 23}, \beta_3, \beta^-_{\tilde 15}, \beta^+_{\tilde 14}$.

We have focused our attention on defects with $SU(k)$ flavor symmetries.  It would of course be interesting to extend this to defects supporting general ADE type flavor symmetries, making use of the known results on elliptic fibrations.

\section{Conclusions and Outlook}

The relation between 6d  $(2,0)$ theory or M5-branes on an elliptic curve and 4d $\mathcal{N}=4$ SYM for fixed coupling constant $\tau$, parametrized by the complex structure of the elliptic curve, makes the $SL(2,\mathbb{Z})$ duality of the 4d theory manifest in the modular action on $\tau$. 
In this paper we studied what happens as we allow the elliptic curve to be fibered non-trivially over a 4d base manifold -- much like the generalization of IIB to F-theory by allowing the axio-dilaton to vary. A concrete setup where this M5-brane configuration can be studied is indeed in the context of M/F-duality, for an elliptic Calabi-Yau four-fold $X_4$, which contains an elliptic three-fold, that is wrapped by the M5-brane. 

The interesting physics appears along singularities in the fiber, when the complex structure $\tau$ diverges, and we defined a suitable topological twist of the 6d theory, which allows for supersymmetry in 4d, including along 3d walls, 2d duality surface defects and 0d point defects. 
The resulting 4d theory for an abelian gauge group agrees with the duality twisted 4d SYM theory of \cite{Martucci:2014ema}, including the walls and surface defects determined therein. In the latter case, this theory was obtained from a D3-D7-brane setup, where the 3-7 string sector gives rise to the surface modes. 
The advantage of the M5-brane approach is that it allows to formulate the theory more generally for non-abelian gauge groups, where the duality $U(1)_\CD$ symmetry is absent. 

The 2d duality defects are labeled by elements of $SL(2,\mathbb{Z})$, which determine the monodromy that $\tau$ undergoes upon encircling the defect. The consistency of the theory is ensured by 3d walls, along which the fields of the 4d theory are glued according to the duality transformations, which are known only for the abelian theory. We determined the spectrum of the 2d theory starting from the M5-brane, by dimensional reduction on singular fibers, which are a source of additional harmonic $(1,1)$ forms. The self-duality of the 3-form on the M5-brane world-volume implies that the localized modes are chiral. The 2d theory preserves two supercharges, couples to the 4d gauge field, and enjoys a flavor symmetry. Furthermore the surface defects generically will intersect at points, which result in local enhancements of the flavor symmetry. 
We determined these properties of the 2d and 0d defects from the singular fibers, in codimension one and two in the base of the elliptic Calabi-Yau four-fold $X_4$ in which the M5-brane world-volume $Y_3$ is embedded. 

There are various directions in which to extend and generalize our analysis: 

\begin{enumerate}
\item Defects in non-abelian theories: \\
We provided the non-abelian generalization of the 4d bulk theory  in section \ref{sec:NonAb}, however we studied the walls/defects in 3d, 2d and 0d only in the abelian theory, where the $SL(2,\bbZ)$ action on the fields in known. It would be extremely interesting to understand the 3d walls, 2d duality defects, and point defects in the non-abelian theory. This could provide some more insights into the electro-magnetic duality of the $\CN=4$ SYM theory.

\item 3-7 strings: \\ 
The present focus has been on the M5-brane theory point of view. The modes arising at the singular fibers were determined using insights into the geometry of the singular fibers. In a IIB/F-theory dual point of view, these insights should be derivable from studying $[p,q]$ strings between the D3 and 7-branes.

\item Lefschetz-fibrations: \\
Another interesting avenue is to consider instead of elliptic curves, a fibration by a higher genus curve. Such Lefschetz-fibrations were 
mentioned in \cite{Gadde:2014wma} and it would be interesting to perform a reduction similar to the one in the present paper\footnote{The compactification on a surface of lower genus, namely on a two-sphere, does not lead to a 4d SCFT, but rather to a 4d sigma model, without complex coupling \cite{Assel:2016lad}.}.
When  the space is a direct product (no fibration) the reduction leads to 4d $\CN=2$ SCFTs \cite{Gaiotto:2009we} sometimes called {\it class $\mathcal{S}$} theories, with the marginal couplings identified with the complex structure moduli of the higher genus curve. The reduction on a non-trivially fibered curve should lead to 4d $\CN=2$ theories with varying couplings and surface defects, around which the theory undergoes duality transformations. However the lack of Lagrangian description for the non-abelian 6d theory, as well as for most of the class $\mathcal{S}$ theories poses a real challenge. 

\end{enumerate}


\subsection*{Acknowledgements}

We thank Andreas Braun, Stefano Cremonesi, Tudor Dimofte, Neil Lambert, Craig Lawrie, Dave Morrison, Cristian Vergu, Timo Weigand, Jenny Wong for discussions.  B.A. acknowledges support by the ERC Starting Grant N. 304806, ``The Gauge/Gravity Duality and Geometry in String Theory". 
 SSN acknowledges support by the ERC Consolidator Grant 682608 ``Higgs bundles: Supersymmetric Gauge Theories and Geometry (HIGGSBNDL)". 
 The work of SSN was also supported in part by STFC grant ST/J002798/1. SSN thanks the Aspen Center for Physics for hospitality during the course of this work. The Aspen Center for Physics is supported by National Science Foundation grant PHY-1066293.




\appendix

\section{Spinor conventions}
\label{app:conventions}

We need to consider spinors in six dimensions and their reduction to four dimensions. We work in Euclidean signature, however we will present the Lorentzian case as well. We follow the conventions of \cite{VanProeyen:1999ni}. 

We denote $\Gamma^A$, $A=0,\cdots,5$, and $\gamma^a$, $a=0,\cdots, 3$, the gamma matrices in six and four dimensions respectively. We use the specific basis
\begin{align*}
\Gamma^0 &= i^{t} \, \sigma_1 \otimes {\bf 1}_2 \otimes {\bf 1}_2 \ \equiv \ \gamma^0 \otimes {\bf 1}_2 \,, \\
\Gamma^1 &= \sigma_2 \otimes {\bf 1}_2 \otimes {\bf 1}_2 \ \equiv \ \gamma^1 \otimes {\bf 1}_2 \,, \\
\Gamma^2 &= \sigma_3 \otimes \sigma_1 \otimes {\bf 1}_2 \ \equiv \ \gamma^2 \otimes {\bf 1}_2 \,, \\
\Gamma^3 &= \sigma_3 \otimes \sigma_2 \otimes {\bf 1}_2 \ \equiv \ \gamma^3 \otimes {\bf 1}_2 \,, \\
\Gamma^4 &= \sigma_3 \otimes \sigma_3 \otimes \sigma_1  \,, \\
\Gamma^5 &= \sigma_3 \otimes \sigma_3 \otimes \sigma_2 \,,
\end{align*}
with $t=0$ in Euclidean signature and $t=1$ in Lorentzian signature. $\sigma_{1,2,3}$ are the usual Pauli matrices
\begin{align*}
\sigma_1 =\left(
\begin{array}{cc} 0 & 1 \\ 
1 & 0\end{array}\right) \,, \quad 
\sigma_2=\left(
\begin{array}{cc} 0 & -i \\
 i & 0\end{array}\right) \, , \quad \sigma_3 =
 \left(\begin{array}{cc} 1 & 0 \\ 
 0 & -1\end{array}\right) \, .
\end{align*}
The chirality matrices in six and four dimensions are
\begin{align}
\Gamma^{\ast} &= (-i)^{3+t} \, \Gamma^0 \cdots \Gamma^5  = \sigma_3 \otimes \sigma_3 \otimes \sigma_3 \, , \quad \gamma^{\ast} = (-i)^{2+t} \, \gamma^0 \cdots \gamma^3 = \sigma_3 \otimes \sigma_3 \, .
\end{align}
Weyl spinors $\lambda^{({\rm 6d})}_{\pm}$, $ \lambda^{({\rm 4d})}_{\pm}$ of positive/negative chirality obey
\begin{align}
\Gamma^{\ast} \lambda^{({\rm 6d})}_{\pm} &= \pm \lambda^{({\rm 6d})}_{\pm} \, , \quad 
\gamma^{\ast} \lambda^{({\rm 4d})}_{\pm} = \pm \lambda^{({\rm 4d})}_{\pm} \, .
\end{align}
We provide some useful identities:
\begin{align}
\Gamma^{A_1 \cdots A_p } =\frac{ i^{3+t}}{(6-p)!} \, \epsilon^{A_1 \cdots A_p}{}_{A_{p+1} \cdots A_6}\Gamma^\ast \Gamma^{A_6 \cdots A_{p+1}}\, , \nn\\
\gamma^{a_1 \cdots a_p } =\frac{ i^{2+t}}{(4-p)!} \, \epsilon^{a_1 \cdots a_p}{}_{a_{p+1} \cdots a_4}\gamma^\ast \gamma^{a_4 \cdots a_{p+1}}\, .
\end{align}
We define
\begin{align}
\CC_{({\rm 6d})} &= \sigma_2 \otimes \sigma_1 \otimes \sigma_2 \ \equiv \ \CC_{({\rm 4d})} \otimes \sigma_2 \,,
\end{align}
satisfying $(\Gamma^A)^{T} = - \CC_{({\rm 6d})} \Gamma^A \CC_{({\rm 6d})}^{-1}$ and $(\gamma^a)^{T} = - \CC_{({\rm 4d})} \gamma^a \CC_{({\rm 4d})}^{-1}$, as well as $\CC_{({\rm 6d})}^{-1} = \CC_{({\rm 6d})}^T = \CC_{({\rm 6d})}$ and $\CC_{({\rm 4d})}^{-1} = - \CC_{({\rm 4d})}^T = \CC_{({\rm 4d})}$.

When we need to use spinor bilinears, we use matrix notations, for instance $\chi_1 ^T \CC \Gamma^A \chi_2$ and we define 
\begin{align}
\bar{\chi}  &\equiv  \chi^T \, \CC \,, \quad  \chi^{\dagger} \equiv  (\chi^\ast)^T   \, .
\end{align}

Spinors $\lambda^m$ in six dimensions transform in the fundamental representation ${\bf 4}$ of $Sp(4)_R$, with $m=1, \cdots,4$.
We denote $\Omega_{mn}$ the invariant anti-symmetric tensor of $Sp(4)_R$. It satisfies the relations
\begin{align}
\Omega_{mn} = -\Omega_{nm} \, , \quad \Omega^{mn} \Omega_{nr} = - \delta^m_r\,.
\end{align}
with $\Omega^{mn} \equiv (\Omega_{mn})^{\ast}$.
Indices are raised and lowered as
\begin{align}
\lambda_m &= \lambda^n \Omega_{nm} \, \quad \lambda^m = \Omega^{mn} \lambda_n \,.
\end{align}
In four dimension the same $\Omega_{mn} (= \Omega_{ \dot m n})$ plays the role of the $SU(4)_R$ invariant tensor, raising and lowering indices, with spinors $\lambda^m$ transforming in the ${\bf 4}$ and spinors $\lambda^{\dot m}$ in the $\bar{\bf 4}$ of $SU(4)_R$.

\subsection{Irreducible spinors}
\label{app:IrredSpinors}

To enforce reality conditions on spinors one also need to define the matrices
\begin{align}
\textrm{Lorentzian:} \quad  \CB_{({\rm 6d})} &= \sigma_3 \otimes \sigma_1 \otimes \sigma_2 \ \equiv \ - \, \CB_{({\rm 4d})} \otimes \sigma_2 \,, \nn\\
\textrm{Euclidean:} \quad  \CB_{({\rm 6d})} &= \sigma_2 \otimes \sigma_1 \otimes \sigma_2 \ \equiv \ - \, \CB_{({\rm 4d})} \otimes \sigma_2 \,,
\end{align}
which satisfy $(\Gamma^A)^{\ast} = (-1)^{t+1} \CB_{({\rm 6d})} \Gamma^A \CB_{({\rm 6d})}^{-1}$ and $(\gamma^a)^{\ast} = (-1)^{t+1} \CB_{({\rm 4d})} \gamma^a \CB_{({\rm 4d})}^{-1}$.

\smallskip

In six dimensions in Euclidean signature Weyl spinors are irreducible and there is no Lorentz invariant reality condition on them.
In Lorentzian signature one can impose a Symplectic-Majorana condition on Weyl spinors
\begin{align}
\lp \lambda^{({\rm 6d})}{}^m \rp^{\ast} &= \Omega_{mn} \CB_{({\rm 6d})}  \lambda^{({\rm 6d})}{}^n \,.
\label{SM6d}
\end{align}
In four dimensions in Euclidean signature one can impose a Symplectic-Majorana condition on Weyl spinors
\begin{align}
\lp \lambda^{({\rm 4d})}{}^m \rp^{\ast} &= \Omega_{mn} \CB_{({\rm 4d})}  \lambda^{({\rm 4d})}{}^n \,.
\label{SM4d}
\end{align}
In Lorentzian signature Weyl spinors are irreducible. Instead one can impose a Majorana condition on Dirac spinors
\begin{align}
\lp \lambda^{({\rm 4d})} \rp^{\ast} &= \CB_{({\rm 4d})}  \lambda^{({\rm 4d})} \,.
\label{Maj4d}
\end{align}

The matrix $\CB$ is also used to define the charge conjugate of a spinor $\chi$:
\begin{align}
\chi^c &= \CB^{-1} \chi^\ast \, ,
\end{align}
which has the opposite chirality compared to $\chi$.

\subsection{Reduction from 6d to 4d}
\label{app:FermionRed}

We consider a spinor $\lambda^{({\rm 6d})}{}^m$ of positive chirality transforming in the $({\bf 4},{\bf 4})$ of $SU(4)_L \times Sp(4)_R$, we decompose it as a tensor product
\begin{align}\label{6d4dSpinors}
\lambda^{({\rm 6d})}{}^m &= \lambda_+^{({\rm 4d})}{}^m \otimes \eta_+  + \lambda_-^{({\rm 4d})}{}^m \otimes \eta_- \, ,
\end{align}
where $\eta_+ = \binom{1}{0}$ and $\eta_- = \binom{0}{1}$. The positive chirality of $\lambda^{({\rm 6d})}{}^m$ implies the chirality on the four-dimensional spinors
\begin{align}
\gamma^{\ast} \lambda^{({\rm 4d})}_{\pm}{}^m &= \pm \lambda^{({\rm 4d})}_{\pm}{}^m \, .
\end{align}
The spinors $\lambda_+^{({\rm 4d})}{}_m = \lambda_+^{({\rm 4d})}{}^n \Omega_{nm}  $ and $\lambda_-^{({\rm 4d})}{}^m$ transform in the  
$({\bf 2}, {\bf 1}, \bar{\bf 4}) \oplus ({\bf 1}, {\bf 2}, {\bf 4}) $ of $SU(2)_1 \times SU(2)_2 \times SU(4)_r$, where $SU(2)_1 \times SU(2)_2$ is the 4d Lorentz group.
In Euclidean signature, one can impose a Symplectic Majorana condition on the 4d spinors, but this does not come from a (Lorentz invariant) reality condition in 6d.

In Lorentzian signature $\lambda^{({\rm 6d})}{}^m$ obeys the Symplectic-Majorana condition \eqref{SM6d}.This translates into the relation
\begin{align}
\lp \lambda_-^{({\rm 4d})}{}^m \rp^{\ast} &= - \CB_{({\rm 4d})}  \lambda_+^{({\rm 4d})}{}_m  \,,
\end{align}
which means that the spinors $ \lambda_+^{({\rm 4d})}{}_m + \lambda_-^{({\rm 4d})}{}^m $ are Majorana.

\subsection{R-symmetry reduction}
\label{app:Rsym}

In six-dimension the R-symmetry is $Sp(4)_R \simeq SO(5)_R$.
We want to consider the $SO(5)_R \rightarrow SO(3)_R \times SO(2)_R \simeq SU(2)_R \times U(1)_R$ symmetry breaking pattern. This leads us to define the $\Gamma^{\wat i}$ matrices, with $\wat i = 1, \cdots , 5$, in the tensor product form :
\be
\ba
\Gamma^{\wat 1} = \sigma^1 \otimes \sigma^3  \quad , \ 
\Gamma^{\wat 2} = \sigma^2 \otimes \sigma^3  \quad , \ 
\Gamma^{\wat 3} = \sigma^3 \otimes \sigma^3  \quad , \ 
\Gamma^{\wat 4} = {\bf 1} \otimes \sigma^2  \quad , \ 
\Gamma^{\wat 5} = {\bf 1} \otimes \sigma^1  \quad , \ 
\ea
\ee
The index $m$ of the ${\bf 4}$ of $Sp(4)_R$ decomposes into the indices $\alpha, \pm 1$ of the ${\bf 2}_{1} \oplus {\bf 2}_{-1}$ of $SU(2)_R \times U(1)_R$.
\begin{align}
\chi^{m} &\rightarrow  \chi_{1}^{\alpha} \,, \ \chi_{-1}^{\alpha} \, .
\end{align}
In matrix-like notations we have 
\begin{align}
\chi &= \chi_1 \otimes \binom{1}{0} +  \chi_{-1} \otimes \binom{0}{1} \,,
\end{align}
where $\chi_{ 1}, \chi_{-1}$ have components $\chi_{1}^{\alpha}, \chi_{-1}^{\alpha} $ respectively. With $i\Gamma^{45}$ generating $U(1)_R$, one can check that $\chi_{1}$ has R-charge $1$ and $\chi_{-1}$ has R-charge $-1$.

Also,  the indices $m, n \leftrightarrow (\alpha, x), ( \beta, y)$ are raised and lowered according to the NW-SE convention with $\Omega^{m n} = \Omega_{m  n} = - \Omega_{n m} = \epsilon_{\alpha \beta} \otimes B_{x y}$, with $B_{x y} =B^{x y} = \sigma^1 = \left(
\begin{array}{cc}
0 & 1 \\ 1 & 0
\end{array}\right)$.
So we have $(\Omega_{m n}) = i \sigma_2 \otimes \sigma_1$.

We also note the relation
\begin{align}
\sum_{k=1,2,3} (\sigma^k)^{\alpha \beta} (\sigma^k)^{\alpha' \beta'} &= - (\epsilon^{\alpha\alpha'} \epsilon^{\beta \beta'} + \epsilon^{\alpha\beta'} \epsilon^{\beta \alpha'}) \,.
\end{align}


\section{Twists of the 6d $(2,0)$ Theory}
\label{sec:6dTwists}

The 6d $(2,0)$ theory  has Lorentz and R-symmetry group
\begin{equation}
  SU(4)_L \times Sp(4)_R \,,
\end{equation}
and supercharges, $Q$, that transform in the $({\bf 4}, {\bf 4})$
representation of this group. If the six-manifold, $M_6$, on which the theory
lives has some reduced holonomy one can ask whether there are supercharges
which are scalars under either the Lorentz group or some twisted Lorentz
group. 

\subsection{K\"ahler Manifolds}

First we consider $M_6$ a K\"ahler manifold.  The holonomy group is reduced $U(3)_L \simeq SU(3)_L \times U(1)_L $
and the supercharges transform under the subgroup $SU(3)_L \times U(1)_L \times Sp(4)_R$ as
\be
({\bf 4}, {\bf 4}) \quad \rightarrow \quad   ({\bf 3}, {\bf 4})_{1} \oplus ({\bf 1}, {\bf 4})_{-3}  \,.
\ee
There are no supercharges
which are scalars under the $SU(3)_L \times U(1)_L$ Lorentz symmetry, however
there are two twists which provide such scalar supercharges.

{\bf Twist 1:} In the first twist consider the decomposition of the R-symmetry
such that
\be
\ba  Sp(4)_R &\quad \rightarrow \quad SU(2)_R \times U(1)_R \cr
  {\bf 4} &\quad \rightarrow \quad  {\bf 2}_1 \oplus {\bf 2}_{-1} \,.
\ea
\ee
The supercharges after this decomposition transform 
as 
\begin{equation}
\ba
SU(4)_L\times  Sp(4)_R &\quad \rightarrow \quad    SU(3)_L \times SU(2)_R \times U(1)_L \times U(1)_R \cr 
({\bf 4}, {\bf 4}) &\quad \rightarrow \quad  ({\bf 3}, {\bf 2})_{1,1} \oplus ({\bf 3}, {\bf 2})_{1,-1} \oplus
  ({\bf 1}, {\bf 2})_{-3,1} \oplus ({\bf 1}, {\bf 2})_{-3,-1}\,,
 \ea
\end{equation}
 where the first entry is the transformation under $SU(3)_L$,
the second under $SU(2)_R$, and the two subscripts the charges under $U(1)_L$
and $U(1)_R$, respectively. One can now choose an alternative Lorentz group
$SU(3)_L \times U(1)_L^\prime$ by embedding $U(1)_L^\prime$ such that it has
generator
\begin{equation}
 T_L^\prime = T_L - 3T_R \,,
\end{equation}
where $T_L$ and $T_R$ are the generators of the Lorentz and R-symmetry
$U(1)$s. The twisted symmetry group decomposition of the supercharges is then
\be
\ba
SU(4)_L\times  Sp(4)_R &\quad \rightarrow \quad  
  SU(3)_L \times SU(2)_R \times U(1)^\prime_L \cr
 ({\bf 4}, {\bf 4})&\quad \rightarrow \quad    ({\bf 3}, {\bf 2})_{-2} \oplus ({\bf 3}, {\bf 2})_{4} \oplus 
  ({\bf 1}, {\bf 2})_{-6} \oplus ({\bf 1}, {\bf 2})_0 \,.
\ea
\ee
The last term corresponds to supercharges that are trivial under the twisted
Lorentz group and transform as a doublet under the residual $SU(2)_R$
R-symmetry group. Therefore there are two scalar supercharges in the twisted theory.

{\bf Twist 2:} There is another decomposition of the R-symmetry group where
the ${\bf 4}$ decomposes differently from above:
\begin{align}
    Sp(4)_R &\rightarrow SU(2)_R \times U(1)_R \cr
    {\bf 4} &\rightarrow {\bf 2}_0 \oplus {\bf 1}_{1} \oplus {\bf 1}_{-1} \,.
\end{align}
In this case the supercharges transform under $SU(3)_L \times SU(2)_R \times U(1)_L \times U(1)_R$ as
\begin{equation}
  ({\bf 3}, {\bf 2})_{1,0} \oplus ({\bf 3}, {\bf 1})_{1,1} \oplus
  ({\bf 3}, {\bf 1})_{1,-1} \oplus ({\bf 1}, {\bf 2})_{-3,0} \oplus
  ({\bf 1}, {\bf 1})_{-3,1} \oplus ({\bf 1}, {\bf 1})_{-3,-1} \,.
\end{equation}
Twisting, similarly to the above, to generate a new Lorentz $U(1)^\prime$, by 
\begin{equation}
  T^\prime = T_L - 3T_R \,,
\end{equation}
gives the representation of the supercharges under the new Lorentz group as
\begin{equation}
  ({\bf 3}, {\bf 2})_{1} \oplus ({\bf 3}, {\bf 1})_{-2} \oplus
  ({\bf 3}, {\bf 1})_{4} \oplus ({\bf 1}, {\bf 2})_{-3} \oplus
  ({\bf 1}, {\bf 1})_{-6} \oplus ({\bf 1}, {\bf 1})_{0} \,.
\end{equation}
The last term is the single and only scalar supercharge of the twisted theory.

{\bf Twist 3:} There is a final distinct decomposition of $Sp(4)$ containing 
$U(1)$, which comes from the special embedding of $SU(2)$ inside $Sp(4)$.
The decomposition is
\begin{align}
  Sp(4)_R &\rightarrow U(1)_R \cr
  {\bf 4} &\rightarrow {\bf 1}_3 \oplus {\bf 1}_1 \oplus {\bf 1}_{-1} \oplus
  {\bf 1}_{-3} \,.
\end{align}
The supercharges, which transformed under $({\bf 3}, {\bf 4})_{1} \oplus ({\bf
1}, {\bf 4})_{-3}$, decompose under the subgroup as 
\begin{equation}
  {\bf 3}_{1,3} \oplus {\bf 3}_{1,1} \oplus {\bf 3}_{1,-1} \oplus {\bf
  3}_{1,-3} \oplus {\bf 1}_{-3,3} \oplus {\bf 1}_{-3,1} \oplus {\bf 1}_{-3,-1}
  \oplus {\bf 1}_{-3,-3} \,.
\end{equation}
It is clear that after this decomposition there are two different twistings,
up to symmetry, that will create scalar supercharges. These are
\begin{enumerate}
  \item[(i)] $T^\prime = T_L - T_R$
  \item[(ii)] $T^\prime = T_L - 3T_R$ .
\end{enumerate}
After the twist $(i)$ the supercharges will transform as
\begin{equation}
  {\bf 3}_{-2} \oplus {\bf 3}_{0} \oplus {\bf 3}_{2} \oplus {\bf
  3}_{4} \oplus {\bf 1}_{-6} \oplus {\bf 1}_{-4} \oplus {\bf 1}_{-2}
  \oplus {\bf 1}_{0} \,,
\end{equation}
and after the twist $(ii)$ they will transform as
\begin{equation}
  {\bf 3}_{-8} \oplus {\bf 3}_{-2} \oplus {\bf 3}_{4} \oplus {\bf
  3}_{10} \oplus {\bf 1}_{-12} \oplus {\bf 1}_{-6} \oplus {\bf 1}_{0}
  \oplus {\bf 1}_{6} \,.
\end{equation}
These twists are equivalent to the first two twists above when the
remaining $SU(2)_R$ symmetry in those theories is further reduced to a $U(1)$ and
that $U(1)$ is twisted with.

As there is not an embedding of $SU(3)$ inside $Sp(4)$ we argue that these are
all of the possible twistings that can be done without assuming that the
$SU(3)_L$ Lorentz group component is further reduced.

\subsection{Calabi-Yau Manifolds}
\label{app:CYsupercharges}

For $M_6$ a Calabi-Yau manifold the holonomy reduces further to $SU(3)$ so
that the total symmetry is
\begin{equation}
  SU(3)_L \times Sp(4)_R \,,
\end{equation}
and the supercharges transform under
\begin{equation}
  ({\bf 3}, {\bf 4}) \oplus ({\bf 1}, {\bf 4}) \,.
\end{equation}
There is immediately, without any requirement of twisting the theory, a set of
four supercharges, which transform in the ${\bf 4}$ of
$Sp(4)_R$ and which are scalars with respect to the Lorentz group
$SU(3)_L$.


\section{Differential Geometry and useful Relations}
\label{app:DiffGeo}

In this appendix we provide some basic identities in differential geometry and some useful self-duality relations on $(m,n)$ forms in 6d and 4d, involving the K\"ahler potentials $J$ and $j$, respectively.

In dimension $d$ with $\omega_{(p)}$ a $p$-form, we define the Hodge star such that
\begin{align}
\star \lp e^{a_1} \wedge \cdots \wedge e^{a_p} \rp &= \frac{1}{(d-p)!} \lp e^{b_1} \wedge \cdots \wedge e^{b_{d-p}} \rp 
\epsilon_{b_1 \cdots b_{d-p}}{}^{ a_1 \cdots a_p } \,, \\
\lp\star \omega_{(p)} \rp_{b_1 \cdots b_{d-p}} &= \frac{1}{p!}\epsilon_{b_1 \cdots b_{d-p}}{}^{ a_1 \cdots a_p } \omega_{a_1 \cdots a_p } \,, \\
\star (\star \omega_{(p)}) &= (-1)^{(d-p)p + t} \omega_{(p)} \,.
\end{align}
with $t=0$ in Euclidean signature and $t=1$ in Lorentzian signature.

With $\omega_1, \omega_2$ two $p$-forms, we have
\begin{align}
\int \star \omega_{1\, (p)} \wedge \omega_{2 \, (p)} &= \int d^dx \sqrt g \, \frac{1}{p!} \, \omega_{1}^{a_1 \cdots a_p} \omega_{2 \, a_1 \cdots a_p}  \,.
\end{align}

Some basic identities:
 \begin{align}
&  (d \omega)_{A_0 A_1 \cdots A_p} = (p+1) \p_{[A_0} \omega_{A_1 \cdots A_p]} \\
 & (\star d \star \omega )_{A_1 \cdots A_{p-1}} = -  D^{A_0} \omega_{A_0 A_1 \cdots A_{p-1}} \\
&  (\omega_1 \wedge \omega_2)_{A_1 \cdots A_p B_1 \cdots B_q} =  \binom{p+q}{p} \, \omega_1{}_{[A_1 \cdots A_p} \, \omega_2{}_{B_1 \cdots B_q]}
 \end{align}

Self-duality properties of $(m,n)$-forms in 6d:
\begin{align}\label{Hodge6d}
& \star \omega_{(3,0)} = i \, \omega_{(3,0)}  \,, \quad  \star \omega_{(0,3)} = - i \, \omega_{(0,3)}  \, \nn\\
& \star \lp  \omega_{(1,0)} \wedge J  \rp =  i \, \omega_{(1,0)} \wedge J  \,, \quad  \star \lp  \omega_{(0,1)} \wedge J  \rp =  -i \, \omega_{(0,1)} \wedge J  \nn\\
&\star \lp  \omega_{(1,2)}  \rp =  i \, \omega_{(1,2)} \quad \textrm{iff} \quad  J \wedge \omega_{(1,2)} = 0 \nn\\
& \star \lp  \omega_{(1,0)} \wedge J \wedge J \rp = 2 i \, \omega_{(1,0)} \nn\\
& \star \lp  \omega_{(0,2)} \wedge J  \rp = - \omega_{(0,2)} \,.
\end{align}
Self-duality properties of $(m,n)$-forms in 4d:
\begin{align}
& \star j = j \,, \nn\\
& \star \omega_{(2,0)} = \omega_{(2,0)}  \,, \quad \star \omega_{(0,2)} = \omega_{(0,2)} \,, \quad
 \star \omega_{(1,1)} = - \omega_{(1,1)} \ {\rm iff} \ \omega_{(1,1)} \wedge j = 0 \, \nn\\
 & \star \omega_{(1,0)} = - i \, \omega_{(1,0)} \wedge j  \,, \quad (\star \omega_{(1,2)} ) \wedge j = i \, \omega_{(1,2)} \,.
\end{align}


\section{Topological Twist and Form-Fields}
\label{app:TwistForms}

We provide some details on the construction of the topologically twisted 6d theory on a K\"ahler three-fold $Y_3$. 
It is useful in particular to understand the decomposition of the spinors. 
To write the spinors in terms of forms on $Y_3$, we make use of the existence in the twisted theory of a covariantly constant spinor of positive chirality that we denote $\eta$
\be
D_{\underline{\mu}} \eta = 0 \,,\qquad D_{\underline{\mu}}= \partial_{\underline{\mu}}+ {1\over 4} \omega_{\underline{\mu}}^{AB} \Gamma_{AB} - i q_R A^{R}_{\underline{\mu}} \,,
\ee
which is covariantized with respect to the non-trivial R-symmetry connection $A^R$ in $U(1)_R$. Here $\underline{\mu}$ is the curved Lorentz index in 6d, while $A,B$ are the flat Lorentz indices. The spinor $\eta$ has charge $q_R =1$. 
We take $\eta$ to be Grassmann-even (commuting spinor) and we normalize it with $\eta^{\dagger}\eta =1$. The complex structure $J$ which satisfies $J^{i}{}_{j} = i \delta^{i}{}_{j}$, $J^{\bar i}{}_{\bar j} = - i \delta^{\bar i}{}_{\bar j}$, can be built (by raising an index) from the K\"ahler $(1,1)$-form
\be
J_{AB} = \  i \,  \eta^{\dagger} \Gamma_{AB} \eta \ = \   i \, \bar{\eta^c} \Gamma_{AB} \eta  \,,
\ee
Importantly the spinor $\eta$ is annihilated by the gamma matrices with holomorphic indices $\Gamma^{i} \eta =0$, $i=1,2,3$, while the charge conjugate spinor $\eta^c = \CC^{-1} \eta^\ast$, which has negative chirality, is annihilated by the gamma matrices with anti-holomorphic indices $\Gamma^{\bar i} \eta^c =0$. This can be rephrased with the help of the projectors 
\begin{align}
P^{A}{}_{B} &= \frac 12 \lp \delta^{A}{}_{B} - i J^{A}{}_{B} \rp \, , \quad  \bar P^{A}{}_{B} =  \frac 12 \lp \delta^{A}{}_{B} + i J^{A}{}_{B} \rp \, ,
\end{align}
which satisfy  $P^{A}{}_{B}P^{B}{}_{C} = P^{A}{}_{C}$, $\bar P^{A}{}_{B} \bar P^{B}{}_{C} = \bar P^{A}{}_{C}$ and $P^{A}{}_{B}\bar P^{B}{}_{C} = 0$, as 
\begin{align}
 P^{A}{}_{B} \gamma^B \eta = 0 \, , \quad  \bar P^{A}{}_{B} \gamma^B \eta^c = 0 \, .
\end{align}
We can also define the holomorphic and anti-holomorphic three-forms:\footnote{In our conventions a holomorphic vector $X$ satisfies $P^A{}_{B} X^B = X^A$, $\bar P^A{}_{B} X^B=0$ and a holomorphic one-form $\omega_{(1,0)}$ satisfies $\omega_B P^B{}_{A} = \omega_A$, $\omega_B \bar P^B{}_{A} =0$. Anti-holomorphic vectors and forms satisfy the opposite conditions.}
\be
\Omega_{ABC} =  \bar{\eta} \Gamma_{ABC} \eta \, , \quad  \bar{\Omega}_{ABC} =  - \bar{\eta}^c \Gamma_{ABC} \eta^c  = (\Omega_{ABC})^\ast \,.
\ee
The three-forms $\Omega, \bar\Omega$ transform under $U(1)_R$ with R-charges $2$ and $-2$ respectively, and therefore are globally defined (anti-)holomorphic three-forms only when the $U(1)_R$ bundle is trival, namely when $Y_3$ is Calabi-Yau (since in the twisted theory the $U(1)_R$ connection is identified with the $U(1)_L$ connection).
We have the relations $J\wedge \Omega = 0$, $- J \wedge J \wedge J = \frac{3i}{4} \Omega \wedge \bar\Omega = 6 \sqrt g \, d^6x$, as well as $\star \, \Omega = i \, \Omega$, $\star \, \bar\Omega = - i \bar \Omega$.
Using the frame $e^A$ we can write explicitly\footnote{The overall sign of $J$ is compatible with $X^0 - i X^1 = X^2 - i X^3 = X^4 - i X^5 = 0$ for a holomorphic vector $\bar P^A{}_{B} X^B = 0$.}
\be\ba
\eta^T &= (1,0,0,0,0,0,0,0) \,, \cr
J \ &= J_{(1,1)} = - \lp e^0 \wedge e^1 +  e^2 \wedge e^3 + e^4 \wedge e^5 \rp  \,, \cr
\Omega_{(3,0)} &=  (e^0 + i e^1)\wedge (e^2 + i e^3) \wedge (e^4 + i e^5) \,.
\label{CanonicalForms}
\ea\ee
For general K\"ahler manifolds that are not Calabi-Yau $d\Omega \not=0$, however we have $D\Omega= D\bar\Omega=0$, with $D = d - i q_R A^R$.

To deal with the spinors $\lambda^m$ we split them into two sets of spinors $\lambda^\alpha_{1}, \lambda^\alpha_{-1}$ corresponding to the decomposition ${\bf 4} \rightarrow {\bf 2}_{1} \oplus {\bf 2}_{-1} $ of the R-symmetry group (see appendix \ref{app:Rsym}), and expand each spinor in the twisted theory in the basis of spinors which is appropriate for the identification of the (anti-)holomorphic forms
\begin{align}
\lambda^\alpha_{-1} &=   \, \Lambda^{\alpha}_i \Gamma^i \eta^c + \, \Lambda^{\alpha}_{ijk} \Gamma^{ijk} \eta^c \,, \quad
\lambda^\alpha_{1} = \Lambda^{\alpha}_{\bar i \bar j} \Gamma^{\bar i \bar j} \eta + \Lambda^{\alpha}  \eta \,,
\end{align}
where we have suppressed the (anti-)holomorphic indices $(1,0), (3,0) , \cdots$. In real indices, with some convenient normalization of the forms, we have
\begin{align}
\lambda^\alpha_{-1} &= \Lambda^{\alpha}_A P^A{}_{B} \Gamma^B \eta^c + \frac{1}{12} \Lambda^{\alpha}_{ABC} P^A{}_{D} P^B{}_{E} P^C{}_{F} \Gamma^{DEF} \eta^c \ = \ \Lambda^{\alpha}_A \Gamma^A \eta^c + \frac{1}{12} \Lambda^{\alpha}_{ABC}  \Gamma^{ABC} \eta^c  \,, \nn\\
\lambda^\alpha_{1} & = \frac{1}{4} \Lambda^{\alpha}_{AB} \bar P^A{}_{C} \bar P^B{}_{D} \Gamma^{CD} \eta + \Lambda^{\alpha}  \eta 
\ = \ \frac{1}{4} \Lambda^{\alpha}_{AB}  \Gamma^{AB} \eta + \Lambda^{\alpha}  \eta \,,
\label{LambdaDecomp}
\end{align}
with the second equalities obtained by making use of the holomorphicity of the forms.  

The spinor equations of motion are $\Gamma^A D_A \lambda^\alpha_{\pm 1} =0$, where $D_{\underline\mu} = \nabla_{\underline\mu} - i q_R A^{R}_{\underline\mu}$ is the covariant derivative including the (non-trivial) $U(1)_R$ connection of the twisted theory.
Plugging the decomposition \eqref{LambdaDecomp} and  using $D \eta  = 0$, the equation $\Gamma^A D_A \lambda^\alpha_{1} =0$ becomes
\begin{align}
0 &= D_A \Lambda^\alpha \, \Gamma^A \eta +\frac{1}{4}  D_A \Lambda^\alpha_{BC} \Gamma^A \Gamma^{BC} \eta \nn\\
&= \lp D_A \Lambda^\alpha  + \frac{1}{2} D^B \Lambda^\alpha_{BA} \rp \Gamma^A \eta + \frac{1}{4} D_{A}\Lambda^\alpha_{BC} \Gamma^{ABC} \eta \,,
\label{eomReduc1}
\end{align}
where we have suppressed the form indices $(m,n)$ and we have used  the identity $\Gamma^A \Gamma^{BC} = \Gamma^{ABC} + 2 \delta^{A[B} \Gamma^{C]}$. 
Contracting with $\bar{\eta}$ and using $\bar{\eta} \Gamma^A\eta = 0$,  we obtain
\begin{align}
0 &= D_{A}\Lambda^\alpha_{BC} \Omega^{ABC} \quad \Rightarrow \quad  (\p + \bar\p) \Lambda^\alpha_{(0,2)} \wedge \star \Omega_{(3,0)} \ = \ 0 \quad \Rightarrow \quad  \bar\p \Lambda^\alpha_{(0,2)} \ = \ 0 \,,
\end{align}
where we have used $\star \Omega_{(3,0)} = i \Omega_{(3,0)}$. 
Using this result one can show that $D_A \Lambda_{BC} \Gamma^{ABC} \eta = 2 D^{B} \Lambda_{BC} \Gamma^C\eta$ and replace the corresponding term in \eqref{eomReduc1}.
Then contracting with $\eta^\dagger \Gamma_C$ we obtain
\begin{align}
0 &= \lp D_A \Lambda^\alpha +  D^B \Lambda^\alpha_{BA} \rp P_{C}{}^{A} \ = \ D_A \Lambda^\alpha \, \bar P^{A}{}_{C} +  D^B \Lambda^\alpha_{BC} \nn\\
& \quad  \Rightarrow \quad  \bar\p \Lambda^\alpha_{(0,0)} -  \star \p \star \Lambda^\alpha_{(0,2)} \ = \ 0 \,,
\end{align}
where we have used $P^{CA} = \bar P^{AC}$ and $D^B \Lambda^\alpha_{BC} = D_A \Lambda^\alpha_{BC} \bar P^{BA} = D_A \Lambda^\alpha_{BC} P^{AB} $. 

The other equations of motion are derived similarly from the equation $\Gamma^A D_A \lambda^\alpha_{-1} =0$, using $D \eta^c = 0$ \footnote{$\eta$ is a constant spinor satisfying $D_A \eta  =  0$.  Taking the complex conjugate and multiplying by $\CB$ we obtain $0 = D_A (\CB^{-1}\eta)= D_A \eta^c $. } and various Gamma matrices identities. The equations of motion of the twisted theory are found to be
\begin{align}
& \bar\p \star \Lambda^\alpha_{(1,0)} \ = \ 0 \,, \quad  \p \Lambda^\alpha_{(1,0)} -  i \star \bar\p \Lambda^\alpha_{(3,0)} \ = \ 0 \,, \nn\\
& \bar\p \Lambda^\alpha_{(0,0)} -  \star \p \star \Lambda^\alpha_{(0,2)} \ = \ 0 \,, \quad  \bar\p \Lambda^\alpha_{(0,2)} \ = \ 0 \, .
\end{align}
In the derivation we used the imaginary self-dual property $\star \Lambda^\alpha_{(3,0)} = i \Lambda^\alpha_{(3,0)}$.



\providecommand{\href}[2]{#2}\begingroup\raggedright\endgroup



\end{document}